\documentclass[twocolumn,trackchanges]{aastex7}
\usepackage{amsfonts,amsmath,amssymb,mathrsfs}
\usepackage{graphicx}
\usepackage{subfigure}
\usepackage{epstopdf}
\usepackage{float}
\usepackage{color}
\usepackage{aastex}
\usepackage{threeparttable}
\usepackage{smartref}
\usepackage{makecell}
\usepackage{multirow}
\usepackage{verbatim}
\usepackage{textcomp}
\usepackage[misc]{ifsym} 
\usepackage{lipsum}
\usepackage{changepage}
\usepackage{tabularx}
\usepackage{booktabs}
\usepackage{cancel}
\usepackage{ulem}

\begin{document}

\title{Theoretical Radio Signals from Radio-Band Gravitational Waves Converted from the Neutron Star Magnetic Field}
\author[0000-0001-7906-0919]{Wei Hong}
\affiliation{Institute for Frontiers in Astronomy and Astrophysics, Beijing Normal University, Beijing 102206, China}
\affiliation{Department of Astronomy, Beijing Normal University, Beijing 100875, China}	
\email{weihong@mail.bnu.edu.cn}

\author[0000-0002-4683-5500]{Zhen-Zhao Tao}
\affiliation{Institute for Astronomical Science, Dezhou University, Dezhou 253023, China}
\affiliation{College of Computer and Information, Dezhou University, Dezhou 253023, China}
\email{taozhenzhao@dzu.edu.cn}

\author{Peng He}
\affiliation{Burerau of Frontier Science and Education, Chinese Academy of Sciences, Beijing 100190, People’s Republic of China}
\email{phe@cashq.ac.cn}

\author[0000-0002-3363-9965]{Tong-Jie Zhang}
\affiliation{Institute for Frontiers in Astronomy and Astrophysics, Beijing Normal University, Beijing 102206, China}
\affiliation{Department of Astronomy, Beijing Normal University, Beijing 100875, China}	
\email[show]{tjzhang@bnu.edu.cn} 

\received{receipt date}\revised{revision date}\accepted{acceptance date}\published{published date}
\submitjournal{ApJ}

\begin{abstract}
Gravitational waves (GWs) can convert into electromagnetic waves in the presence of a magnetic field via the Gertsenshtein-Zeldovich (GZ) effect. The characteristics of the magnetic field substantially affect this conversion probability. This paper confirms that strong magnetic fields in neutron stars significantly enhance the conversion probability, facilitating detectable radio signatures of very high-frequency (VHF, $\left(10^6-10^{11}\mathrm{~Hz}\right)$) gravitational waves. We theoretically identify two distinct signatures using single-dish telescopes (FAST, TMRT, QTT, GBT) and interferometers (SKA1/2-MID): transient signals from burst-like gravitational wave sources and persistent signals from cosmological background gravitational wave sources. These signatures are mapped to graviton spectral lines derived from quantum field theory by incorporating spin-2 and mass constraints, resulting in smooth, featureless profiles that are critical for distinguishing gravitational wave signals from astrophysical foregrounds. FAST attains a characteristic strain bound of $h_c<10^{-23}$, approaching $10^{-24}$ in the frequency range of $1-3\mathrm{~GHz}$ with a 6-hour observation period. This performance exceeds the $5 \sigma$ detection thresholds for GWs originating from primordial black holes (PBHs) and nears the limits set by Big Bang nucleosynthesis. Additionally, projections for SKA2-MID indicate even greater sensitivity. Detecting such gravitational waves would improve our comprehension of cosmological models, refine the parameter spaces for primordial black holes, and function as a test for quantum field theory. This approach addresses significant deficiencies in VHF GW research, improving detection sensitivity and facilitating the advancement of next-generation radio telescopes such as FASTA and SKA, which feature larger fields of view and enhanced gain.
\end{abstract}


\section{Introduction}
\label{introduction}
LIGO has observed the first binary black hole GW event \citep{LIGOScientific:2016aoc,LIGOScientific:2018mvr,LIGOScientific:2020ibl}, heralding the beginning of GW astronomy and opening up new opportunities for cosmic exploration. GWs are expected to be detectable across the entire frequency spectrum, revealing a wide range of discoveries and exhibiting unique physical processes similar to electromagnetic waves \citep{Aggarwal:2020olq}. One of the detections of extremely low-frequency GWs is planned through B-mode polarization of the cosmic microwave background (CMB) by the Ali project \citep{Li:2017drr}. The very low-frequency GWs have been detected via pulsar timing arrays \citep{1990ApJ...361..300F,2008AIPC..983..584M,2013CQGra..30v4010M,2022ApJ...932L..22G,2013CQGra..30v4009K,2022MNRAS.509.5538C,2009arXiv0909.1058J,2020ApJ...905L..34A,2022MNRAS.512.1234N,2022PASA...39...27S,2016ASPC..502...19L,2023RAA....23g5024X,NANOGrav:2023bts,Srivastava:2023mht,Main:2023mgp,EPTA:2023fyk,Reardon:2023gzh,NANOGrav:2023pdq}. Detections at low frequency are going to employ space-based GW interferometers \citep{TianQin:2015yph,Luo:2019zal,10.1093/nsr/nwx116,2012CQGra..29l4016A}, while deci-hertz interferometer detectors are going to be utilized to detect the intermediate-frequency GWs \citep{Kawamura:2006up,Kawamura:2011zz,Kawamura:2018esd,Crowder:2005nr}. And ground-based laser interferometers have been served for high-frequency detection and won the Nobel Prize \citep{Barish:1999vh,LIGOScientific:2007fwp,LIGOScientific:2014pky,VIRGO:2012dcp,VIRGO:2014yos,KAGRA:2013rdx,Aso:2013eba,KAGRA:2018plz}. 

However, plans for exploration at very high-frequencies ($10^6-10^{12}\mathrm{~Hz}$) and ultra-high frequencies (over $10^{12}\mathrm{~Hz}$) are presently deficient. The weakness of GW signals in these frequencies, coupled with exceedingly low photon conversion probabilities, renders detection challenging. Nonetheless, recent studies suggest potential advancements. From an observational perspective, radio telescopes \citep{Domcke:2020yzq, Domcke:2023qle, Ito:2023nkq, Herman:2022fau, Dandoy:2024oqg}, laboratory microwave cavities \citep{Stephenson:2009zz, Li:2014bma, Li:2013fna, Li:2011zzl, Li:2009zzy, Tong:2008rz, Li:2008qr, Li:2004df, Li:2003tv, Li:2000du, Berlin:2021txa, Bernard:2001kp, Ballantini:2003nt, Cruise:2000za}, X/$\gamma$-ray satellites \citep{Ramazanov:2023nxz, Liu:2023mll, Dandoy:2024oqg}, and other detection methods \citep{ShenShen:2024qhi,Panasenko:2024kjp,Schnabel:2024hem,Valero:2024ncz,Antusch:2024ypp,Ratzinger:2024spd,Barrau:2024kcb} are being explored as potential VHF GW detectors for the future detection. Concurrently, the theoretical origins of VHF GWs are notably abundant: the high-frequency band of the primordial GWs \citep{Grishchuk:2005qe,Tong:2009vk,Gasperini:2002bn,Giovannini:1999bh,Giovannini:2008tm,Giovannini:2014jca,Ito:2016aai,Vagnozzi:2022qmc}, inflaton annihilation into gravitons \citep{Bartolo:2015qvr,Barnaby:2010vf,Sorbo:2011rz,Peloso:2015dsa,Kim:2004rp,Cannone:2014uqa,Ricciardone:2016lym,Ananda:2006af,Baumann:2007zm,Cai:2018dig}, (p)reheating after inflation \citep{Caprini:2018mtu,Barman:2023ymn,Ema:2020ggo,Kofman:1997yn,Kofman:1994rk,Khlebnikov:1997di,Garcia-Bellido:1998qth,Garcia-Bellido:2007nns,Garcia-Bellido:2007fiu,Easther:2006gt,Dufaux:2007pt,Dufaux:2008dn}, KK-gravitons from the braneworld scenarios \citep{Nishizawa:2013eqa,Andriot:2017oaz,Clarkson:2006pq,Servin:2003cf} and so on. Therefore, the identification of VHF GWs holds the potential to provide us profound novel insights into the universe, particularly the very early universe, as GWs decouple essentially immediately after being generated \citep{Roshan:2024qnv}.

In this paper, our focus is on detecting VHF GWs within the radio band $10^{6}-10^{11}\mathrm{~Hz}$, by utilizing the inverse GZ effect. This effect delineates the conversion of GWs into electromagnetic waves in the presence of a magnetic field \citep{Gertsenshtein:1962,Cruise:2012zz,Ejlli:2019bqj,Boccaletti:1970pxw,DeLogi:1977qe,Raffelt:1987im,Macedo:1983wcr,Fargion:1995mm,Dolgov:2012be,Ejlli:2020fpt}. Strong magnetic fields convert VHF GWs into photons more significantly. Furthermore, the GZ effect is also widely applied in the research of axion-photon conversion and is used for detecting axions \citep{Hook:2018iia,Walters:2024vaw,Bhura:2024jjt}. They traverse interstellar magnetic fields to our solar system, eventually reaching the radio telescope receiver. Along this trajectory, VHF GWs and their conversed photons pervade the universe, thereby improving the quality of our observations through enhanced time-series data and heightened resolution. Our detection outcomes will provide ground-based laboratories with accessible VHF GW sources, thereby constituting collaborative observations. And, we utilize neutron stars with magnetic fields to calculate the entire conversion process and estimate the sensitivity of detecting GWs. In this work, we adopt a simplified theoretical model that assumes alignment between the neutron star's rotational and magnetic axes, thereby neglecting the inclination angle inferred from observational measurements. This approximation is made for the purpose of initial investigation. However, certain physical phenomena can influence the conversion probability, such as the diffraction of GWs by the neutron star. We use toy modeling in this paper to provide a brief discussion of the diffraction of GWs by the neutron star. In addition, we will examine the mechanisms that affect the conversion probability in additional articles. These mechanisms include diffraction and refraction of electromagnetic waves in the plasma surrounding the neutron star, as well as possible interference caused by speed differences between gravitational waves and electromagnetic waves in the magnetic field.

Employing pulsars or magnetars to detect GWs of different frequencies seems to improve the observation efficiency of pulsars or magnetars. Nanohertz GWs have been found in pulsar's ``Fold-mode" data, and VHF GWs can be identified in baseband data due to our reliance on comprehensive electromagnetic observations around neutron stars. The observations of nanohertz and VHF GWs can be made at the same time, and the observations are not affected. In this paper, we investigate the potential of combining strong magnetic fields of celestial bodies with radio telescopes to observe VHF GWs, especially from mergers of primordial black holes, where detection sensitivity has been significantly improved. 

The rest of paper is organized as follows: In Section \ref{methods}, we analyze the conversion of radio signals from GWs in the strong magnetic fields of a single neutron star and investigate the diffraction of GWs by the neutron star on the photon-specific intensity. In Section \ref{sec:sslcafc}, we calculated the spectral line broadening of the graviton and the frequency-dependent variation of the conversed electromagnetic waves. In Section \ref{sec:stop}, we calculated the equivalent system flow and the signal-to-noise ratio of the utilized telescope. In Section \ref{Results}, we present the calculated expected gravitational wave radio signals and the detection sensitivities of six different telescopes, including the anticipated radio signals on two time scales derived from our calculations. In Section \ref{conclusion}, we present a summary and discussion of this paper. Finally, the Appendix contains essential computational procedures, unit conversions, telescope parameters, and derived results. This paper employs the natural unit system where $c=\hbar=\epsilon_{0}=\mu_{0}=1$.

\section{Gravitational wave conversion probability}
\label{methods}
In magnetic fields, gravitational waves are converted to electromagnetic waves by the GZ effect. In this section, we will demonstrate in turn the magnitude of this conversion probability in the magnetic field of an isolated neutron star and the periodic modulation of the gravitational wave intensity caused by the gravitational wave diffraction of this neutron star, which ultimately results in the same level of periodic modulation of the electromagnetic wave conversed by the GZ effect.

\subsection{Gravitational-electromagnetic wave mixing and photon specific intensity}
\label{subsec:gw-photon mixing and photon specific intensity}
Now, we show the probability of converting GWs into electromagnetic waves in a typical neutron star magnetic field. In this paper, we only consider the simpler, rapidly rotating magnetic field structures. In our subsequent paper, we will simulate the entire neutron star’s magnetic field using the particle-in-cell method and refine the simulated field with current observational data \citep{Hong:2025tbd}. In the context of rapidly rotating magnetic fields, pulsars and magnetars can often be approximated as rotating magnetic dipoles, with the magnetic dipole axis aligned to the rotation axis, represented as $B(r) \sim B\left(r_0\right)\left(r_0 / r\right)^3$, where $r_0=10\mathrm{~km}$ is the radius of neutron stars and the magnetic dipole axis is aligned with the rotation axis $B(\theta)=\frac{B_0}{2}\left(3 \cos^2 \theta+1\right)^{1/2}$ where $\theta$ represents the polar angle from the rotation and magnetic axis and $B_0$ is the surface magnetic field at the poles. While the Goldreich-Julian model (GJ model) was first suggested for aligned neutron stars with $\theta_m=0$, it works just as well for oblique neutron stars \citep{Goldreich:1969sb}. It gives a charge density of $n_c=\frac{2 \boldsymbol{\Omega} \cdot \mathbf{B}}{e} \frac{1}{1-\Omega^2 r^2 \sin ^2 \theta}$, where $\Omega$ is the neutron star's spin period and $\theta$ is its polar angle with respect to the axis of rotation. We shall use the charge density as an approximate measure of the electron number density: $n_e=\left|n_c\right|$. Therefore, the plasma frequency is denoted as $\omega_\mathrm{plasma}=\left(1.5\times 10^{11}\right) \sqrt{\left(\frac{B_\perp x_s(r)}{10^{14} \mathrm{~Gauss}}\right)\left(\frac{1 \mathrm{~sec}}{P}\right)}\mathrm{~Hz}$ where $B_\perp x_s=\left(B_0 / 2\right)\left(r_0 / r\right)^3\left[3 \cos \theta\mathbf{m}\cdot\mathbf{r}-\cos \theta_m\right]$ is the component of the magnetic field along the perpendicular direction of traveling direction $+x_s$ of GWs, and $\mathbf{m}\cdot \mathbf{r}=\cos \theta_m \cos \theta+\sin \theta_m \sin \theta \cos (\Omega t)$ depends on time due to the rotation of the neutron stars. As we only consider the cold plasma scenario in order to simplify the model, relativistic effects on the plasma frequency correction are not relevant. Electromagnetic wave propagation in plasma is based on the premise that the frequency of the electromagnetic wave is higher than that of the plasma. This allows us to calculate the minimum spatial scale of the inverse GZ effect occurring radially using the formula
\begin{equation}
	\begin{aligned}
&r_{\mathrm{occur}}=2.24\times 10^4\left|3 \cos \theta \mathbf{m} \cdot \mathbf{r}-\cos\theta_m\right|^{1 / 3}\\
&\times\left(\frac{r_0}{10 \mathrm{~km}}\right)\left( \frac{B_0}{10^{14} \mathrm{~Gauss}}\right)^{1/3}\left(\frac{1 \mathrm{~sec}}{P}\right)^{1/3} \left(\frac{10^6\mathrm{~Hz}}{\omega}\right)^{2/3}\mathrm{~km}.
	\end{aligned}
\end{equation}
It is important to note that when the GW frequency is higher, the calculated result of $r_{\mathrm{occur}}$ will be less than the radius of the neutron star $r_0$, which is not physical. Therefore, when the calculated result of $r_{\mathrm{occur}}$ is less than the neutron star radius $r_0$, we fix $r_{\mathrm{occur}}$ to $r_0=10\mathrm{~km}$. Obviously, when the radius is greater than $r_{\mathrm{occurs}}$, the GZ effect still exists, but the conversion probability decreases as the neutron star's magnetic field strength decreases.

GWs are converted into photons in neutron stars' magnetic fields. However, some neutron stars' magnetic fields are stronger than the critical magnetic field $B_\mathrm{critical}\equiv m_e^2 c^3/\hbar|q_f|=m_e^2/|q_f|=4 \times 10^{13} \mathrm{~Gauss}$ \citep{Lai:2000at,Lai:1995eq}, which will lead to some additional physical processes, where $|q_f|$ denotes the elementary charge. In the presence of an external field, the interaction of observables must be analyzed due to the resummation of higher-order diagrams, leading to nonlinear dependencies known as ``nonlinear QED". Within the frequency range of $10^{6}-10^{11} \mathrm{~Hz}$, the photon's wavelength is $\lambda_{\gamma}\approx3 \times 10^{-4}-300 \mathrm{m}$, which is much larger than the electron's Compton wavelength: $\lambda_{e}=\frac{\hbar}{m_e c}=\frac{1}{m_e}=2.426 \times 10^{-12} \mathrm{m}$. Additionally, their energy $E_{\gamma}\approx 10^{-10}-10^{-5} \mathrm{eV}$, significantly smaller than the electron's mass $m_{e}=0.511 \mathrm{MeV}$, thereby validating the Heisenberg-Euler effective Lagrangian \citep{Heisenberg:1936nmg,Schwinger:1951nm}. Consequently, the action of GW-photon conversion within the proper time integral is \citep{Tsai:1974ap,Urrutia:1977xb,Adler:1971wn,Tsai:1974fa,Tsai:1975iz,Melrose:1976dr,Dittrich:1985yb,Dittrich:2000zu}
\begin{equation}
S= \frac{1}{16\pi G}\int d^4 x \sqrt{-g}\left[R+L_{\mathrm{eff}}\right],\label{effect-action}
\end{equation}
where $R$ represents the Ricci scalar and $g$ denotes the determinant of the metric $g_{\mu \nu}$. The effective Lagrangian is given by $L_{\mathrm{eff}}=L^{(0)}+L^{(1)}$, where $L^{(0)}=-\frac{1}{4}F_{\mu\nu}F^{\mu\nu}$ is the original Maxwell Lagrangian and $L^{(1)}$ is equal to $\frac{im_e^4}{2\hbar^3}\int_0^{\infty}\frac{d\tau}{\tau} \mathrm{e}^{-\epsilon\tau} \mathrm{e}^{-i\tau m_e^2} \operatorname{tr}\left[\left\langle x\left|\mathrm{e}^{-i \hat{H}\tau}\right| x\right\rangle-\left\langle x\left|\mathrm{e}^{-i \hat{H}_0 \tau}\right| x\right\rangle\right]$, and represents the effective action for the gauge field $A_{\mu}$. In $L^{(1')}$, $\tau$ represents the proper time, $\epsilon>0$ is an infinitesimal parameter used for the resummed of the effective action, $tr$ signifies the remaining trace over the Dirac spinor space, and $\hat{H}=D^2+1/2q_fF^{\mu\nu}\sigma_{\mu\nu}$ denotes the Hamiltonian. $D^{\mu}=\partial^{\mu}+iq_fA^{\mu}(x)$, $\sigma_{\mu\nu}=i/2\left[\gamma_\mu,\gamma_\nu\right]$, and $\hat{H}_0=\partial^{2}$ indicates the free Hamiltonian \citep{Tsai:1974fa,Tsai:1975iz,Hattori:2012je,Hattori:2012ny}. Under the one-loop effective action assumption and the Bianchi identity with the relation $F^4+2\mathcal{F} F^2-\mathcal{G}^2=0$, we can expand the Lagrangian $L^{(1)}$ to first order in the form with the heat-kernel method $L^{(1)}=\partial_\sigma F_{\alpha \beta} \partial_\rho F_{\gamma \delta} \mathcal{L}_1^{\sigma \alpha \beta \gamma \delta \rho}\left(F_{\mu \nu}\right)$, where the $\mathcal{L}_k$'s denote local functions of the field strength tensor $F_{\mu \nu}$, and the background gauge is assumed \citep{Gusynin:1995bc,Gusynin:1998bt,Vassilevich:2003xt}. Moreover, the two Lorentz invariants of mass-dimension four are $\mathcal{F}\equiv \frac{1}{4} F_{\mu \nu} F^{\mu \nu}=\frac{1}{2}\left(\boldsymbol{B}^2-\boldsymbol{E}^2\right)$ and $\mathcal{G}\equiv\frac{1}{4} F_{\mu \nu} \tilde{F}^{\mu \nu}=-\boldsymbol{B} \cdot \boldsymbol{E}$. Then, the remaining trace term in $L^{(1)}$ can be obtained as $\frac{\mathcal{G}}{4 \pi^2}\cot \left(\tau f_1\right) \cot \left(\tau f_2\right)\left[1+\mathcal{H}^{\mu \nu}+\cdots\right]$, where $\mathcal{H}$ is $F^i F_\mu F^j F_\nu F^k Y_1^{i j k}+F^i F_\nu F^j F_\mu Y_2^{i j k}+F^i\mathrm{tr}\left[ F_\mu^* F_v F^j\right] Y_3^{i j}$. In term $\mathcal{H}$, $F_\lambda$ denotes the matrix form of the tensor $\nabla_\lambda F_{\mu \nu}$, $Y_n^{i j k}$ and $Y_m^{i j}(i, j, k=0,1,2,3)$ are functions of $\mathcal{F}$, $\mathcal{G}$, and $\tau$, and summation over $i, j$, and $k$ is assumed. Finally, we can express the gauge potential in terms of the electromagnetic fields from the field strength tensor as $A_\nu^{\mathrm{FS}}(x) =\sum_{n=0}^{\infty} \frac{\left(x^\mu-x_0^\mu\right)\left(x^{\sigma_1}-x_0^{\sigma_1}\right) \cdots\left(x^{\sigma_n}-x_0^{\sigma_n}\right)}{n!(n+2)} \partial_{\sigma_1} \partial_{\sigma_2} \cdots \partial_{\sigma_n} F_{\mu \nu}\left(x_0\right)$. Combining with the electrodynamics equations in curved spacetime, we can calculate the linearized equations of motion for GWs and photons \citep{Boccaletti:1970pxw}. At the same time, some enlightening calculations of the GZ effect can also be referred to in these papers that discuss the conversion of axions into photons \citep{Hook:2018iia,Walters:2024vaw,Bhura:2024jjt}.

Given that the inverse GZ effect is most pronounced when the direction of the magnetic field is perpendicular to the direction of the GW propagation. As is well known, GWs can be expressed with the traceless transverse gauge, which means $h_{0 i}=0$, $\partial^j h_{i j}=0$, and $h_i^i=0$. The Euler Lagrange equations of motion from the Heisenberg-Euler effective Lagrangian (\ref{effect-action}) for the propagating photon and graviton fields components, $A^\mu$ and $h_{i j}$ propagating in the external magnetic field, are generally read as $\nabla^2 A^0=0$, $\boldsymbol{A}^i-L^{(1)}+\partial^i \partial_\mu A^\mu=\sqrt{16\pi G}\partial_\mu\left[h^{\mu \beta} F_\beta^i-h^{i \beta} F_\beta^\mu\right]$, and $\square h_{i j} =-\sqrt{16\pi G}B_i B_j\partial_{i} A_j$. When we consider the case where gravitational waves pass through along the rotational axis, the above equation can be further simplified. The linearized equations of motion for GWs and photons propagating along the direction $x_s$, perpendicular to the background magnetic field, are derived as follows as the Coulomb gauge condition makes $\partial_i \boldsymbol{A}^i=0$ and we can also choose $A^0=0$ \citep{Raffelt:1987im, Dolgov:2012be, Maggiore:1999vm}
\begin{equation}
	\begin{aligned}
		&\left(\partial_t^2-\partial_s^2\right) h_{i j}^{\mathrm{TT}}(t, s) =16\pi G\left( F_{i\rho}F^{\rho}_{j}-\frac{\eta_{ij}}{4}F_{\alpha\beta}F^{\alpha\beta}\right), \\
		&\left(\partial_t^2-\partial_s^2+\Delta_\omega^2\right) A_i(t, s) =\delta^{k j}\delta^{l m}\left(\partial_s h_{i j}^{\mathrm{TT}}\right) \epsilon_{ksl} B_m,\label{linearized-eq}
	\end{aligned}
\end{equation}
where TT means transverse and traceless, so $h_{\times}=h_{12}^{\mathrm{TT}}=h_{21}^{\mathrm{TT}}$ and $h_{+}=h_{11}^{\mathrm{TT}}=-h_{22}^{\mathrm{TT}}$, similar to vector potential $A_i$. The total frequency term $\Delta_\omega^2=\omega_{\mathrm{plasma}}^2-\omega_{\mathrm{QED}}^2$ involves the plasma and QED effect frequencies \citep{Adler:1971wn,Heisenberg:1936nmg,Itzykson:1980rh,1960ecm..book.....L}. Moreover, we also consider the cyclotron frequency $\omega_{\mathrm{cyc}}$ within magnetar. In this paper, we consider cold plasma, and its frequency is $\omega_{\mathrm{plasma}}=\sqrt{\frac{4 \pi \alpha n_e}{m_e}}$ where $\alpha=\frac{q_f^2}{4\pi \epsilon_{0}\hbar c}=\frac{q_f^2}{4\pi}$ is the fine structure constant and $n_{e}$ is the electron number density. It is worth noting that the frequency of photons $\omega_{\gamma}$ needs to be greater than the plasma frequency $\omega_{\mathrm{plasma}}$ to propagate in the plasma, indicating a minimum detectable frequency $\omega_{\mathrm{GW}}>\omega_{\mathrm{plasma}}$. The QED effect frequency is derived from the vacuum polarization tensor and corresponds to the refractive indices. For an ordinary magnetic field, the QED effect frequencies are $\omega_{\mathrm{QED},\times}^{\mathrm{normal}}=\sqrt{\frac{28\alpha^2 B^2\omega^2 }{45 m_e^4}}$, and $\omega_{\mathrm{QED},+}^{\mathrm{normal}}=\sqrt{\frac{16\alpha^2 B^2\omega^2 }{45 m_e^4}}$. For a strong magnetic field whose $B>B_\mathrm{critical}$, the Coulomb force on electrons is a small perturbation compared to the magnetic force \citep{Lai:2000at}. Therefore, we calculate the QED effect frequency under a wrenchless field state with a pure magnetic field \citep{Heyl:1996dt,Heyl:1997hr,Dunne:2004nc,Andersen:2014xxa}. These frequencies are denoted as $\omega_{\mathrm{QED},\times}^{\mathrm{beyond}}=\sqrt{\frac{2\alpha\left\{\rho^2 \tilde{X}^{(2)}(\rho)-\rho \tilde{X}^{(1)}(\rho)\right\} \omega^2}{4\pi}}$ and $\omega_{\mathrm{QED},+}^{\mathrm{beyond}}=\sqrt{\frac{2\alpha\left\{2 \tilde{X}(\rho)-\rho \tilde{X}^{(1)}(\rho)-\frac{2}{3} \tilde{X}^{(2)}(\rho)+\frac{2}{9 \rho^2}\right\} \omega^2}{4\pi}}$, where $\tilde{X}(\rho)=\left(\frac{\rho}{2}\right)^2-\frac{1}{3}+4 \zeta\left(-1, \frac{\rho}{2}\right) \ln \frac{\rho}{2}+4 \zeta^{\prime}\left(-1, \frac{\rho}{2}\right)$, $\tilde{X}^{(n)}(\rho) \equiv d^n \tilde{X}(\rho) / d \rho^n$, $\rho=B_\mathrm{critical}/|B|$, and $\zeta(z, a)=\sum_{n=0}^{\infty}(n+a)^{-z}$ is the Hurwitz zeta function \citep{Elizalde:1986}. It is worth noting that for weaker magnetic fields and lower detection frequencies, it is calculated that the QED effect can be ignored. But with an increase of magnetic field intensity and detection frequency, the QED effect gradually becomes significant. In the analysis of the inhomogeneous wave equation system (\ref{linearized-eq}), one can typically determine the eigenmodes of the associated homogeneous problem by employing the method of separation of variables in conjunction with Duhamel’s principle. Accordingly, we first assume that the photon propagates a short distance $s$ along the $+x_s$ direction, which is perpendicular to the background magnetic field. Under this assumption, the magnetic field within this region can be regarded as slowly varying and nearly uniform. That is, the temporal variation of the neutron star's magnetic field at this short distance is much slower than the characteristic frequency of the photon. The solution of Eq. (\ref{linearized-eq}) with a single frequency mode $\omega$ can be expressed as $A_{\omega,\lambda}(t, s)=\hat{A}_{\omega,\lambda}(s) e^{-i \omega t}, \quad h_{\omega,\lambda}(t,s)=\hat{h}_{\omega,\lambda}(s) e^{-i \omega t}$ where $\lambda=\times$ and $+$. Hence, Eq. (\ref{linearized-eq}) is restated as
\begin{equation}
	\left[\omega^2+\partial_s^2+\mathcal{M}\right]\left(\begin{array}{c}
		\hat{A}_{\omega,\lambda}(s) \\
		\hat{h}_{\omega,\lambda}(s)
	\end{array}\right)=0,\label{rewritten-linearize-eq}
\end{equation}
where $\mathcal{M}$ is the mixing mass matrix \citep{Raffelt:1987im,Domcke:2020yzq,Liu:2023mll,Ejlli:2020fpt}
\begin{equation}
\mathcal{M}\equiv\left(\begin{array}{cc}
		\frac{-\Delta_\omega^2}{ 2 \omega} & \frac{\sqrt{16\pi G}B_\mathrm{eff}}{2}\\
	\frac{\sqrt{16\pi G}B_\mathrm{eff}}{2}& 0
	\end{array}\right),
\end{equation}
where $B_\mathrm{eff}$ represents the effective magnetic field strength related to the magnetic field distribution at the local level. Similar to the WKB limit \citep{1948RvMP...20..399E,1926ZPhy...38..518W}, Eq. (\ref{rewritten-linearize-eq}) is reduced to a linearized system $\psi \equiv\left(\begin{array}{l}\psi_1 \\ \psi_2\end{array}\right) \equiv U\left(\begin{array}{c}\hat{A}_{\omega,\lambda}(s) \\ \hat{h}_{\omega,\lambda}(s)\end{array}\right)$ with $\left[\omega-i \partial_s+m_j\right] \psi_j=0$ for $j=1,2$. This system yields an exact solution $\psi_j(s)=e^{i\left(\omega+m_j\right) s} \psi_j(s=0)$, where $m_j$ represents the eigenvalue of $\mathcal{M}$ and $U$ denotes the eigenvector matrix that is associated with the scattering cross section from the particle collider. Then, the GW-photon conversion process is obtained by solving $\hat{A}_{\omega,\lambda}(s)$ with the condition $\hat{A}_{\omega,\lambda}(0)=0$, so
\begin{equation}
	\begin{aligned}
		\hat{A}_{\omega,\lambda}(s)&=\frac{i 4\sqrt{G \pi} B_\mathrm{eff}\omega  \left[e^{i(\omega+m_1)s}-e^{i(\omega+m_2)s}\right]}{\sqrt{64G\pi B_\mathrm{eff}^2 \omega^2+\Delta_\omega^4}} \hat{h}_{\omega,\lambda}(0)\\
		&=\frac{8\sqrt{G \pi} B_\mathrm{eff}\omega}{\sqrt{64G\pi B_\mathrm{eff}^2 \omega^2+\Delta_\omega^4}}\sin\left(\frac{m_1-m_2}{2}s\right)\\
		&\times e^{i\hat{\theta}(s)}e^{i\omega s}\hat{h}_{\omega,\lambda}(0),
	\end{aligned}\label{sp-comb}
\end{equation}
where $\hat{\theta}(s)=\arctan\left[\tan\left(\frac{m_1+m_2}{2}s\right)+\pi\right]$.

\begin{figure}
	\centering
	\includegraphics[width = 0.3\textwidth]{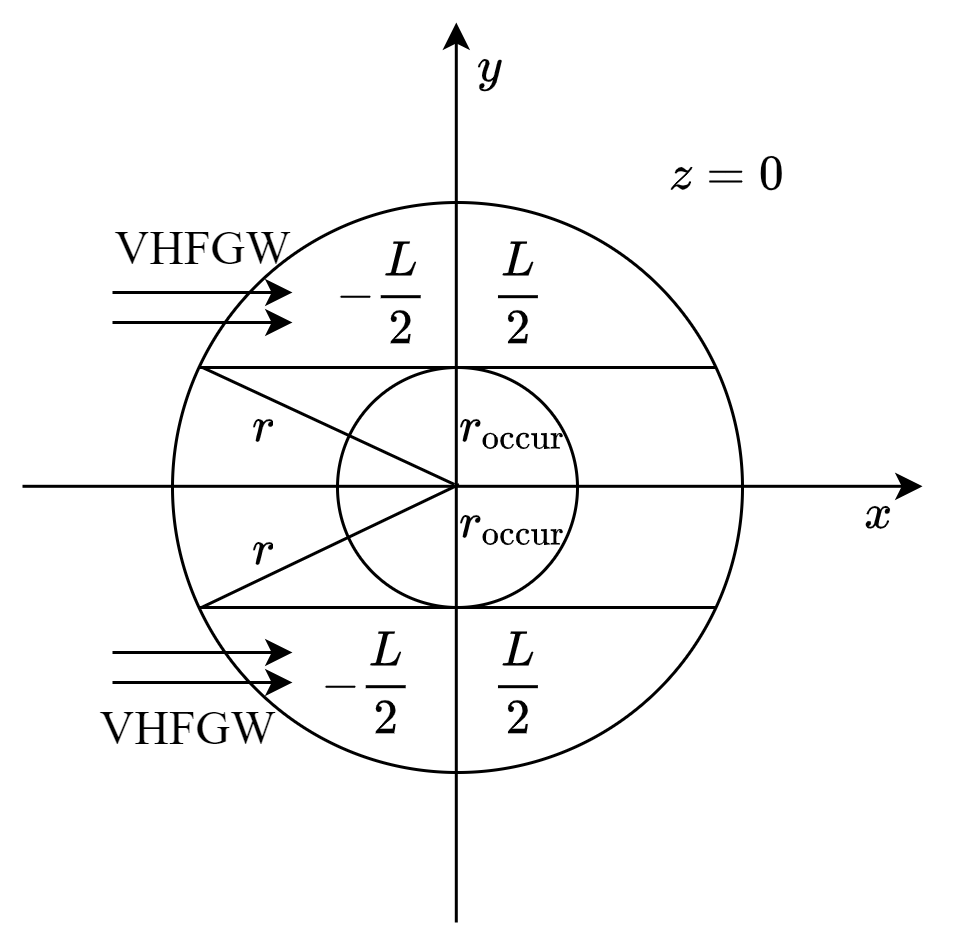}
	\caption{The top view illustrates the movement of GWs across a magnetar or pulsar equator in three-dimensional coordinates.}
	\label{fig:direction-dig}
\end{figure}

Furthermore, the distance over which gravitational waves propagate through the magnetosphere of a neutron star is evidently much larger than the infinitesimal displacement $s$, rendering the approximation of a homogeneous magnetic field invalid. The spatial inhomogeneities in both the magnetic field and the electron density lead to position-dependent coefficients in the wave equation, necessitating a perturbative treatment of Eq. (\ref{linearized-eq}). This can be achieved by transforming to the interaction picture via $\hat{A}_{\mathrm{int }}=U^{\dagger} \hat{A}_{\omega,\lambda}$, allowing the full solution to be systematically constructed through iterative expansion from usual iteration. In practice, a first-order approximation using the distorted wave function approach proves sufficient, wherein conversion probabilities between regions of locally homogeneous magnetic field and electron density are incorporated. Finally, the probability that a GW with frequency $\omega$ traveling a distance $L$ at the polar angle $\theta$ of a neutron star is converted into photons is
\begin{equation}
\begin{aligned}
&P_{g \rightarrow \gamma}(L,\omega,\theta) =\left|\left\langle\hat{A}_{\omega,\lambda}(L) \mid \hat{h}_{\omega,\lambda}(0)\right\rangle\right|^2\\
&=\left|\int_{-L/2}^{L/2} d l' \frac{\sqrt{2}B_\mathrm{eff}(l',\theta)}{2M_{\mathrm{planck}}}\exp \left(-i \int_{-L/2}^{l'} d l'' \frac{-\Delta_\omega^2(l'')}{ 2 \omega}\right)\right|^2.
\label{convertion-probability}
\end{aligned}
\end{equation}
For readability, please refer to the Appendix \ref{sec:function-detail} for the detailed calculation of Eq. (\ref{convertion-probability}). In theory, the typical distance $L$ should be influenced by the absorption and scattering of matter in the magnetic field. However, given that the interaction cross-section of gravitons is $\sigma\approx l_p^2$ with four-dimensional Planck length $l_p$ \citep{Palessandro:2019tmj,Boughn:2006st}, and this value is generally unaffected by the internal structure of gravitons \citep{Sawyer:2019plp}, coupled with the absence of extreme gravitational objects like black holes in our observation path, we infer that the absorption and scattering coefficients of gravitons are minimal and can be disregarded. Therefore, the distance $L$ is solely determined by the graviton's energy and the intensity, orientation, and magnitude of the magnetic field. Moreover, due to the absorption and scattering effects of the medium on photons, the conversion probability from a single GW to a photon is not equivalent to the conversion probability from a single photon to a GW. So we need to calculate the length of this typical distance $L$. In Fig. \ref{fig:direction-dig}, we show the minimum radius $r_{\mathrm{occur}}$ at which GWs cross the magnetar equator and radio-observable converted photons. Start with the simplest, we can now determine the conversion probability of neutron stars for a particular path at the location of $r_{\mathrm{occur}}$ at their equators $\theta=0$. 

For electromagnetic waves travelling through plasma, the propagation speed of the wave energy is described as the group velocity $v_g =c \mu=c \sqrt{1-\omega_{\mathrm{plasma}}^2 / \omega^2}<c$. Obviously, in the neutron star magnetosphere, the speed of the GW and the generated electromagnetic wave are not consistent, so the electromagnetic wave generated by the GZ effect will cause interference in the path of the GW. Instead of calculating complex interference phenomena, we can turn to calculating the coherence lengths $L_{\mathrm{coherence}}$ of electromagnetic waves in different regions of the neutron star's magnetic field \citep{Fargion:1995mm}
\begin{equation}
	L_{\mathrm{coherence}} \approx 3 \times 10^{17}\left(\frac{\omega}{10^{6} \mathrm{~Hz}}\right)\left(\frac{n_e}{10^{-7} \mathrm{~cm}^{-3}}\right)^{-1} \mathrm{~cm}.
\end{equation}
In the magnetosphere and the surrounding magnetic field of a neutron star, the coherence length $L_{\mathrm{coherence}}$ is always significantly smaller than the travel distance $L$ of the GW. Therefore, in the travel distance of GWs with length $L$, the conversion probability $P_{g \rightarrow \gamma}^{\mathrm{coherence}}(L,\omega,\theta)$ considering electromagnetic wave interference phenomenon can be approximated by the total number of coherent domains $\eta_\mathrm{coherence}=L/L_{\mathrm{coherence}}$ and the conversion probability $P_{g \rightarrow \gamma}(L,\omega,\theta)$ without considering interference conditions
\begin{equation}
	P_{g \rightarrow \gamma}^{\mathrm{coherence}}(L,\omega,\theta) \approx \eta_\mathrm{coherence} P_{g \rightarrow \gamma}(L,\omega,\theta).
\end{equation}

\begin{figure*}
	\centering
	\includegraphics[width=0.48\linewidth]{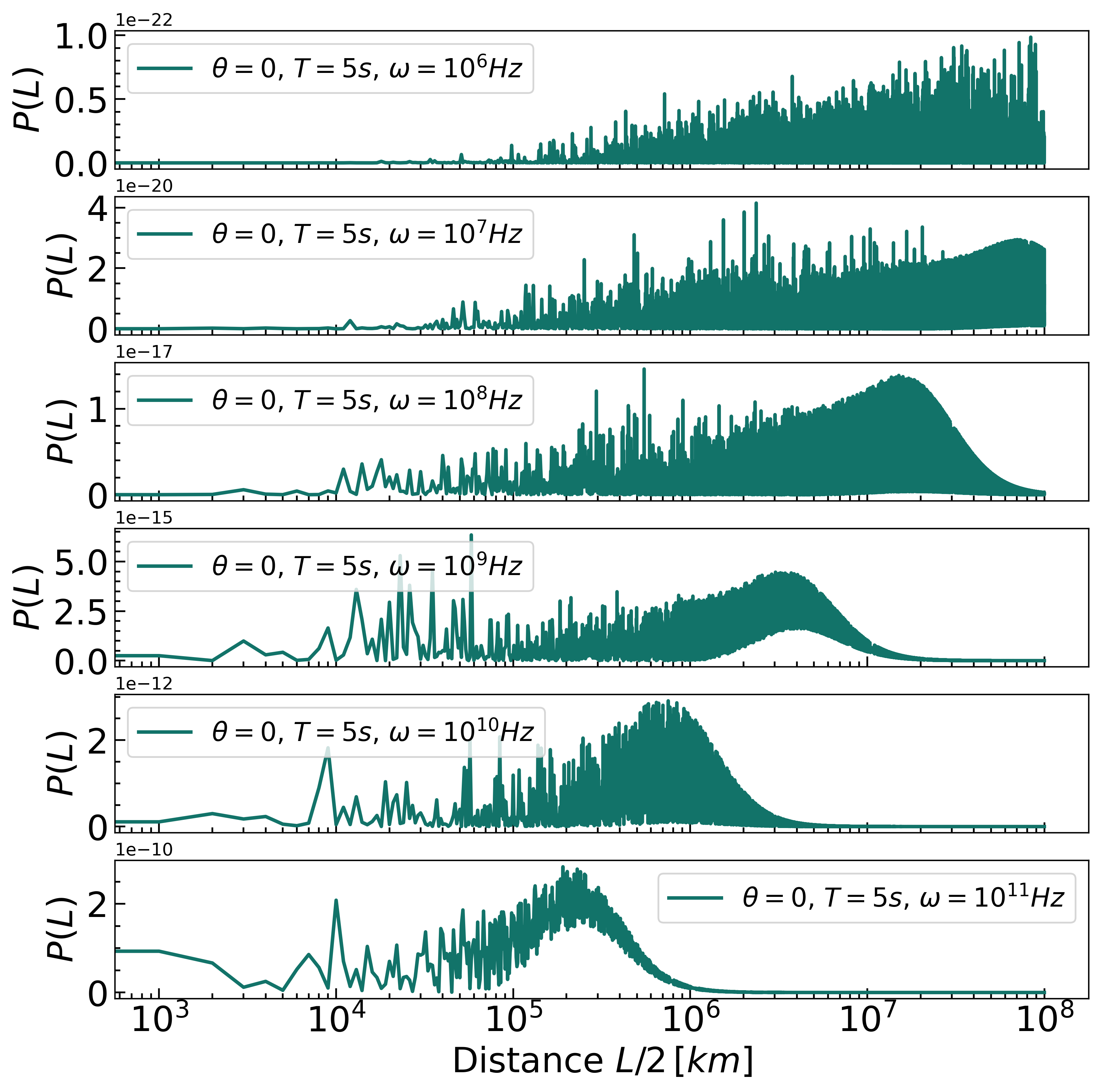}
	\caption{The conversion probability $P_{g \rightarrow \gamma}^{\mathrm{coherence}}(L,\omega,\theta=0)$ pertains to the cross-polarization (``$\times$"-polarization) of gravitational waves as they traverse varying distances within a neutron star's magnetic field. The assumed parameters of the magnetic field are $B=10^{14}\mathrm{~Gauss}$ and $T=5\mathrm{~sec}$. The panel illustrates, from top to bottom, the outcome of a 10-fold increase in the frequency of the radio telescope over time, alongside a gradual increase in the overall conversion probability. Simultaneously, it is evident that as frequency increases, the location of the maximum conversion probability shifts closer to the neutron star radius}
	\label{fig:conversion-probability-times}
\end{figure*}

\begin{figure*}
	\centering
	\includegraphics[width=0.48\linewidth]{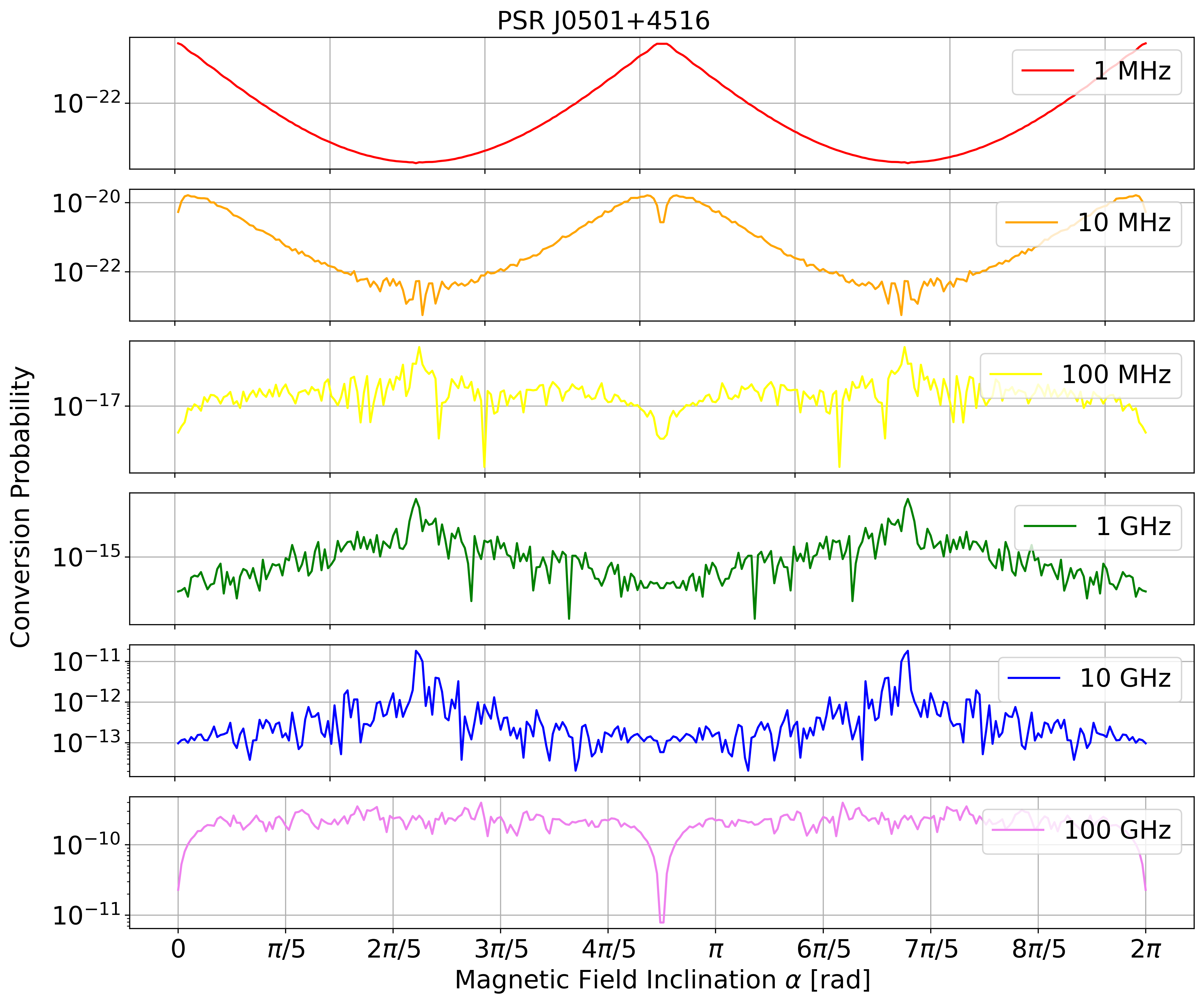}
	\caption{The conversion probability $P_{g \rightarrow \gamma}^{\mathrm{coherence}}(L,\omega,\theta=0)$ pertains to the cross-polarization (``$\times$"-polarization) of gravitational waves as they traverse varying inclinations within a neutron star's magnetic field. The assumed parameters of the magnetic field are $B=10^{14}\mathrm{~Gauss}$ and $T=5\mathrm{~sec}$. The panel illustrates, from top to bottom, the outcome of a 10-fold increase in the frequency of the radio telescope over a specific period, revealing a clear periodic variation in the conversion probability.}
	\label{fig:conversion-probability-times-inclinations}
\end{figure*}

To estimate the order of magnitude of the conversion probability $P_{g \rightarrow \gamma}^{\mathrm{coherence}}(L,\omega,\theta=0)$ for a single GW traveling along the equator, we preselected neutron star data from the ATNF Pulsar Catalogue \citep{2005AJ....129.1993M,anft} and the McGill Online Magnetar Catalog \citep{2014ApJS..212....6O,mcgill}. It is listed in Table \ref{tab:pulsar-list}. Our selection of these pulsars or magnetars is based on the distance of the neutron stars and the strength of the magnetic field. Estimating the order of magnitude of the conversion probability can be divided into two steps. Firstly, we use this table to select neutron stars with magnetic fields $B_\mathrm{eff}$ of $10^{15}\mathrm{~Gauss}$, $10^{14}\mathrm{~Gauss}$, $10^{13}\mathrm{~Gauss}$, spin periods $P$ of 10 seconds, 5 seconds, and 1 second, and distances of $2\mathrm{~kpc}$, $1\mathrm{~kpc}$, $0.5\mathrm{~kpc}$. We then use the information to calculate the radio signal conversed by GWs in the neutron star's magnetic field along this path and the simple radiation mechanism. We have categorized the results of our calculations into cross-polarization and plus-polarization of conversion probability, as fully shown in Figs. \ref{figapp:conversion-probability-times} and \ref{figapp:conversion-probability-plus} respectively, in Appendix \ref{sec:more-figures} for readability, and only shown one diagram for comparison in Fig. \ref{fig:conversion-probability-times}. Meanwhile, under identical parameter settings, we computed the conversion probabilities of gravitational waves at various deflection angles at the radius $r=r_{\mathrm{occur}}$ of the neutron star. The calculation results for various frequencies are presented in Fig. \ref{fig:conversion-probability-times-inclinations}. Figure \ref{fig:conversion-probability-times-inclinations} illustrates that the conversion probability exhibits periodic variation in relation to the magnetic field distribution of neutron stars.

 Because of the space-accumulation effect of the GZ effect and the change in the magnetar's magnetic field's direction, we can see from the two figures that the conversion probability first rises and then falls as the GW passes through the magnetar for a specific distance \citep{Li:2017jcz}. These two figures also show that the distance with the highest probability of GW-photon conversion is between $\sim10^{6}$ and $\sim10^{8}\mathrm{~km}$ when the frequency of the GW is between $10^{6}\mathrm{~Hz}$ and $10^{11}\mathrm{~Hz}$. This distance exceeds the radius of a neutron star's light cylinder, which is $R_{\mathrm{LC}}=c P / 2 \pi \simeq 5 \times 10^9 \mathrm{~cm} P_0$ \citep{2020ApJ...901L..13Y}. Additionally, as the magnetar's spin period and GW frequency increase, this distance gets smaller. It is important to note that some of the finer magnetic field structures in the neutron star magnetosphere cannot be reflected in the simple estimate because we consider the aligned GJ model in this paper. This is the main paper we will do in the next paper and is highly worthy of further investigation.
\begin{table*}
	\caption{\label{tab:pulsar-list}The list of typical neutron stars for estimation the order of magnitude of the conversion probability in this paper.}
	\resizebox{1\linewidth}{!}{
		\begin{tabular}{cccccc}
			\toprule
			Pulsar Name&Right Ascension&Declination&Barycentric&Pulsar&Surface Magnetic\\
			J2000&J2000&J2000&Period&Distance&Flux Density\\
			&(hh:mm:ss.s)&(+dd:mm:ss)&(s)&(kpc)&(Gauss)\\
			\midrule
			PSR J1808-2024 \citep{Kouveliotou:1998ze} &18:08:39.337&-20:24:39.85&7.55592&13.000&2.06e+15\\
			PSR J0501+4516 \citep{2008GCN..8118....1G} &05:01:06.76&+45:16:33.92& 5.76209653&2.000&1.85e+14\\
			PSR J1809-1943 \citep{2004ApJ...609L..21I} &18:09:51.08696&-19:43:51.9315&5.540742829&3.600&1.27e+14\\
			PSR J1550-5418 \citep{2007ApJ...666L..93C} &15:50:54.12386&-54:18:24.1141&2.06983302&4.000&2.22e+14\\
			PSR J0736-6304 \citep{2010MNRAS.402..855B} & 07:36:20.01&-63:04:16&4.8628739612&0.104&2.75e+13\\
			PSR J1856-3754 \citep{2007ApJ...657L.101T} &18:56:35.41&-37:54:35.8&7.05520287&0.160&1.47e+13\\
			PSR J0720-3125 \citep{1997AA...326..662H} &07:20:24.9620&-31:25:50.083&8.391115532&0.400&2.45e+13\\
			PSR J1740-3015 \citep{1986Natur.320...43C} &17:40:33.82&-30:15:43.5&0.60688662425&0.400&1.7e+13\\
			PSR J1731-4744 \citep{Large:1968mi} &17:31:42.160&-47:44:36.26&0.82982878524&0.700&1.18e+13\\
			PSR J1848-1952 \citep{1978MNRAS.185..409M} &18:48:18.03&-19:52:31&4.30818959857&0.751&1.01e+13\\
			SGR J0501+4516 \citep{2014MNRAS.438.3291C} &05:01:06.76&+45:16:33.92&5.7620695&2&1.9e+14\\
			SGR J0418+5729 \citep{2013ApJ...770...65R} &04:18:33.867&+57:32:22.91&9.07838822&2&6.1e+12\\
			SGR J1935+2154 \citep{2016MNRAS.457.3448I} &19:34:55.598&+21:53:47.79&3.2450650&-&2.2e+14\\
			\bottomrule
		\end{tabular}
	}
\end{table*}

As aforementioned, the determination of the specific intensity of the converted photon $I_{\gamma,\omega}$ at frequency $\omega$ relies on the total conversion rate $\left\langle P_{g \rightarrow \gamma}^{\mathrm{coherence}}(\omega,L,\theta)\right\rangle=\int_{\Delta\theta}P_{g \rightarrow \gamma}^{\mathrm{coherence}}(\omega,L,\theta)d\theta$ \citep{Raffelt:1987im}, denoting the integrated conversion probability per unit time and solid angle, alongside the VHF GW's differential energy fraction $\Omega_{\mathrm{GW}}(\omega)$. Considering the radio telescope's frequency binning $\int_\omega^{\omega+\Delta\omega} d(\ln \omega)\Omega_{\mathrm{GW}}(\omega) \approx \frac{\Delta \omega}{\omega} \Omega_{\mathrm{GW}}(\omega)$, the final specific intensity generated in the magnetic field can be expressed as
\begin{equation}
	\begin{aligned}
	I_{\gamma,\omega}&=\frac{dE_\omega}{dtdAd\Omega d\omega}\\
	&=\frac{3 H_0^2M_{\mathrm{planck}}^2}{4 \pi\omega}\Omega_{\mathrm{GW}} (\omega)\left\langle P_{g \rightarrow \gamma}^{\mathrm{coherence}}(\omega,L,\theta)\right\rangle,
		\label{specific-intensity}
	\end{aligned}
\end{equation}
where $H_0$ is the Hubble constant \citep{2020AA...641A...6P} and $M_{\mathrm{planck}}=\sqrt{\frac{1}{8\pi G}}$ represents the reduced Planck mass. Now let's determine the magnitude of this solid angle $\Delta\theta$. Considering a neutron star with a distance $d$ from us, the region where the VHF GWs are converted into photons is spherical and has a radius of $R_{tot}$. This radius $R_{tot}$ can be a local point of the neutron star's magnetic field, a region, or the entire magnetic field. The size and choice of this radius will greatly affect the size of the final conversion probability. To determine the size of a $R_{tot}$, it is most convenient and optimistic to analyze an integral region of infinite size. This allows for the calculation of the maximum conversion probability and the optimal utilization of the telescope's field of view. However, this is actually not a rational assumption. When the region is of considerable size, the magnetic field of the neutron star will weaken significantly, eventually being superseded by other magnetic fields in astrophysics. Consequently, we can no longer solely attribute the observational effect we are currently investigating to the neutron star. Hence, two factors influence the magnitude of a signal's field of view: First, the strength of the magnetic field must be higher than the average magnetic field of the universe $B_0\lesssim 47\mathrm{~pGauss}$ which allows the magnetic field region to extend to a radius of up to $10^9\mathrm{~km}$ \citep{Neronov:2010gir,Tavecchio:2010mk,Takahashi:2013lba,Durrer:2013pga,Planck:2015zrl,Pshirkov:2015tua,Jedamzik:2018itu,Domcke:2020yzq}. Secondly, it is evident that the signal's effective field of vision cannot surpass the telescope's field of view (see our discussion of this in Sec. \ref{sec:quantities-rewritten-in-the-nautural-lorentz-heaviside-untis} in the appendix). Therefore, we can consider the following simple model: The telescope is centered on a neutron star, and the data at the back end of the observation is actually the integrated total voltage of all radio signals across the telescope's field of view \citep{2017isra.book.....T}. It is clear that $R_{tot}$ is greater than $r_{\mathrm{occur}}$ for actual observation. Therefore, we can determine the total energy flux reaching Earth
\begin{equation}
	F_{\gamma,\omega}=\pi I_{\gamma,\omega}\left(\frac{R_{tot}}{d}\right)^2.
	\label{total-energy-flux}
\end{equation}

\begin{figure*}
	\centering
	\includegraphics[width=0.5\linewidth]{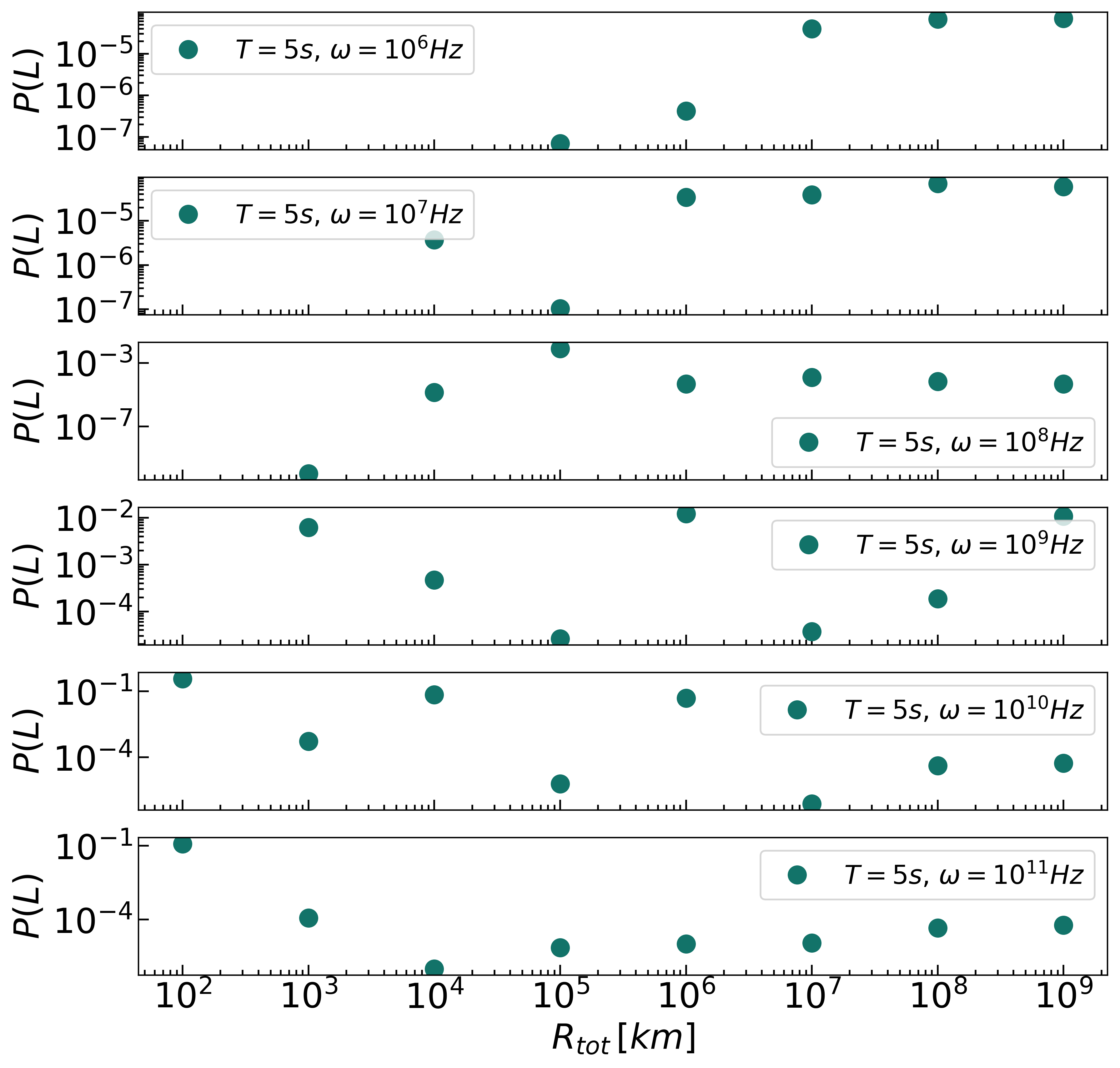}
	\caption{The total conversion probability $P_{g \rightarrow \gamma}^{\mathrm{coherence}}(L)$ of the GWs traveling at different $R_{tot}$ in neutron star magnetic field. The parameters of the magnetic field assumed are  $B=10^{14}\mathrm{~Gauss}$, and $T=5\mathrm{~sec}$. The figure shows from top to bottom the result of a 10-fold increase in the frequency of the radio telescope at one time, and the conversion probability as a whole gradually increases.}
	\label{fig:conversion-probability-total}
\end{figure*}

Combining these points, we calculated the integral conversion probability $\left\langle P_{g \rightarrow \gamma}^{\mathrm{coherence}}(\omega,L,\theta)\right\rangle$ with the change of size of the conversion region $R_{tot}$. The results are presented in Fig. \ref{fig:conversion-probability-total}, utilizing the same parameters as in Fig. \ref{fig:conversion-probability-times}. The complete results are available in Fig. \ref{figapp:conversion-probability-total} in Appendix \ref{sec:more-figures}. The absence of data points in the figure at lower frequencies is attributed to the radius at which conversion occurs, $r_{\mathrm{occur}}$, being greater than the radius of the conversion region, $R_{tot}$, assumed in our calculation. The figure demonstrates that the decrease in total conversion probability near the neutron star's light cylinder, $R_{CL}\simeq 2.5\times 10^{5}\mathrm{~km}$, is due to changes in the topological configuration of the magnetic field in that region. Simultaneously, it is observed that when the radius of this region exceeds $1\times 10^{8}\mathrm{~km}$, the total conversion probability curve approaches a flat trend. This phenomenon occurs because the magnetic field of the neutron star decays with the third power of the radius, leading to the superposition of the conversion probability from the larger region being considered a higher-order term. Finally, we selected the radius of the conversion region as $R_{tot}=10^9\mathrm{~km}$. The radius is less than the field of view radius of the telescope, and the magnetic field strength exceeds 47 pGauss. The selected conversion region represents an optimistic estimate, and actual observations are likely to differ significantly: (a) Due to the magnetic field crossing of the surrounding objects, the magnetic field of the neutron star does not have such a clear dividing line. (b) We assumed that the physical process of converting VHF GWs into electromagnetic waves occurs uniformly throughout the conversion region. However, VHF GWs may be localized to a limited area within this conversion region, necessitating observations to accurately identify the emission region of electromagnetic waves. 

\subsection{Diffraction of gravitational waves by single neutron star}
Recent studies have revealed that when nanohertz GWs pass through the universe, they will produce weak interference modulation due to the diffraction effect of a large number of galactic disks, and the strain amplitude will change about $\Delta h / h \sim 10^{-3} \lambda_{1 \mathrm{pc}}^{-1}$ \citep{Jow:2024bwq}. Similarly, when a VHF GW passes through a neutron star's magnetic field, the GW is also diffracted by the neutron star, which we discuss in this subsection and give the amount of change in the modulated strain amplitude.

A solar mass neutron star with a radius of $10\mathrm{~km}$ has a Schwarzschild radius $R_{s}=\frac{2GM}{c^2}$ of about $3\mathrm{~km}<r_{0}$, and the Fresnel length can be calculated to determine whether GWs will be diffracted as they pass by the neutron star. The Fresnel length, which relates to the GW frequency and the distance between the neutron star and radio telescope, can be described as \citep{Takahashi:2003ix,Macquart:2004sh,Choi:2021bkx,2017mcp..book.....T} 
\begin{equation}
	r_{Fre}=\sqrt{\frac{d_{\mathrm{eff}}}{2 \pi \omega}} \approx 545.78 \sqrt{\left(\frac{d_L}{1 \mathrm{kpc}}\right)\left(\frac{10^6 \mathrm{~Hz}}{\omega}\right)}\mathrm{~km}.
\end{equation}
Here, $d_{\mathrm{eff}}=d_L d_{L S} / d_S \approx d_L$ denotes the effective angular diameter distance to the lens, where $d_L$ is the distance from the neutron star to the Earth. This approximation holds under the assumption that the incident gravitational waves can be treated as plane waves, implying that their sources lie either beyond the Milky Way or originate from the very early universe. Then, the condition for diffraction is $r_{Fre}^2\gtrsim R_s^2$ \citep{Choi:2021bkx,Takahashi:2005ug,Oguri:2020ldf}. It is clear that GWs in the radio band $10^6-10^{10}\mathrm{~Hz}$ combined with distant neutron stars satisfy the diffraction condition. The frequency dependence in the diffraction regime is measured by the shear $\gamma(x)$ from the complex amplification factor with Kirchhoff-Fresnel integral $F_{amp}(\omega)=\frac{\omega}{i d_{\mathrm{eff}}} \int d^2 \boldsymbol{r} \exp \left[i 2 \pi \omega T_d\left(\boldsymbol{r}, \boldsymbol{r}_s\right)\right]$ \citep{Nakamura:1999uwi}. Here, $\boldsymbol{r}$ denotes the physical displacement vector on the lens plane, measured from the center of the lens, and $\boldsymbol{r}_s$ represents the projected position of the source onto the same plane. The quantity $T_d$ is the arrival-time difference between the deflected trajectory passing through $\boldsymbol{r}$ under the influence of the lens and the unperturbed, straight-line path that would occur in the absence of the lens. In analogy with the energy density spectrum of gravitational waves, reformulating the Kirchhoff–Fresnel integral in terms of a logarithmic frequency interval proves advantageous for the analysis that follows \citep{Choi:2021bkx} 
\begin{equation}
	\frac{d|F_{amp}(\omega)|}{d \ln\omega} \approx \gamma\left(r=r_{Fre} e^{i \frac{\pi}{4}}\right).
\end{equation}

To obtain a fast estimate of the cumulative diffraction amplification factor, we consider only the single diffraction flux formed by the contribution of the isolated neutron star. The diffraction amplification factor, which is the energy amplification by the GW due to the diffraction of the neutron star, is the bridge between before and after GWs are diffracted
\begin{equation}
	h_{\mathrm{diff}}(\omega)=F_{amp}(\omega) h(\omega),
\end{equation}
where $h_{\mathrm{diff}}(\omega)$ denotes the lensed waveform, while $h(\omega)$ represents the unlensed waveform. And both are expressed in the frequency domain with respect to frequency $\omega$. Therefore, the GW sensitivity we observe needs to be divided by the diffraction amplification factor $F_{amp}(\omega)$ \citep{Jow:2024bwq}. Given that the spatial scale of the neutron star magnetosphere under consideration greatly exceeds the Fresnel scale, the characteristic transverse size of the lensing structure is much larger than the Fresnel radius. Under such conditions, the lens can be effectively treated as one-dimensional, and the resulting diffracted flux is well approximated by the diffraction pattern of a one-dimensional Gaussian lens
\begin{equation}
	F_{amp}(\omega)=64 \pi^4\left(\frac{R_s r_{0}}{\lambda r_{Fre}}\right)^2\sqrt{\frac{r_{0}^4}{4\pi^2r_{0}^4+r_{Fre}^4}}.
\end{equation}
For example, if the distance parameter $d_L=2\mathrm{~kpc}$ of PSR J0501+4516 is selected and the frequency $\omega$ is $10^6\mathrm{Hz}$, $F_{amp}$ is equal to 0.176. It should be noted that GWs in the radio band are still tensor waves, and the diffraction formula we use is a reasonable approximation \citep{Dolan:2017zgu,Hou:2019wdg,Li:2022izh,He:2020qpf,Jow:2020rcy}. In principle, a more precise tensor diffraction formula is needed to describe the properties of tensor waves, which we will discuss in a subsequent paper.

\section{Signal spectral line characteristics and frequency characteristics}
\label{sec:sslcafc}
VHF GWs are transformed into electromagnetic waves through the GZ effect in the magnetic fields of neutron stars. The signal spectral lines and frequency characteristics of these waves are the keys to detection and physical analysis. This section systematically expounds the spectral line characteristics of graviton-photon conversion and the frequency dependence of signal propagation.

\subsection{The graviton spectral line broadening.}
\label{subsec:the-graviton-spectral-line-broadening}
In this subsection, we will discuss the line-broadening mechanism of gravitons. The order of discussion is collision broadening, the Zeeman effect in a magnetic field, and natural broadening.

The Lorentz transformation allows us to represent the GW amplitudes $h_{+}$ and $h_{\times}$ under the GW's transverse and traceless gauge
\begin{equation}
	\begin{aligned}
		& h_{+}^{\prime}=h_{+} \cos 2 \psi-h_{\times} \sin 2 \psi, \\
		& h_{\times}^{\prime}=h_{+} \sin 2 \psi+h_{\times} \cos 2 \psi.
	\end{aligned}
\end{equation}
If we assume rotations around the axis, the combinations $h_{\times} \pm i h_{+}$ undergo a transformation
\begin{equation}
	\left(h_{\times} \pm i h_{+}\right) \rightarrow e^{\mp 2 i \psi}\left(h_{\times} \pm i h_{+}\right).\label{gw-trans}
\end{equation}
Another characteristic of GWs is that they adhere to the massless Klein-Gordon field equation
\begin{equation}
	\square h_{i j}^{\mathrm{TT}}=0.
\end{equation}
Therefore, we can examine the massless, one-dimensional representation of the Poincar\'{e} group, which is defined by the condition of four-momentum $P_\mu P^\mu=0$ and is characterized by a specific value of helicity $h$. Helicity $h$ is defined as the projection of total angular momentum onto the direction of motion, expressed as $h=\mathbf{J} \cdot \hat{\mathbf{n}}$, where $\hat{\mathbf{n}}$ represents the propagation direction of the gravitational wave (GW) and $\mathbf{J}=\mathbf{L}+\mathbf{S}$ denotes the total angular momentum, which is the sum of orbital angular momentum $\mathbf{L}$ and spin angular momentum $\mathbf{S}$. Upon rotating the direction of motion by an angle $\psi$, the helicity eigenstate $|h\rangle$ is transformed as follows:
\begin{equation}
	|h\rangle \rightarrow e^{h i \psi}|h\rangle.\label{gw-helicity}
\end{equation}
By combining Eqs. (\ref{gw-trans}) and (\ref{gw-helicity}), the helicity of gravitational waves, as well as that of massless gravitons, can be determined to be $h=\pm 2$. In conclusion, gravitons, massless particles with helicity $\pm 2$ that propagate along null geodesics, are compatible with the linearization of general relativity and contemporary quantum field theory. This indicates that their theoretical speed ought to equal the speed of light, denoted as $c$. Observations have verified this, indicating that the speed of gravitational waves is nearly equivalent to the speed of light \citep{Romano:2023bge}.

\begin{figure*}
	\centering
	\includegraphics[width = 0.47\textwidth]{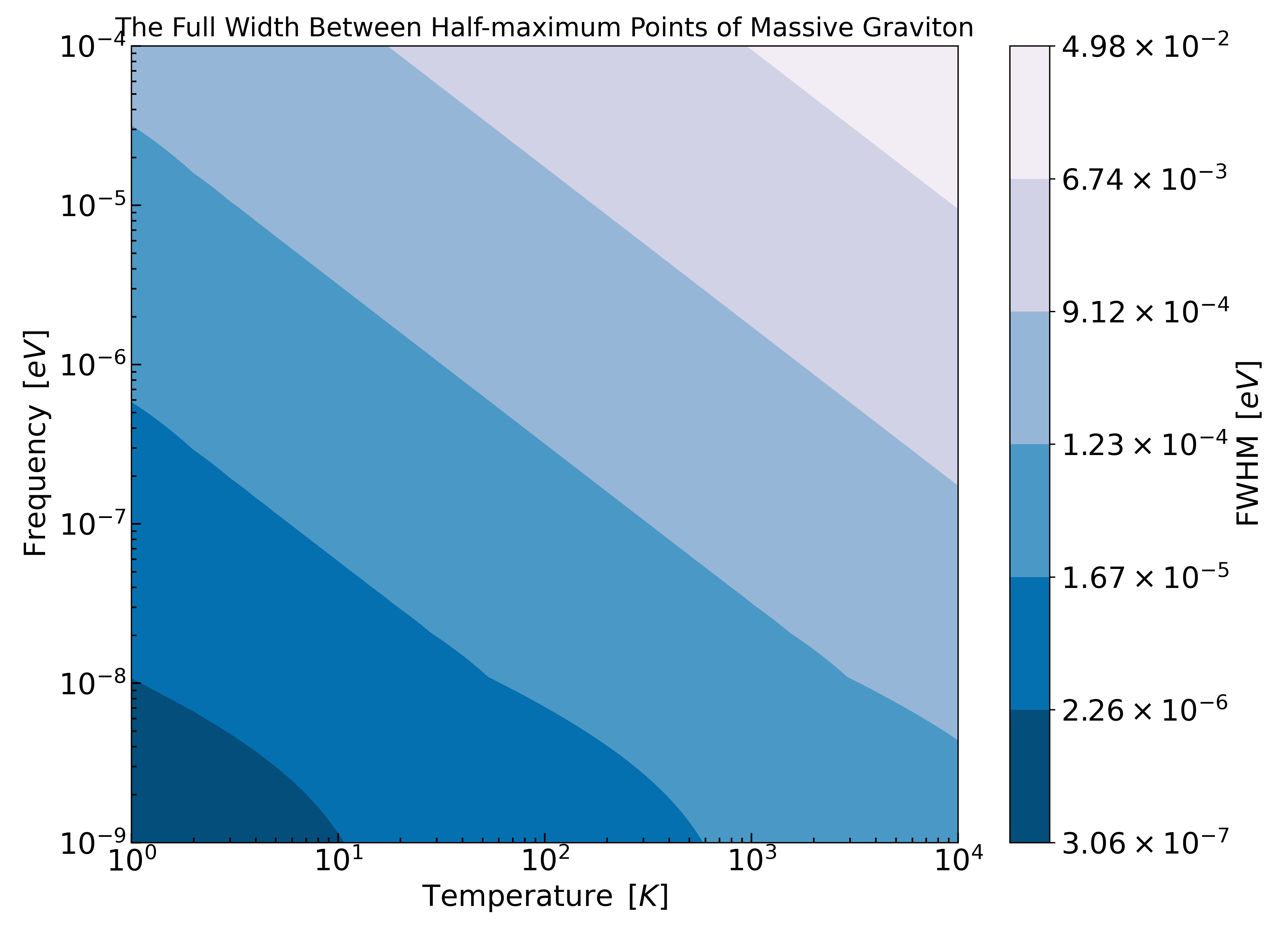}
	\caption{The FWHM of massive gravitons at different observational frequencies and temperatures is derived from collisions of spectral lines $f_d(\omega)$. In the panel, the FWHM is represented by different colored filled contours. Moreover, $f_d(\omega)$ satisfies the normalization condition $\int f_d(\omega)d\omega=1$.}
	\label{figapp:collision}
\end{figure*}

\begin{figure*}
	\centering
	\includegraphics[width = 0.6\textwidth]{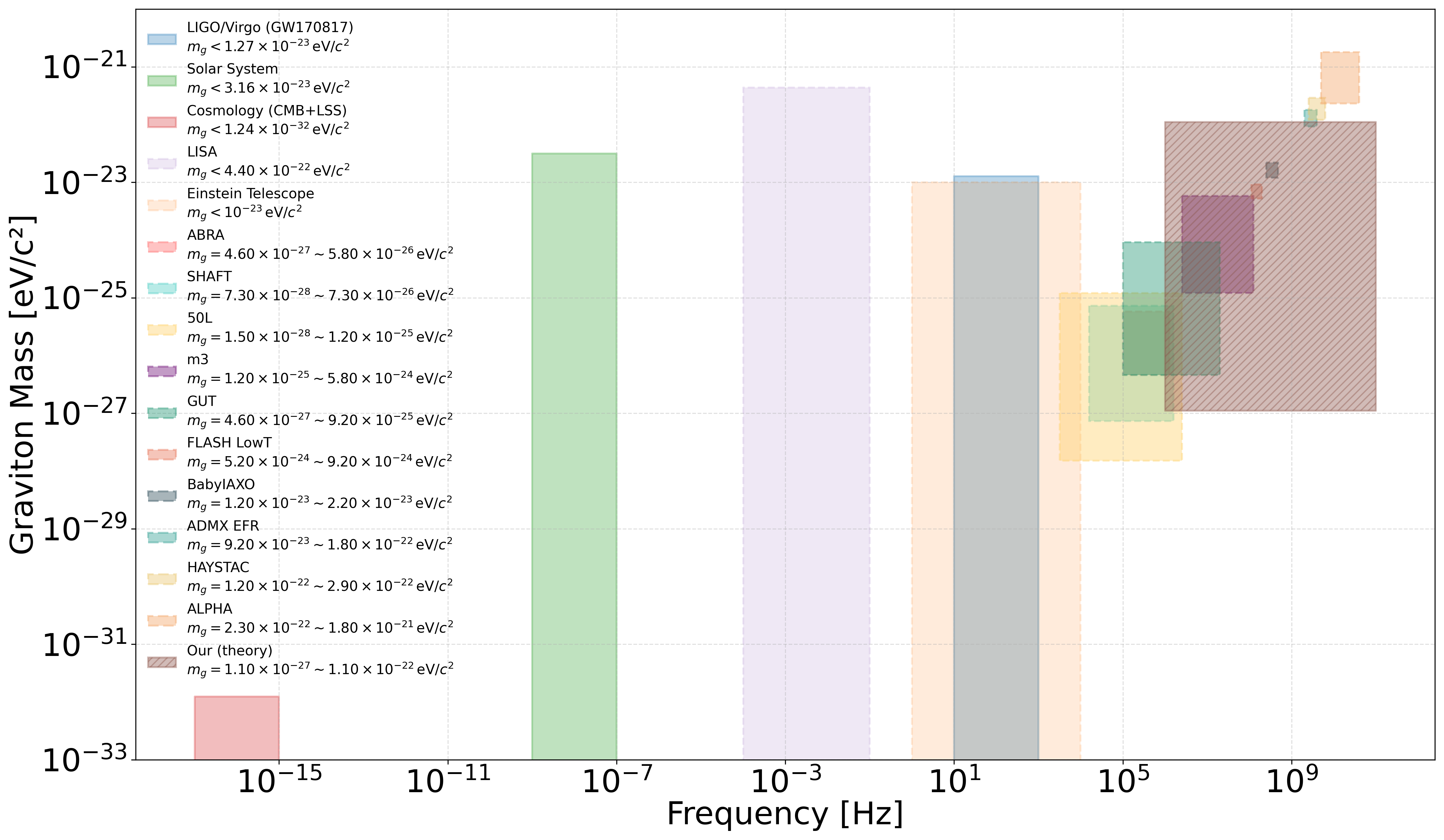}
	\caption{Current and prospective experimental constraints on graviton mass. The various colored regions in the panel indicate the frequency ranges identified by different experiments in relation to graviton constraints. Various experiments indicate the mass of the graviton, as reported in these studies: \citep{LIGOScientific:2021sio, Bernus:2020szc, Bessada:2009np, Cutler:2002ef, Arun:2009pq, Punturo:2010zz, ET:2019dnz, Salatino:2020skr,Domcke:2022rgu,Salemi:2021gck,Gramolin:2020ict,Gatti:2024mde,ADMX:2018gho,ADMX:2019uok,ADMX:2021nhd,Alesini:2023qed,Lawson:2019brd,ALPHA:2022rxj,Brubaker:2017ohw,IAXO:2020wwp,https://doi.org/10.1002/andp.202300326,Goryachev:2014yra}. Brown diagonal lines occupy the region indicative of our potentially observable mass range.}
	\label{figapp:mass}
\end{figure*}

Because the massless gravitons propagate along null geodesics, there is no change in their velocity, and therefore no Maxwell velocity distribution law for them. Hence, the frequency distribution $f_{d}(\omega)$ at temperature $T$ broadened by the collision of spectral lines is invalid
\begin{equation}
	f_{d}(\omega)=\frac{1}{(2\pi\sigma^2)^{\frac{1}{2}}}e^{-\frac{(\omega-\omega_0)^2}{2\sigma^2}},\label{collision-board}
\end{equation}
where $\sigma=\omega_0\left(\frac{k_B T}{mc^2}\right)^{\frac{1}{2}}$, where $\omega_0$ represents the fundamental frequency, and the most probable speed is given by $v_{\mathrm{m}}=\sqrt{\frac{2k_B T}{m}}$. Simultaneously, we observe that several studies suggest that gravitons can have mass \citep{Payne:2023kwj,Hatta:2023fqc,Floss:2023nod,LIGOScientific:2016lio,LIGOScientific:2017bnn,LIGOScientific:2017ycc,LIGOScientific:2018dkp,Eardley:1973br,Isi:2019asy}. The distribution of collisional broadening can be derived in the presence of mass using Eq. (\ref{collision-board}). Then, the full width between half-maximum points (FWHM) of massive gravitons can be determined.
\begin{equation}
	\Delta \omega=\left[\frac{8 \ln (2) k_B}{c^2}\right]^{1 / 2}\left(\frac{T_e}{m_g}\right)^{1 / 2} \omega_0,
\end{equation}
where $k_B$ signifies the Boltzmann constant, and $T_e$ indicates the temperature of the neutron star magnetosphere. Figure \ref{figapp:collision} presents the full width at half maximum (FWHM) for massive gravitons. The mass of gravitons is determined to be $m_g=10^{-9}-10^{-4}\mathrm{~eV}$, corresponding to frequencies of $\omega\approx10^{6}-10^{11}\mathrm{~Hz}$. Furthermore, in Fig. \ref{figapp:mass}, we present the constraints imposed by current graviton observations and experiments on its mass range. The quality range limit of the method discussed in this paper is denoted by a brown slash. The quality limit range of our method is somewhat narrower than that of other experiments, featuring a marginally higher upper limit. The final observational results reveal distinct characteristics that enable the determination of various graviton masses and the analysis of gravitational theories.

Furthermore, the helicity of massless gravitons causes the Zeeman effect to inhibit the spectral lines from splitting into multiple transitions among their magnetic sublevels, as transitions with $\Delta m=0$ and $\Delta m= \pm 1$ are permitted according to selection rules. The shifted transitional frequency $\Delta \omega_Z$ between magnetic sublevels equals zero
\begin{equation}
	\Delta \omega_Z\left(\mathbf{J}_1, m_1, \mathbf{J}_2, m_2\right)=\frac{\mu_B}{2\pi \hbar} B\left(g_1 m_1-g_2 m_2\right),
\end{equation}
where $\mathbf{J}_1$ represents the angular momentum of the upper state and $\mathbf{J}_2$ denotes the angular momentum of the lower state. $m_1$ and $m_2$ represent the magnetic quantum numbers. Besides, $\mu_B$ denotes the Bohr magneton; $\hbar$ represents the reduced Planck's constant; $g_1$ and $g_2$ signify the g-factors of the two states; and $B$ indicates the magnetic field. Next, we can calculate the natural line width using the uncertainty principle
\begin{equation}
	\Delta E \Delta t \approx \hbar, 
\end{equation}
where $\Delta E$ is represented as $h \Delta \omega_{\mathrm{nat}}$, while $\Delta t$ relates to the spontaneous emission rate of gravitons as per Bohr’s correspondence principle. Given the current lack of knowledge regarding the internal structure of gravitons, and considering that gravitons exhibit helicity of $\pm 2$, it can be inferred that $\Delta t$ approaches infinity. Consequently, the natural frequency, $\Delta \omega_{\mathrm{nat}}$, tends toward zero, rendering the natural broadening negligible.

In conclusion, we believe that in traditional general relativity, the broadening of massive gravitons is negligible, and massless gravitons exhibit no spectral line broadening. Consequently, the spectral line broadening of the photon signal transformed by the inverse GZ effect is minimal and can be effectively considered negligible. We can differentiate the radio signal from other typical astrophysical processes once we ascertain its shape and specific spectral line attributes. In the subsequent subsection, we will examine the frequency-dependent characteristics of this signal.

\subsection{Signal frequency characteristics}
\label{subsec:the-three-criteria}
\begin{figure}
	\centering
	\includegraphics[width = 0.4\textwidth]{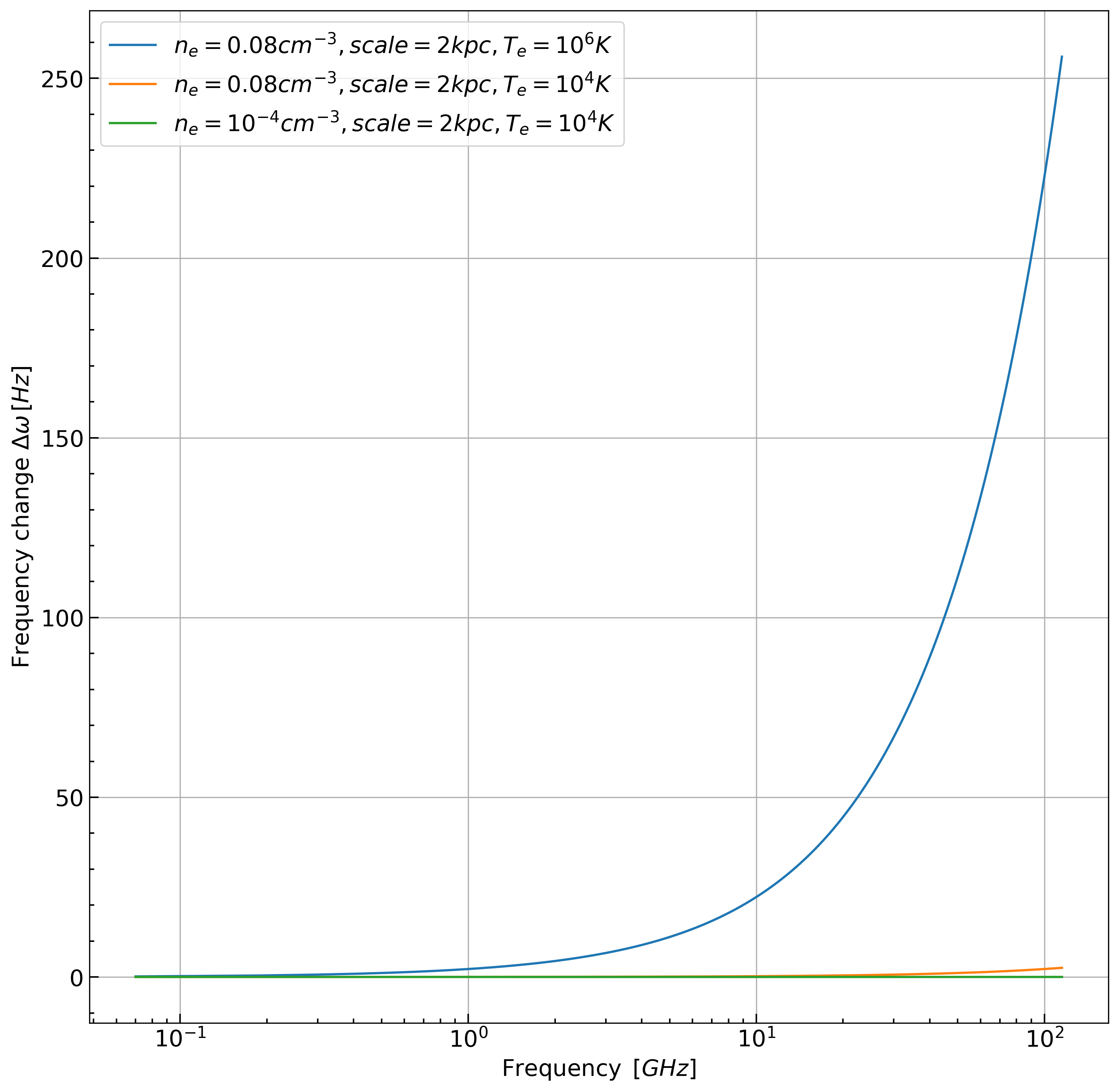}
	\caption{Compton scattering of electrons results in a change in the frequency of photons. The blue line represents an electron number density of $n_e=0.08\mathrm{~cm}^{-3}$, a medium scale of $R=2\mathrm{~kpc}$, and an electron temperature of $T_e=10^6\mathrm{~K}$. The yellow line represents an electron number density of $n_e=0.08\mathrm{~cm}^{-3}$, a medium scale of $R=2\mathrm{~kpc}$, and an electron temperature of $T_e=10^4\mathrm{~K}$. The green line represents an electron number density of $n_e=10^{-4}\mathrm{~cm}^{-3}$, medium scale $R=2\mathrm{~kpc}$, electron temperature $T_e=10^4\mathrm{~K}$.}
	\label{fig:frequncy-criteria}
\end{figure}

In contrast to gravitons, which exhibit minimal interaction with matter due to their exceedingly small cross section, photons traversing the universe undergo scattering by matter or celestial entities, including electrons and protons. Therefore, it is essential to consider the energy depletion of photons in radio observations. To maintain generality, we concentrate exclusively on the scattering of photons converted by gravitational waves through electrons during their propagation, as the energy variation of an electromagnetic wave is directly associated with its trajectory. In a medium of small optical thickness, the probability of photon scattering through the medium is given by $1-e^{\tau_{\mathrm{es}}}\approx\tau_{\mathrm{es}}$. According to the law of large numbers, when a photon traverses the same medium $M$ times ($M\rightarrow\infty$), the total number of scattering events is $M\tau_{\mathrm{es}}$. Consequently, the number of scattering events occurring through the medium only once is $\frac{M\tau_{\mathrm{es}}}{M}=\tau_{\mathrm{es}}$. In an optically thick medium, the number of times photons are scattered, denoted as $N$, can be calculated using random walk theory as $N=\tau_{\mathrm{es}}^2$, where $\tau_{\mathrm{es}}=n_e\sigma_{T}R$. Moreover, $n_e$ represents the electron density. The Thomson scattering cross section is given by $\sigma_{T}=\frac{8\pi}{3}r_{e}^2$, where $r_{e}=\frac{q_f^2}{m_{e}c^2}$ is the classical electron radius, and $R$ signifies the medium scale. Consequently, for a medium with any optical thickness, the typical scattering number can be represented as $N\approx\mathrm{max}\left\lbrace\tau_{\mathrm{es}},\tau_{\mathrm{es}}^2 \right\rbrace$.  The Compton y-parameter serves as a metric for assessing the significance of scattering.  The total energy change is determined by multiplying the average energy change per scattering event by the total number of scatterings. $y_{\mathrm{com,eff}}=N\left[\left(\frac{4k_B T_e}{m_ec^2}\right)^2+\frac{4k_B T_e}{m_ec^2}-\frac{\hbar\omega}{m_ec^2}\right]$. Subsequent articles will analyze the interstellar scintillation of a pulsar to develop a more accurate model of energy loss in signal propagation. We choose a representative electron densities $n_e=0.08\mathrm{~cm}^{-3}$ and $n_e=10^{-4}\mathrm{~cm}^{-3}$ with electron temperatures for a diffuse ionized component of $T_e=10^4\mathrm{~K}$ and $T_e=10^6\mathrm{~K}$. This data is utilized to plot Fig. \ref{fig:frequncy-criteria}, illustrating the total frequency variation of converted photons from the magnetar to the telescope as a result of Compton scattering. This suggests that targeting neutron stars with elevated electron temperatures and densities along the path to Earth can enhance the distinguishability of the GW signal.

Radio data processing typically entails the extraction of signals from radio observation data and the differentiation of various target sources through the utilization of polarization information from radiation sources. Four Stokes parameters can characterize the polarization features of a quasi-monochromatic electromagnetic wave
\begin{equation}
	\begin{aligned}
		I=&\frac{1}{2}\left(\left\langle E_{+} \mid E_{+}\right\rangle+\left\langle E_{\times} \mid E_{\times}\right\rangle\right)\\=&\frac{\omega^2}{2}\left( \left\langle \hat{A}_{\omega,+}(s) \mid \hat{A}_{\omega,+}(s)\right\rangle+\left\langle \hat{A}_{\omega,\times}(s) \mid \hat{A}_{\omega,\times}(s)\right\rangle\right) ,\\
		Q=&\frac{1}{2}\left(\left\langle E_{+} \mid E_{+}\right\rangle-\left\langle E_{\times} \mid E_{\times}\right\rangle\right)\\=&\frac{\omega^2}{2}\left(\left\langle \hat{A}_{\omega,+}(s) \mid \hat{A}_{\omega,+}(s)\right\rangle-\left\langle \hat{A}_{\omega,\times}(s) \mid \hat{A}_{\omega,\times}(s)\right\rangle\right),\\ 
		U=&\operatorname{Re}\left(\left\langle E_{+} \mid E_{\times}\right\rangle\right)=\omega^2\operatorname{Re}\left(\left\langle \hat{A}_{\omega,+}(s) \mid \hat{A}_{\omega,\times}(s)\right\rangle\right),\\ 
		V=&\operatorname{Im}\left(\left\langle E_{+} \mid E_{\times}\right\rangle\right)=\omega^2\operatorname{Im}\left(\left\langle \hat{A}_{\omega,+}(s) \mid \hat{A}_{\omega,\times}(s)\right\rangle\right),
	\end{aligned}
\end{equation}
where $\left|E_{+}\right\rangle=E_+(t, s)=-\partial_t A_{\omega,+}(t, s)$ and $\left|E_{\times}\right\rangle=E_{\times}(t, s)=-\partial_t A_{\omega,\times}(t, s)$, assuming the scalar potentials are set to zero. The parameter $I$ denotes total intensity, while $Q$ and $U$ represent linear polarization and its position angle, respectively. The symbol $V$ signifies circular polarization. Utilizing the four Stokes parameters, the observed polarization angle is defined as $\Phi=\arctan\frac{U}{Q}$, the linear polarization is given by $\Pi_{L}=\frac{\sqrt{Q^2+U^2}}{I}$, the circular polarization is expressed as $\Pi_{V}=\frac{V}{I}$, and the overall degree of polarization is represented by $\Pi_{P}=\frac{\sqrt{Q^2+U^2+V^2}}{I}$. Additionally, a four-vector can be defined using the Stokes parameters, which can be transformed by the generalized Faraday rotation tensor $\rho_{\alpha\beta}$ \citep{1991epdm.book.....M,1969SvA....13..396S}, incorporating Faraday rotation and conversion coefficients. And the absorption tensor $\eta_{\alpha\beta}$ in relation to the absorption coefficients of Stokes parameters \citep{1991epdm.book.....M,1969SvA....13..396S}, respectively. This results in the general radiation transfer equation
\begin{equation}
	\frac{d}{ds}S_{\mathrm{t},i}^\mathrm{para}=\epsilon_{i}-\sum_{j=1}^{j=4}\left[ \left(\eta_{\alpha\beta}-\rho_{\alpha\beta}\right)_{ij}S_{\mathrm{t},j}^\mathrm{para}\right],
\end{equation}
where $S_{\mathrm{t}}^\mathrm{para}=\left(I,Q,U,V\right)^T$ represent the vector of four Stokes parameters. The indices $i=j=1,2,3,4$ correspond to the components labeled as $I,Q,U,V$, while $\epsilon_{i}$ denotes the coefficients of spontaneous emission.

However, the polarization angle $\Phi$ is subject to the influence of Faraday rotation, which is characterized by the rotation measure (RM). Faraday rotation is an optical effect where the polarization plane undergoes a linear rotation with the square of the wavelength $\tilde{\lambda}$, expressed as $\mathrm{RM}=\frac{d \Phi}{d \tilde{\lambda}^2}$. The RM is directly linked to the magnetic field aligned with the line-of-sight (LOS), considering the free electron density integrated along the path from the source to the observer. There are several methods for measuring the Faraday rotation, or RM of polarized astrophysical signals, such as RM synthesis, wavelet analysis, compressive sampling, and QU-fitting \citep{2015AJ....149...60S}. In this paper, we employ RM synthesis along with Faraday synthesis. RM synthesis \citep{1966MNRAS.133...67B,2005AA...441.1217B} is a robust technique for quantifying Faraday rotation, akin to a Fourier transformation
\begin{equation}
	F_{rot}(\phi)=\int_{-\infty}^{\infty} \left[Q\left(\tilde{\lambda}^2\right)+i U\left(\tilde{\lambda}^2\right)\right]\left(\tilde{\lambda}^2\right) e^{-2 i \phi \tilde{\lambda}^2} d \tilde{\lambda}^2,
\end{equation}
where $\phi$ represents the Faraday depth, an extension of RM is used when the polarized signal is subject to varying degrees of Faraday rotation. Repeating this process for various $\phi$ values produces a Faraday dispersion function that shows the polarized intensity at different test levels. When RM synthesis is applied to emission spread over a large area of space, it often results in a complicated Faraday dispersion function with significant polarized emission at various Faraday depths \citep{2016ApJ...825...59A,2019ApJ...871..106D}. At times, the polarization effect from the detected source varies over time, requiring the use of the polarization position angle to analyze the results. The polarization position angle differs from the observed polarization angle $\Phi$ since it describes the orientation of the polarized signal before being affected by Faraday rotation. We can use the observed RM as a multiplicative phase factor to counteract the rotation of the spectrum and eliminate the impact of Faraday rotation
\begin{equation}
\begin{aligned}
 \left[ Q+i U_{\mathrm{source}}\right] (\tilde{\lambda}, t)&=[Q+i U]_{\mathrm{obs}}(\tilde{\lambda}, t)\\
&\times\exp \left\lbrace  2i\left[ \operatorname{RM}\left(\tilde{\lambda}^2-\tilde{\lambda}_0^2\right)+\Phi_0(t)\right] \right\rbrace ,
\end{aligned}
\end{equation}
where $[Q+i U]_{\mathrm{obs}}$ is the observed spectrum and $[Q+i U]_{\mathrm{source}}$ is the intrinsic polarization vector at the source, while $\mathrm{RM}$ and $\Phi_0$ are fitted parameters. $\Phi_0$ is the polarization position angle at a reference wavelength $\tilde{\lambda}_0$. In the case of calibrated polarized observations, $\Phi_0$ is often referenced at infinite frequency where Faraday rotation is zero. In principle, any time dependence of $\Phi_0$ can be determined by fitting the polarized signal through the burst duration. This time-resolved analysis is hard to do in practice because of S/N limitations. It's also not good for an automated pipeline that needs reliable ways to describe the polarized signal. An alternative method for characterizing time dependence in $\Phi_0$ is to apply the observed polarization angle $\Phi=\arctan\frac{U}{Q}$ to the burst profiles of the de-rotated Stokes $Q$ and $U$ parameters $\Phi_0(t)=\arctan\left(\frac{U_{\mathrm{integ}}(t)}{Q_{\mathrm{integ}}(t)}\right)$, where $Q_{\mathrm{integ}}$ and $U_{\mathrm{integ}}$ are integrated over frequency to optimize the signal-to-noise of the $\Phi_0$ measurement under the assumption that there is no frequency dependence in the intrinsic polarization angle at the source. Calculating the $\Phi_0(t)$ curve in this way makes it less sensitive to measurement errors associated with Stokes $Q$ and $U$, yielding a more stable curve through the observational duration. In addition, it should be noted that the wind of magnetars can affect RM \citep{2022ApJ...933L...6L}, but because it requires a relatively dense and slow wind, we will not consider this effect in this paper for the purpose of simplifying the model.

\begin{figure}
	\centering
	\includegraphics[width=0.6\linewidth]{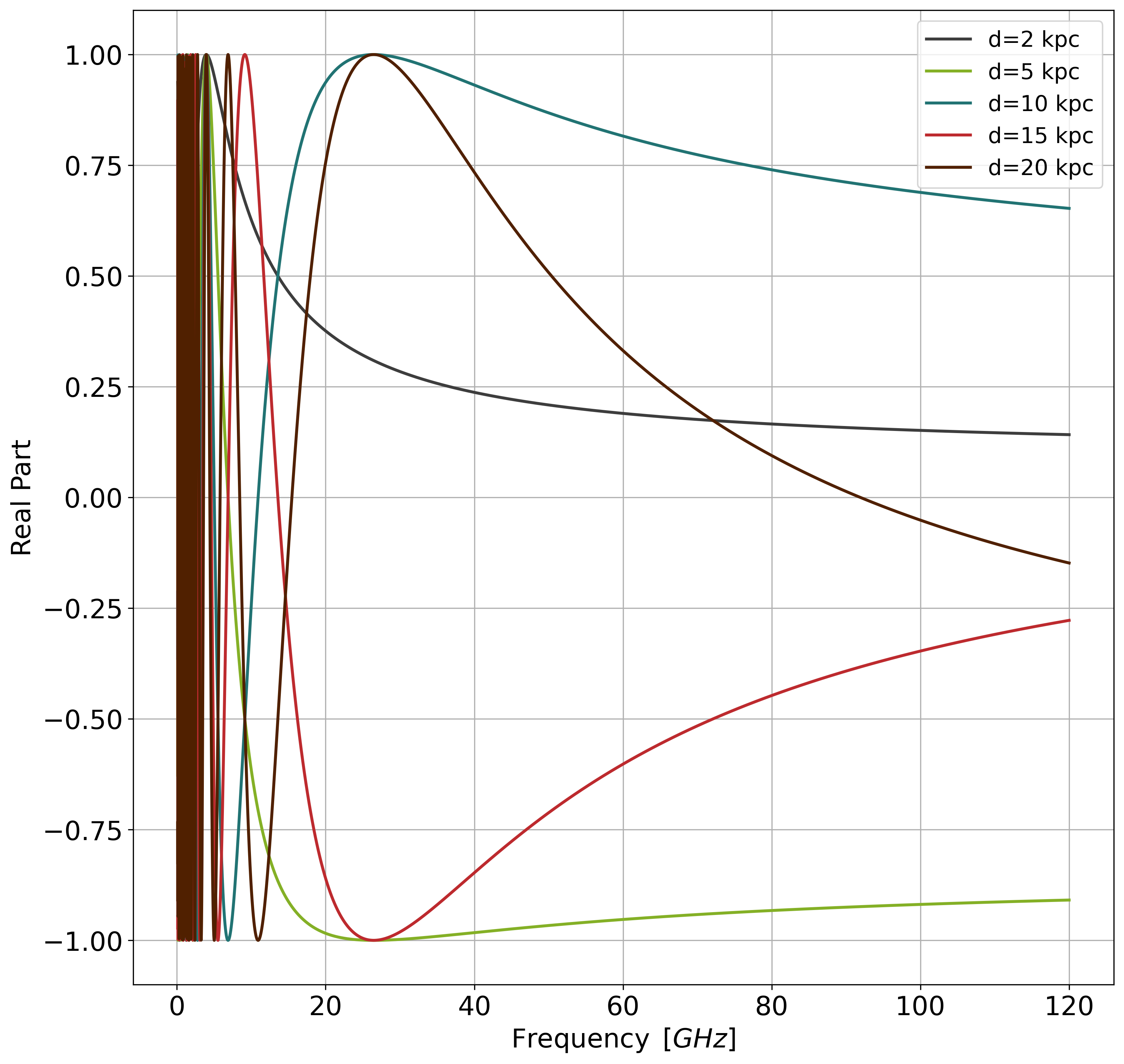}
	\includegraphics[width=0.6\linewidth]{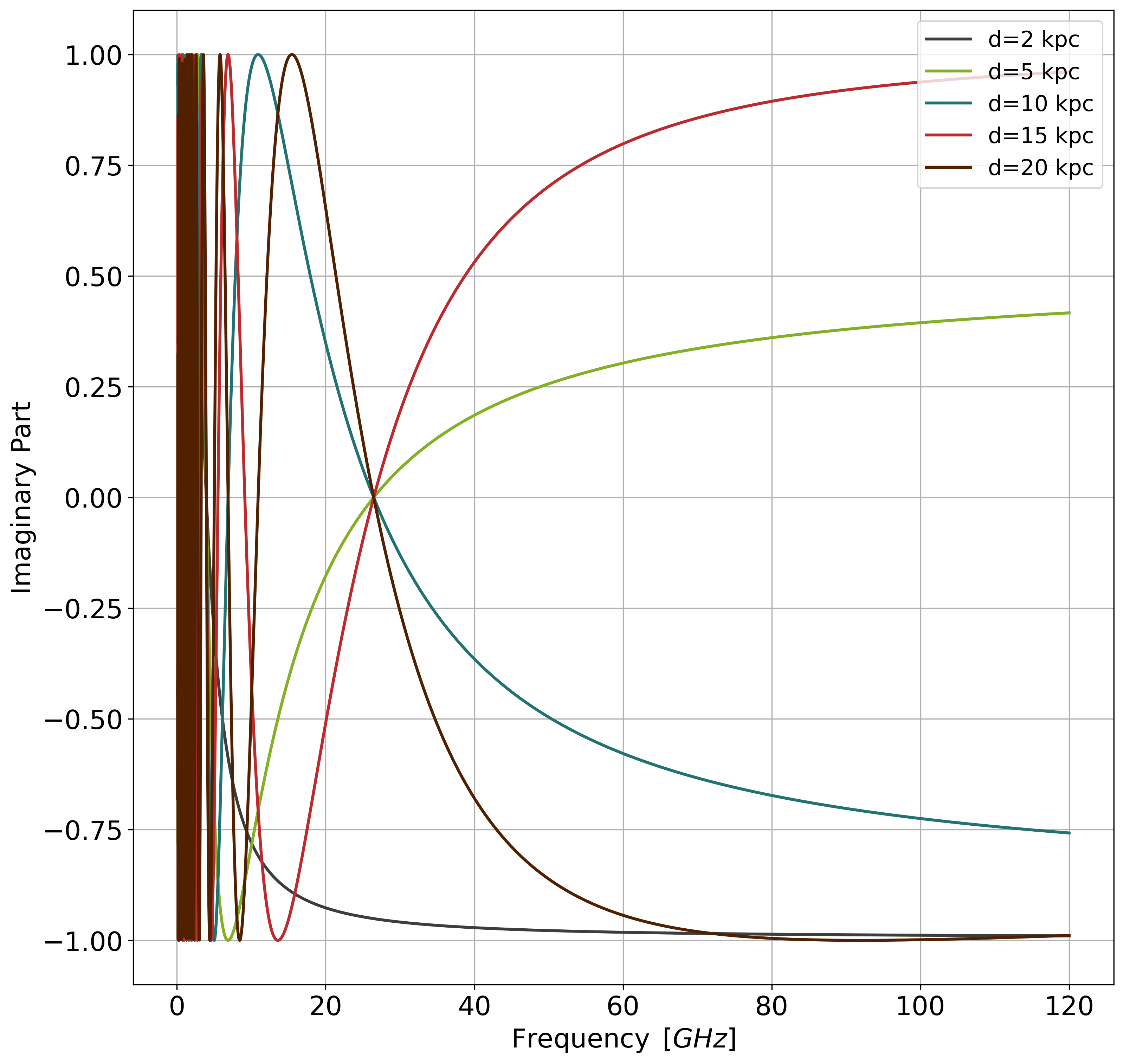}
	\caption{The real and imaginary components of the ratio relationship between $[Q+i U]_{\mathrm{source}}/[Q+i U]_{\mathrm{obs}}$ across the frequency range $\omega\in[0.1,120]\mathrm{GHz}$. The electron density is set to $n_e \sim 10^{-7}\mathrm{~cm}^{-3}$, and the magnetic field strength along the line of sight is $B_{\|} \sim 10^{-9}\mathrm{~Gauss}$. Top panel The real parts of the de-rotation ratio are presented, corresponding to the distance that radio signals traverse from the source to the radio telescopes, as indicated in the bottom panel for the imaginary parts of the de-rotation ratio. The various colored lines represent the distinct distances between neutron stars and radio telescopes, ranging from $2\mathrm{kpc}$ to $20\mathrm{kpc}$.}
	\label{fig:rotation-ratio}
\end{figure}

By employing the relationship between $A_{\omega,\lambda}(t, s)$ and $h_{\omega,\lambda}(t, s)$ in conjunction with RM synthesis, we can discern various sources of VHF GWs within the signal and eliminate some astrophysical signals resembling GW signals. This constitutes our polarization criterion. Since these radio signals are essentially conversed in resonance with GWs, the initial polarization of radio signals is exactly the same as that of GWs, resulting in linear polarization. By measuring the electron number density and magnetic field strength of the neutron star in the direction of the radio telescope's line of sight, we can obtain the rotation measure of signal $\mathrm{RM}=\frac{e^3}{2 \pi m_{\mathrm{e}}^2 c^4} \int_0^d n_{\mathrm{e}} B_{\|} \mathrm{d} l=0.812\left[ \int_0^d \left( \frac{n_{\mathrm{e}}}{1\mathrm{~cm}^{-3}}\right)  \left( \frac{B_{\|}}{10^{-6}\mathrm{~Gauss}}\right)  \left( \frac{\mathrm{d} l}{1\mathrm{~pc}}\right)\right] \mathrm{rad}\mathrm{~m}^{-2}$. We can de-rotate the spectrum by multiplying the phase factor, which helps to restore its shape prior to Faraday rotation. From this, we can find the ratio relationship between $[Q+i U]_{\mathrm{source}}$ and $[Q+i U]_{\mathrm{obs}}$ that is $[Q+i U]_{\mathrm{source}}/[Q+i U]_{\mathrm{obs}}=\exp \left\lbrace  2i\left[ \operatorname{RM}\left(\tilde{\lambda}'^2-\tilde{\lambda}^2\right)\right] \right\rbrace$. Where $\tilde{\lambda}'$ is the frequency after the frequency shift. Taking into account the values of the parameters in Appendix \ref{sec:quantities-rewritten-in-the-nautural-lorentz-heaviside-untis} and the calculation results of frequency drift shown in Fig. \ref{fig:frequncy-criteria}, we can plot the result of the change of the real and imaginary parts of this ratio with the distance and frequency of the observing source in Fig. \ref{fig:rotation-ratio}, respectively. Fig. \ref{fig:rotation-ratio} illustrates that at lower observation frequencies, the vibrations of the real and imaginary components of $[Q+i U]_{\mathrm{source}}/[Q+i U]_{\mathrm{obs}}$ are pronounced.  Furthermore, increasing the distance does not mitigate this vibration, and the calculated results for varying distances are challenging to differentiate.  At higher observation frequencies, an increase in distance results in more pronounced changes in the real and imaginary components of $[Q+i U]_{\mathrm{source}}/[Q+i U]_{\mathrm{obs}}$ with respect to frequency. This analysis indicates that a higher frequency should be selected for the polarization observation of VHF GWs.

\section{Simulation of telescope observation parameters}
\label{sec:stop}
The detection of VHF GWs being converted into radio signals through the GZ effect is highly dependent on the sensitivity of the telescope and system parameters. The weakness of such signals requires radio telescopes to have extremely low noise levels and high gain. Therefore, in this section, the system simulates the key observation parameters of the radio telescope: the equivalent flux density of the system, providing a theoretical framework for the subsequent sensitivity analysis and signal-to-noise ratio calculation.

\subsection{The system-equivalent flux density.}
\label{sec:the-system-equivalent-flux-density}
The system equivalent flux density (SEFD) is a figure of merit that characterizes the sensitivity of a radio telescope within a certain frequency band $\mathrm{SEFD}=\frac{T_\mathrm{sys}(\omega)}{\mathrm{Gain}}$, where $T_\mathrm{sys}(\omega)=T_{\mathrm{AST}}(\omega)+T_{\mathrm{ATM}}(\omega)+T_{\mathrm{RT}}(\omega)$ represents the frequency-dependent system temperature, and $\mathrm{Gain}=\frac{A_{\mathrm{eff}}}{2k_{B}}$ denotes the telescope gain, where $A_{\mathrm{eff}}=\eta_{\mathrm{A}} A_{\mathrm{phys}}$ is the effective collecting area. The effective collecting area is determined by multiplying the physical antenna aperture $A_{\mathrm{phys}}$ with an aperture efficiency $\eta_{\mathrm{A}}=\eta_{\mathrm{sf}}\eta_{\mathrm{bl}} \eta_{\mathrm{s}} \eta_{\mathrm{t}} \eta_{\mathrm{misc}} \eta_{\mathrm{sloss}}$. Here, $\eta_{\mathrm{sf}}$ denotes the reflection efficiency of the main reflector, which is governed by the Ruze equation $\eta_{\mathrm{sf}}=\exp \left[-(4 \pi \varepsilon / \lambda)^2\right]$, where $\varepsilon$ is the RMS surface error and $\lambda$ is the observational wavelength. The blockage efficiency $\eta_{\mathrm{bl}}$, accounting for the feed cabin obstruction. $\eta_{\mathrm{s}}$ represents the spillover efficiency, while $\eta_{\mathrm{t}}$ describes the illumination efficiency. $\eta_{\mathrm{misc}}$ includes various minor loss factors such as phase center offset and impedance mismatch. The factor most sensitive to the telescope pointing is $\eta_{\mathrm{sloss}}$, which reflects the reduction in effective aperture area with zenith angle $\theta_{\mathrm{ZA}}$. Since the other efficiency terms remain nearly constant during observations, the variation in total aperture efficiency $\eta$ is primarily driven by $\eta_{\mathrm{sloss}}$.

The main system temperature receives contributions from (a) astrophysical backgrounds $T_{\mathrm{AST}}(\omega)=\left[ T_{\mathrm{cmb}}(\omega)+T_{\mathrm{rsb}}(\omega)+\Delta T_{\mathrm{source}}(\omega)\right]e^{-\csc(z_{e}) \tau(\omega)}$, where $\tau(\omega)$ is the atmospheric zenith opacity, $z_{e}$ is the elevation of the telescope, (b) the atmosphere $T_{\mathrm{ATM}}(\omega)=T_{\mathrm{atm}}\left[1-e^{-\csc(z_{e}) \tau(\omega)} \right]$, and (c) the telescope electronics $T_{\mathrm{rt}}(\omega)=T_{\mathrm{spill}}(\omega)+T_{\mathrm{radio}}(\omega)$. In this paper, we consider the following radio telescopes: FAST \citep{Jiang:2019rnj,2020RAA....20...64J,2020Innov...100053Q,fastref}, TMRT \citep{tmrtref,tmrtsr,tmrtintr,tmrtpar}, QTT \citep{Wang:2023occ,qttref}, SKA1-MID \citep{skaref,2019arXiv191212699B}, SKA2-MID \citep{skaref}, and GBT \citep{gbtref,GBT_Proposer_Guide}. We use several single-dish telescopes and radio arrays, which can be superimposed by considering the beam width $2\theta_b$ and baseline lengths $B_{\max}$ of a single-dish telescope. The beam width of a radio telescope is proportional to the wavelength of radiation under consideration, with the full width at half maximum $\theta_b$ obeying the ratio $\theta_b \approx 1.25 \lambda / D=12.5^{\prime}\left(\frac{1 \mathrm{GHz}}{\omega}\right)\left(\frac{100 \mathrm{~m}}{D}\right)$, where $D$ represents the telescope diameter. The individual beam elements, which are composed of dishes of diameter $D$, have approximate gains $
\mathrm{Gain}(D) \approx 2 \frac{\mathrm{K}}{\mathrm{Jy}}\left(\frac{D}{100 \mathrm{~m}}\right)^2$. In our simplified treatment, the primary beam radius is simply $r_{\mathrm{prim}} \approx \theta_b / 2$. On the other hand, the synthesized beam diameter $\theta_{\mathrm{synth}}=2 r_{\mathrm{synth}}$ is determined by the maximum baseline $B_{\max}$ of the array $
\theta_{\mathrm{synth}}\approx \lambda/B_{\max} \approx 62^{\prime \prime}\left(\frac{1 \mathrm{GHz}}{\omega}\right)\left(\frac{1 \mathrm{~km}}{B_{\max}}\right)$. We summarize these parameters of the telescope, as well as the number of antennas and the altitude of the telescope station site, in Table \ref{tab:telescope-parameters-1}.

\begin{table*}
	\caption{\label{tab:telescope-parameters-1}The list of radio telescope configurations used in this paper.}
	\resizebox{1.0\linewidth}{!}{
		\begin{threeparttable}
			\begin{tabular}{ccccccccc}
				\toprule
				Telescope&Working&Telescope&Antennas&Antennas&Primary Beam&Synthesized Beam&Altitude of the&Atmosphere\\
				Name&Frequency&Diameter&Gain&Number&Radius&Diameter&Station Site&Pressure\\
				&(GHz)&(m)&(K/Jy)&(counts)&($'$)\tnote{a}&($''$)\tnote{b}&(m)&(mbar)\\ 
				\midrule
				FAST&0.07-3&300&16&1&2.1&-&1000&$\sim$1321-1325\\
				TMRT&1-50&65&0.845&1&9.6&-&10&$\sim$1000\\
				SKA1-MID&0.35-15&15&0.045&197&41.7&0.2&1100&$\sim$895\\
				SKA2-MID&0.05-24&15&0.045&2000&41.7&TBD&TBD&TBD\\
				GBT&0.1-116&100&1-2&1&6.3&-&807.43&$\sim$1025-1035\\
				QTT&0.15-115&110&2.42&1&5.7&-&1800&TBD\\
				\bottomrule
			\end{tabular}
			\begin{tablenotes}
				\footnotesize
				\item[a] The unit of primary beam radius is expressed in arcminutes $'$ under 1GHz.
				\item[b] The unit of synthesized beam diameter is expressed in arcseconds $''$ under 1GHz.
			\end{tablenotes}
		\end{threeparttable}
	}
\end{table*}

For astrophysical backgrounds $T_{\mathrm{AST}}(\omega)=T_{\mathrm{ast}}(\omega)e^{-\csc(z_{e}) \tau(\omega)}$, and $T_{\mathrm{ast}}= T_{\mathrm{cmb}}(\omega)+T_{\mathrm{rsb}}(\omega)+\Delta T_{\mathrm{source}}(\omega)$, it usually consists of the nearly isotropic cosmic microwave background $T_{\mathrm{cmb}}\approx 2.73\mathrm{K}$, the average sky brightness temperature $T_{\mathrm{rsb}}(\omega)$ which can be estimated from the Haslam map \citep{1981AA...100..209H,1982AAS...47....1H}, WMAP \citep{2013ApJS..208...19H}, Planck \citep{Planck:2018nkj,2020AA...641A...2P,2020AA...641A...3P,2020AA...643A..42P} and Beyond Planck data \citep{Svalheim:2020zud} with assuming a power-law spectrum $T_{\mathrm{rsb}}(\omega)=T_{408}(408 \mathrm{MHz}/\omega)^\beta$ with $T_{408}=17.1$, $25.2$, $54.8\mathrm{K}$ for the $10^{th}$, $50^{th}$ and $90^{th}$ percentiles of the all-sky distribution when gridded in an equal area projection at zero zenith angle, and $\Delta T_{\mathrm{source}}(\omega)$ is from the astronomical source being observed, which is related to the target source in the actual observation, and the value is so small that it can be ignored in theory. For example, in the $v_{\mathrm{RF}} \approx 4.85 \mathrm{GHz}$ sky survey made with the 300 -foot telescope, the system noise was $T_{\mathrm{sys}} \approx 60 \mathrm{~K}$, but the faintest detected sources added only $\Delta T_{\mathrm{source}} \approx 0.01 \mathrm{~K}$. Moreover, taking the FAST telescope as an example. For a representative pulsar with an average flux density of $0.1 \mathrm{~Jy}$, the resulting brightness temperature contribution is approximately $1.54\mathrm{~K}$. This is significantly smaller than the typical system temperature of several tens of kelvin, and even for exceptionally bright pulses, the time-averaged effect remains minor. Therefore, it is justified in most cases to assume $\Delta T_{\mathrm{source }} \ll T_{\mathrm{sys}}$ when evaluating the system sensitivity in pulsar or continuum observations. For LFC's telescopes like FAST, we can take into account sky brightness temperature with the parameter $\beta=2.76$ with the frequency from 500MHz to 3GHz \citep{1998ApJ...505..473P} and the parameter $\beta=2.51$ with the frequency from 1MHz to 500MHz since the Galactic plane is a bright diffuse source at low frequencies \citep{2021ApJ...914..128C,2022MNRAS.509.4923I}. One can use the Ultralong-wavelength Sky Model (ULSA), which can simulate the radio sky at frequencies below 10 MHz to obtain the low-frequency sky brightness temperature \citep{2021ApJ...914..128C}. For MFC's telescopes like the TMRT telescope, we can take into account the sky brightness temperature with the parameter $\beta=2.76$ at frequency $\omega\in[3,10]\mathrm{~GHz}$ \citep{1998ApJ...505..473P} and $\beta=3.00$ at frequency $\omega\in[10,14]\mathrm{~GHz}$ \citep{Weiland:2022hjd}. For HFC's telescopes like SKA1-MID, we can take into account sky brightness temperature with the parameter $\beta=3.25$ \citep{Weiland:2022hjd}. For VHFC's telescopes like GBT, we can take into account sky brightness temperature with the parameter $\beta=3.25$ at frequency $\omega\in[30,70]\mathrm{~GHz}$ \citep{Svalheim:2020zud} and $\beta=3.30$ at frequency $\omega\in[70,120]\mathrm{~GHz}$ \citep{Svalheim:2020zud}.

\begin{figure*}
	\centering
	\includegraphics[width=0.3\linewidth]{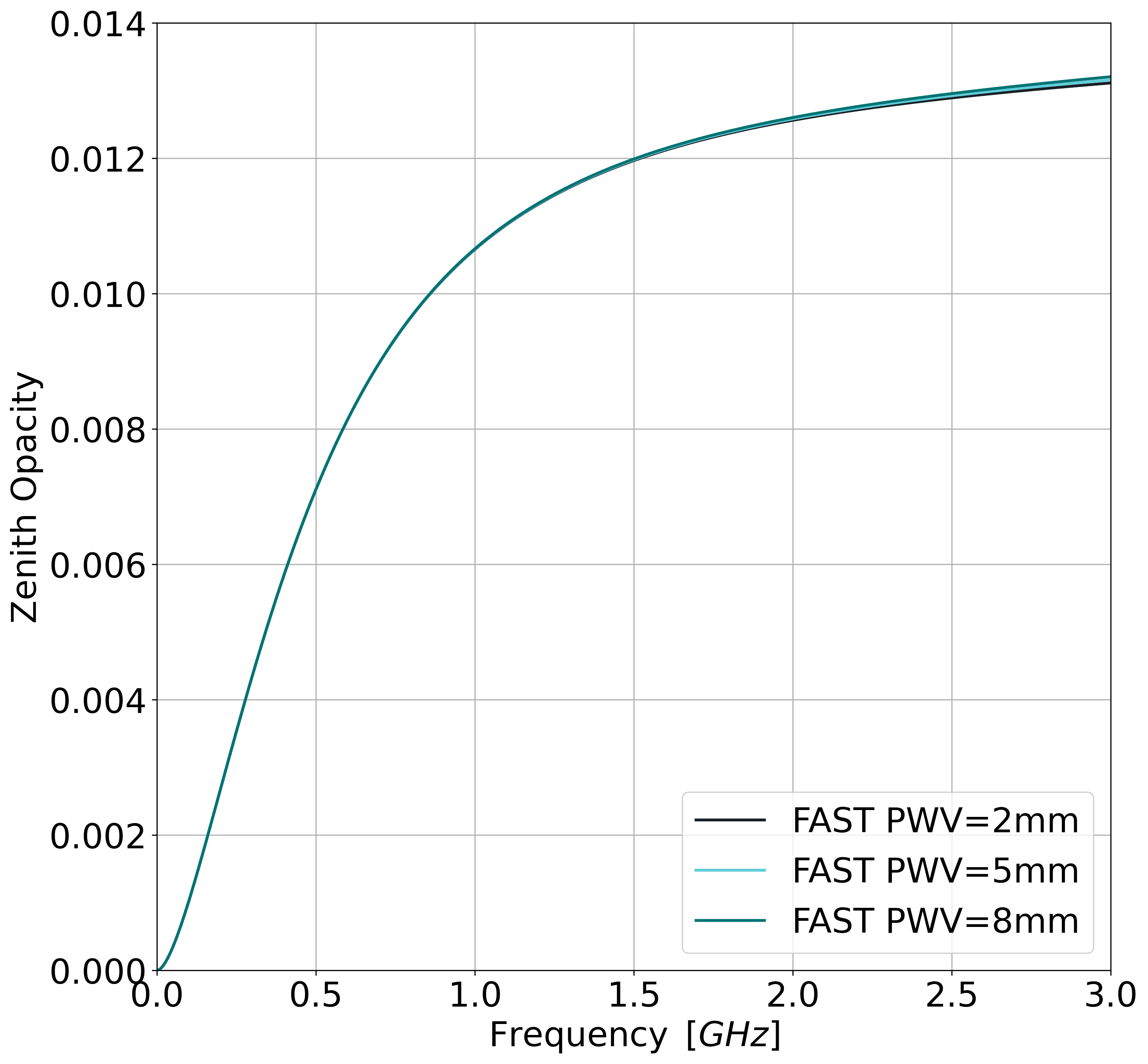}
	\includegraphics[width=0.3\linewidth]{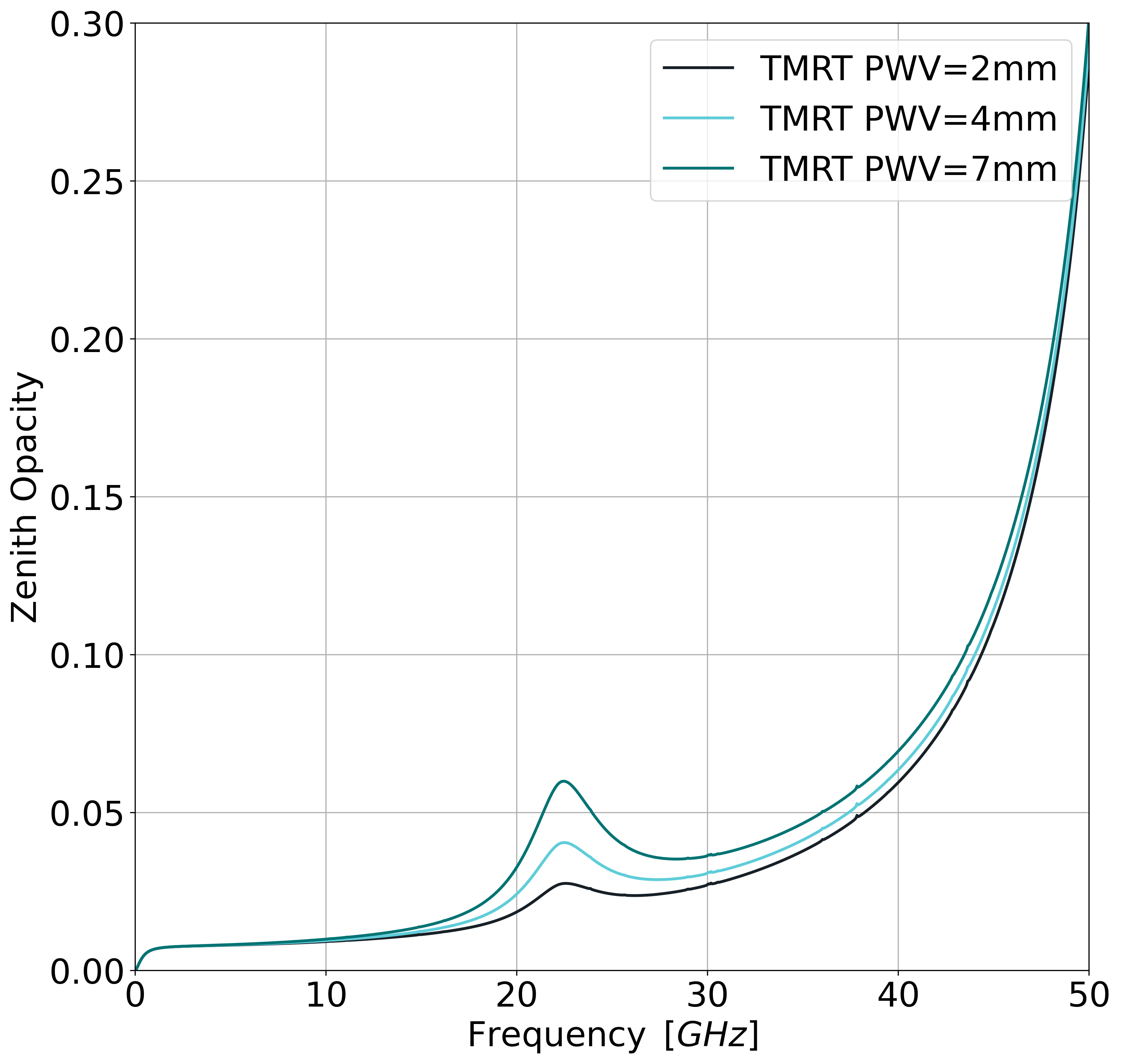}
	\includegraphics[width=0.3\linewidth]{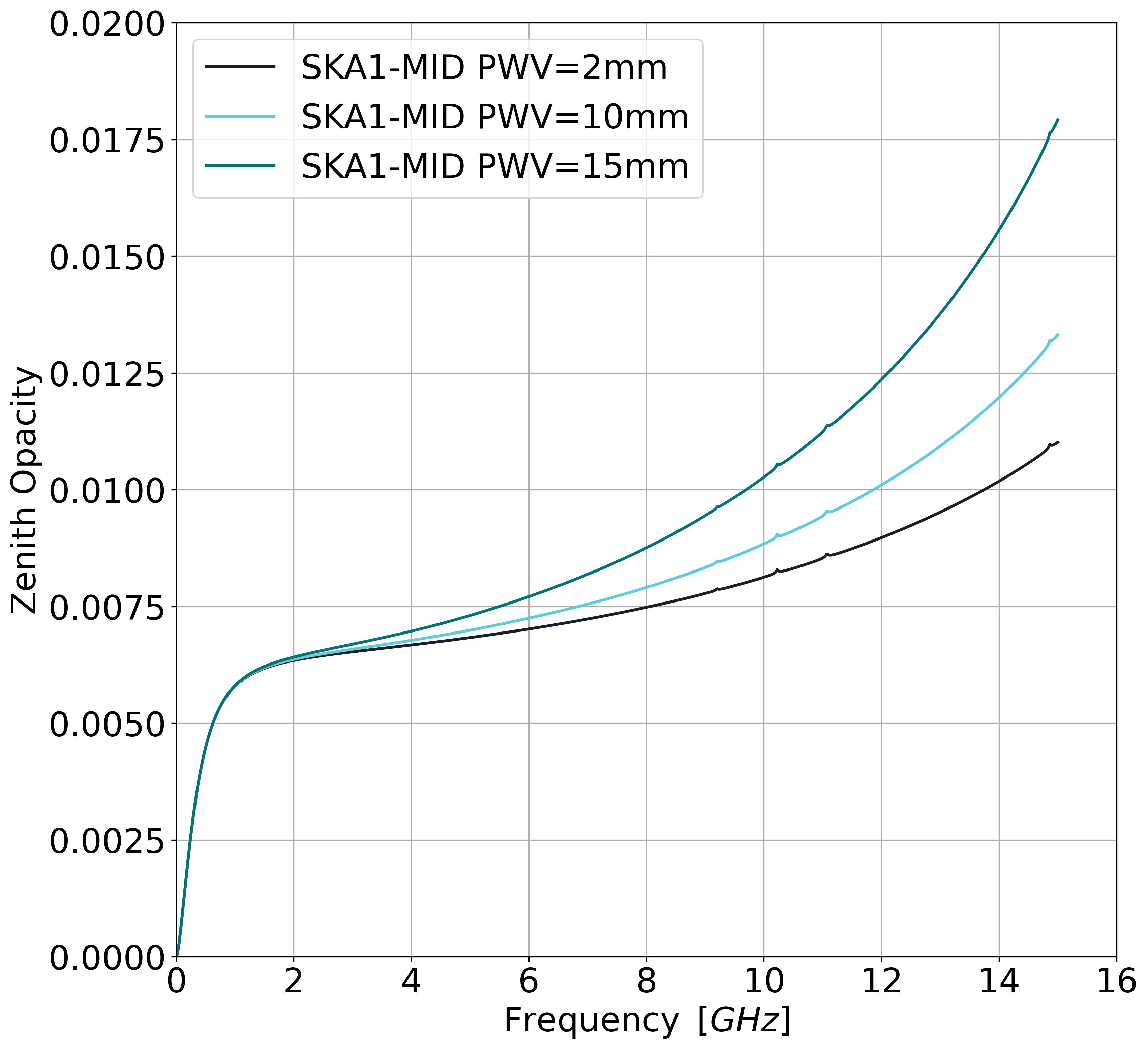}
	\includegraphics[width=0.3\linewidth]{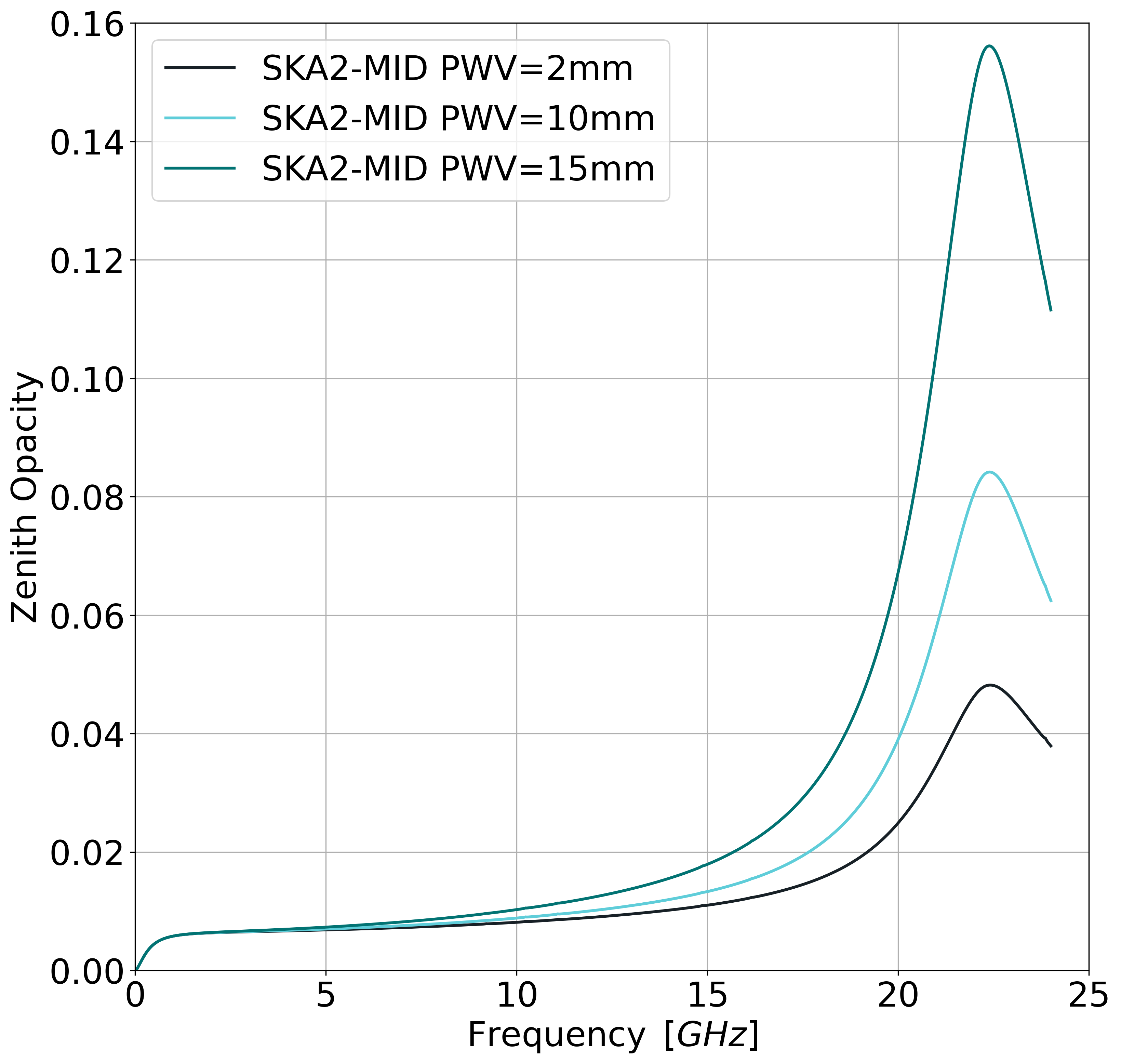}
	\includegraphics[width=0.3\linewidth]{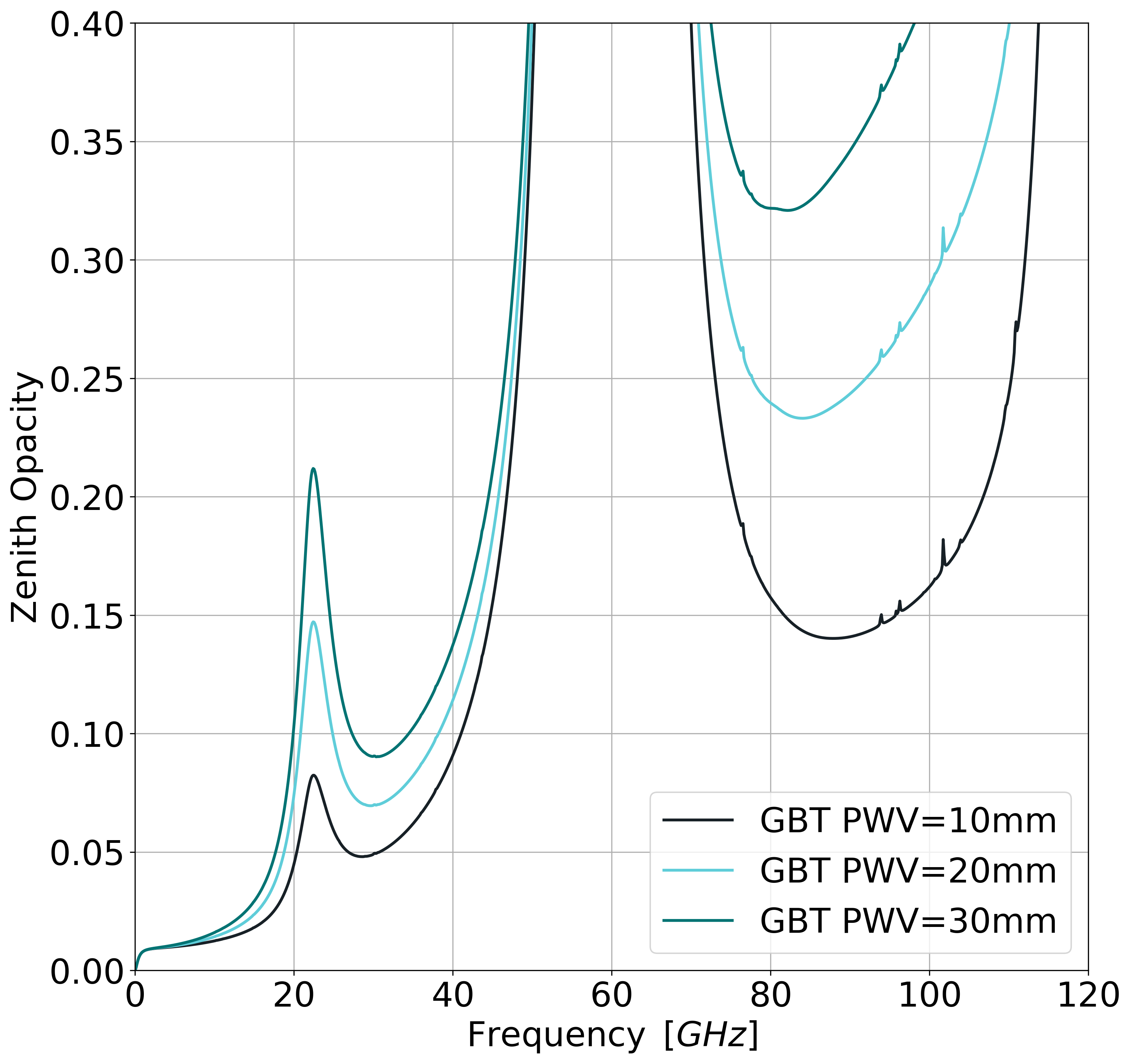}
	\includegraphics[width=0.3\linewidth]{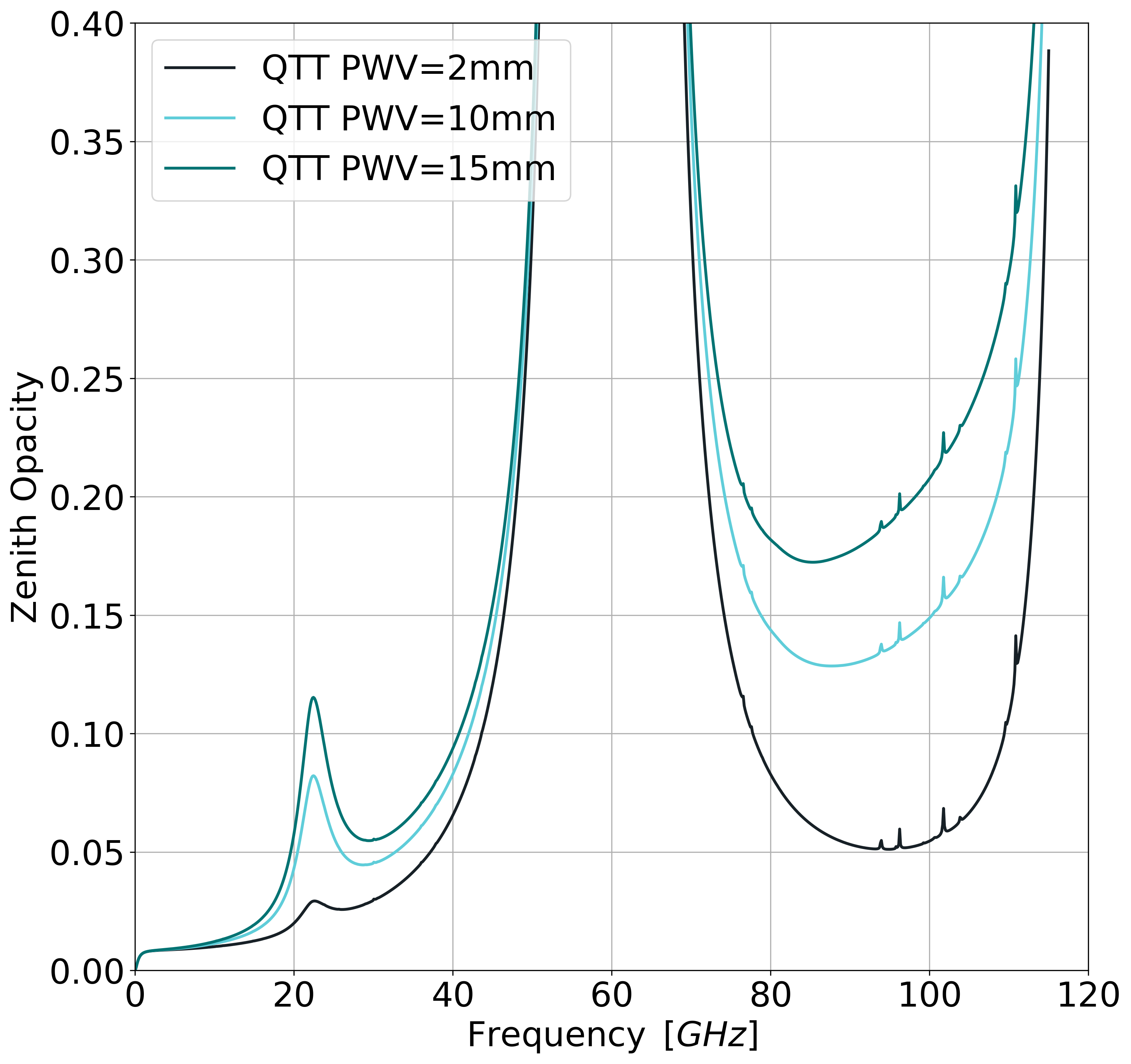}
	\caption{The simulation results of the zenith atmospheric opacity with a frequency change of six telescopes. The telescopes are (top panel) FAST, TMRT and SKA1-MID; (bottom panel) SKA2-MID, GBT, and QTT, from left to right in order. The black, wathet blue, and dark blue lines correspond to the $10^{th}$, $50^{th}$, and $90^{th}$ percentiles of those distributions, respectively.}
	\label{fig:zenith-opacity}
\end{figure*}

The $T_{\mathrm{ATM}}(\omega)=T_{\mathrm{atm}}\left[1-e^{-\csc(z_{e}) \tau(\omega)} \right]$ is the surface air temperature in the telescope beam, which is related to the atmospheric environment of the real-time observation day. However, we can estimate it theoretically. We consider a layer of water vapor at a temperature $T_{\mathrm{atm}}$ and an atmospheric zenith opacity $\tau(\omega)$ at a given observing frequency. The total optical depth is usually influenced by hydrosols $\tau_{\mathrm{hy}}$, oxygen $\tau_{\mathrm{ox}}$, dry air $\tau_{\mathrm{da}}$, and water vapor $\tau_{\mathrm{wv}}$. And the oxygen and dry-air opacities are nearly constant, while the water vapor and hydrosol contributions vary significantly with weather. We use the Atmospheric Transmission at Microwaves model (ATM) \citep{2001ITAP...49.1683P} code as implemented within the Common Astronomy Software Applications (CASA) \citep{2022PASP..134k4501C} package to obtain atmospheric zenith opacity $\tau(\omega)$. For all the telescopes, we set the water vapor scale height at 2 km without loss of generality. For the telescopes that have been built, we use local weather stations to obtain $10^{th}$, $50^{th}$, and $90^{th}$ percentiles of data on average annual precipitable water vapor (pwv), average annual temperature, average annual atmospheric pressure, and average annual humidity as our predictions. For the incompletely built SKA2-MID and QTT telescopes, we use the atmospheric prediction values consistent with SKA1-MID and the atmospheric data of Qitai Observatory, respectively. We present our simulations of the zenith atmospheric opacity of these telescopes as a function of observation frequency in Fig. \ref{fig:zenith-opacity}. And the three different colors show three different average annual precipitable water vapor. In Section Result, we present the final results using the $50^{th}$ average annual precipitable water vapor.

For the telescope electronics temperature $T_{\mathrm{rt}}(\omega)$, we can estimate from the spillover radiation $T_{\mathrm{spill}}(\omega)$ that the feed picks up in directions beyond the edge of the reflector, primarily from the ground. This contribution originates from side lobe of radio telescope. Unlike other single dish with fixed surface and horn, the illumination area of horn varies for different zenith angle, leading to different value of $T_{\mathrm{spill}}(\omega)$. And the radiometer noise temperature $T_{\mathrm{radio}}(\omega)$ attributes to noise generated by the radiometer itself. For example, we can set $T_{\mathrm{spill}}(\omega)=3\mathrm{K}$ at all frequencies at zero zenith angle and $T_{\mathrm{radio}}=15+30\left(\omega_{\mathrm{GHz}}-0.75\right)^2 \mathrm{K}$ from 0.35GHz to 1.05GHz for SKA1-MID \citep{2019arXiv191212699B}.

\begin{figure*}
	\centering
	\includegraphics[width=0.3\linewidth]{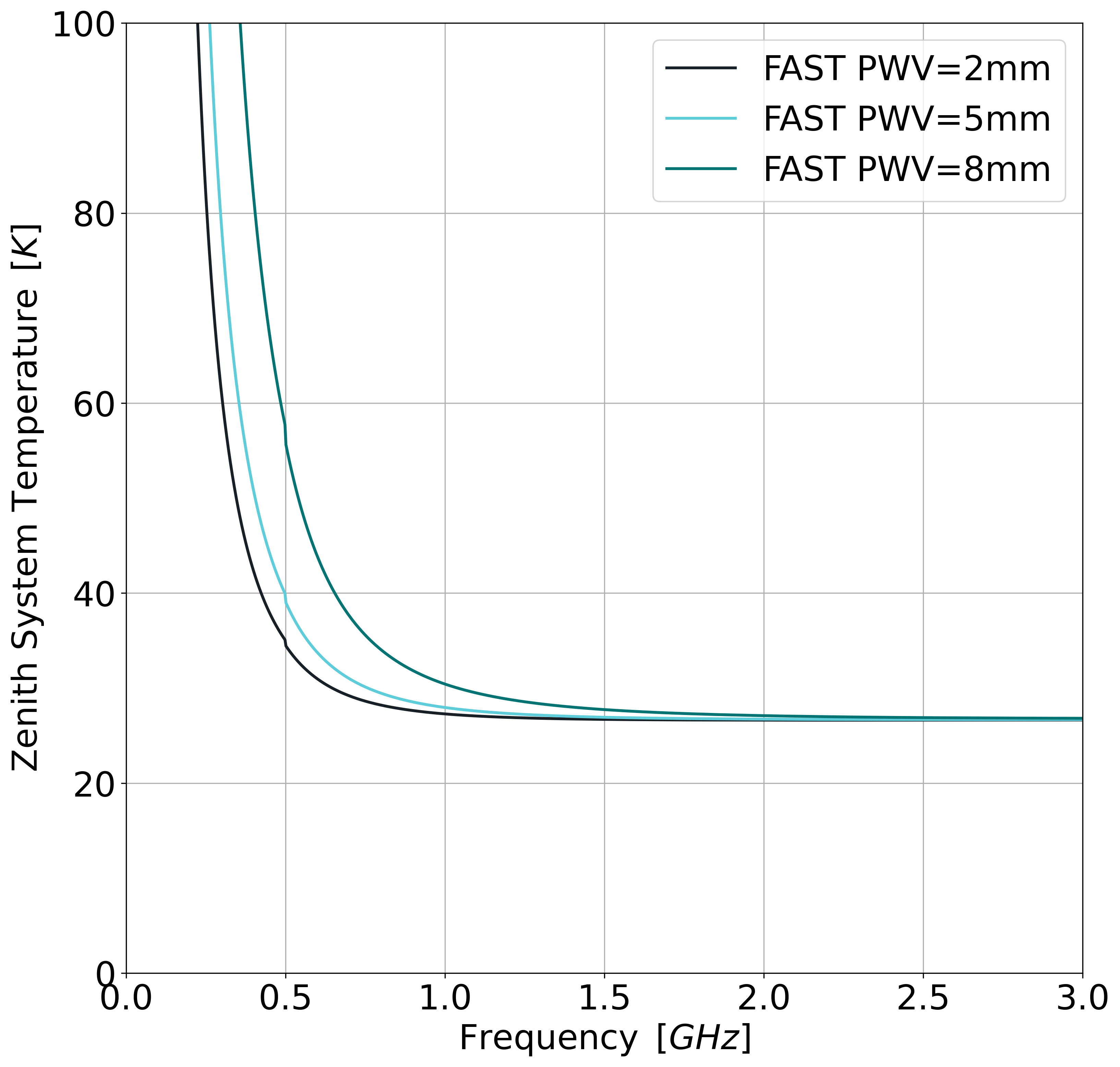}
	\includegraphics[width=0.3\linewidth]{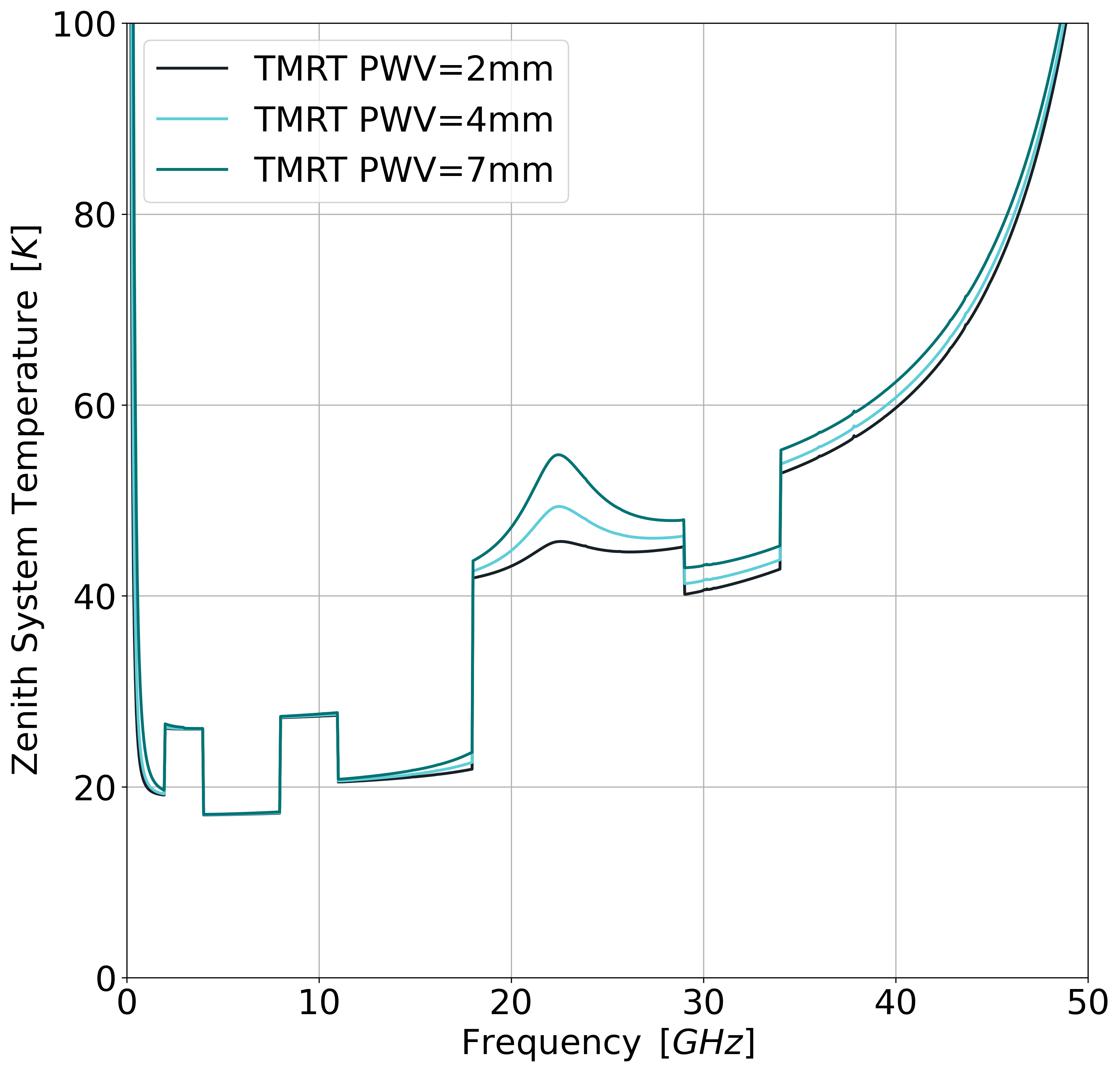}
	\includegraphics[width=0.3\linewidth]{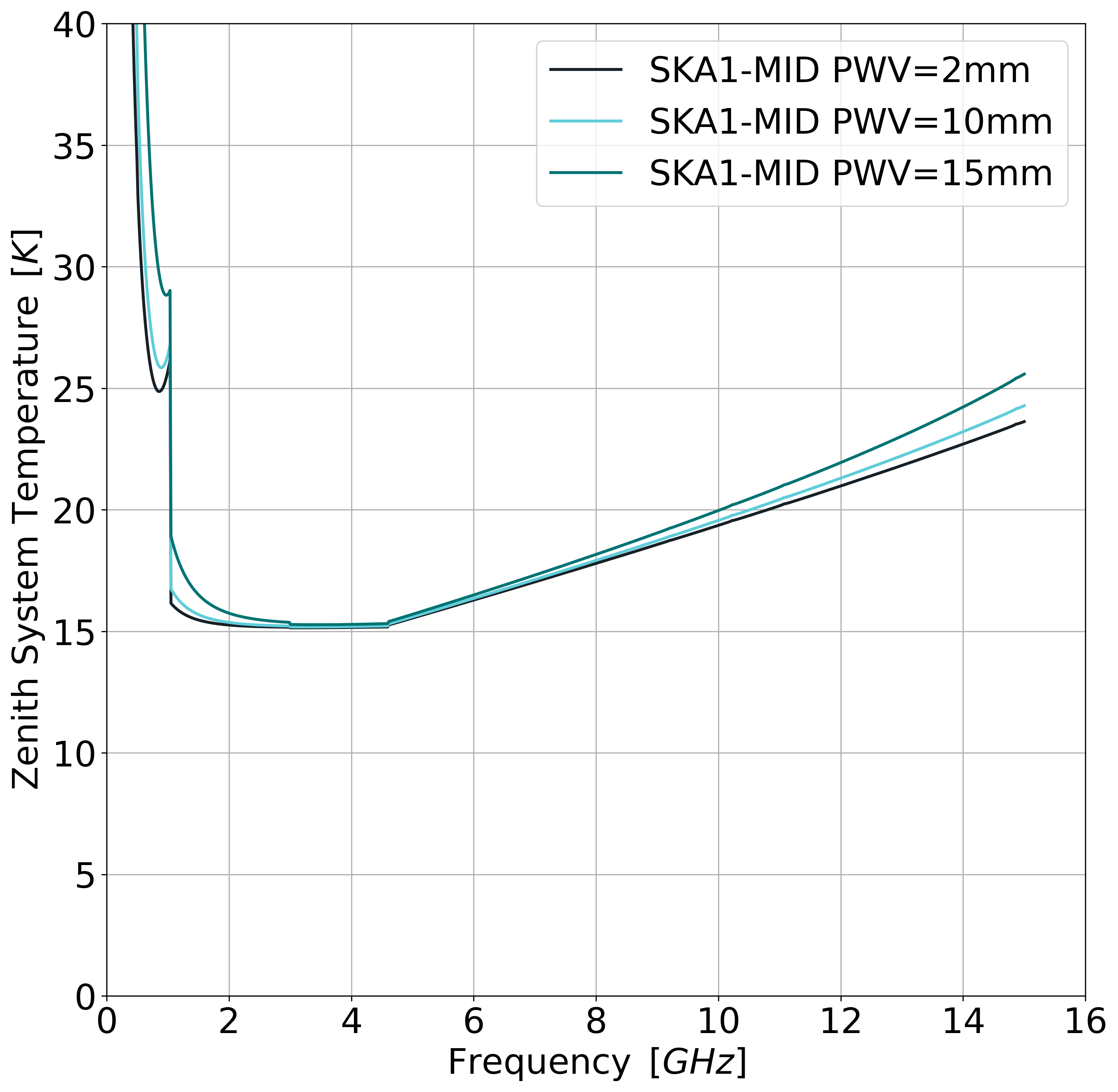}
	\includegraphics[width=0.3\linewidth]{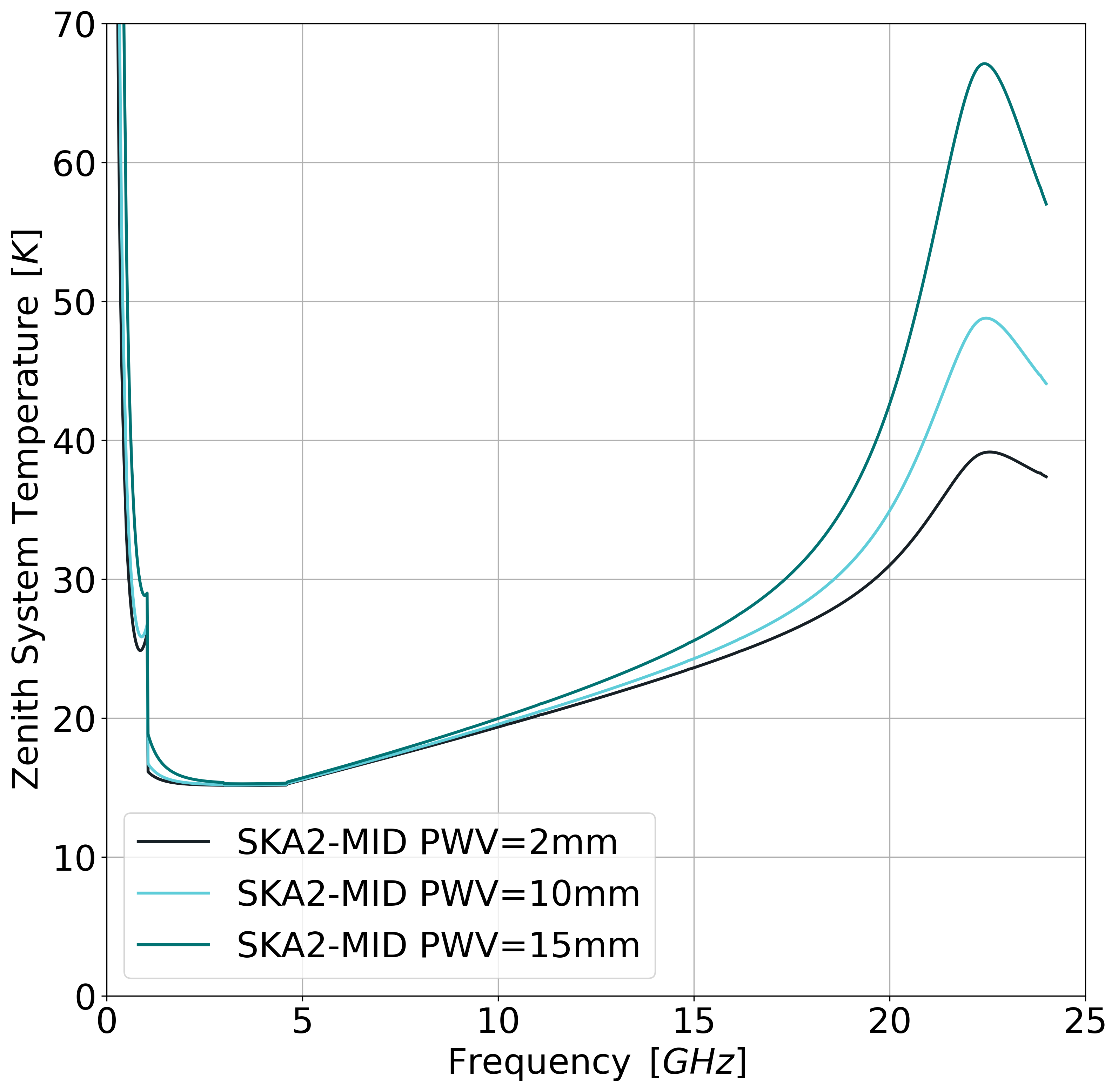}
	\includegraphics[width=0.3\linewidth]{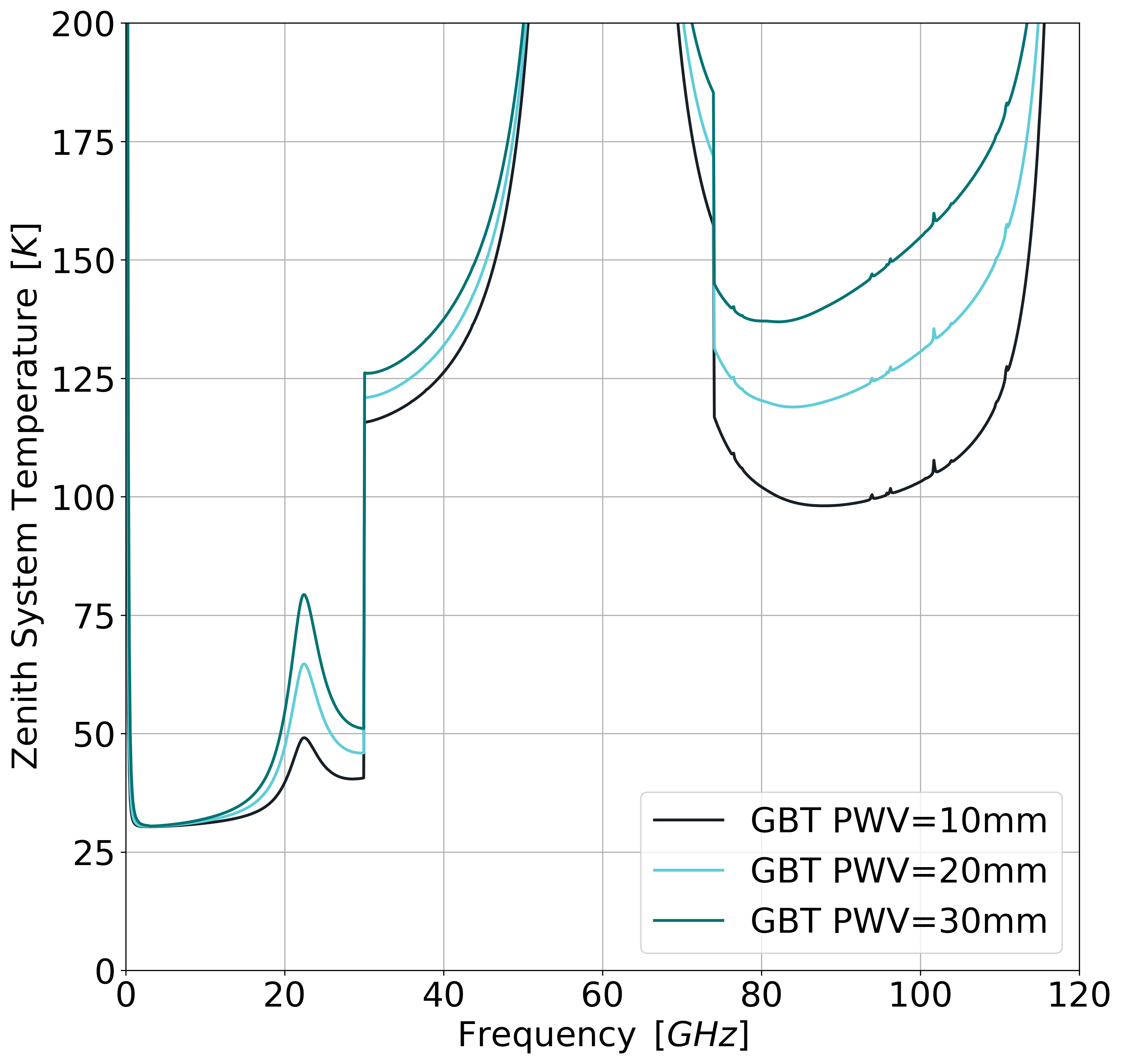}
	\includegraphics[width=0.3\linewidth]{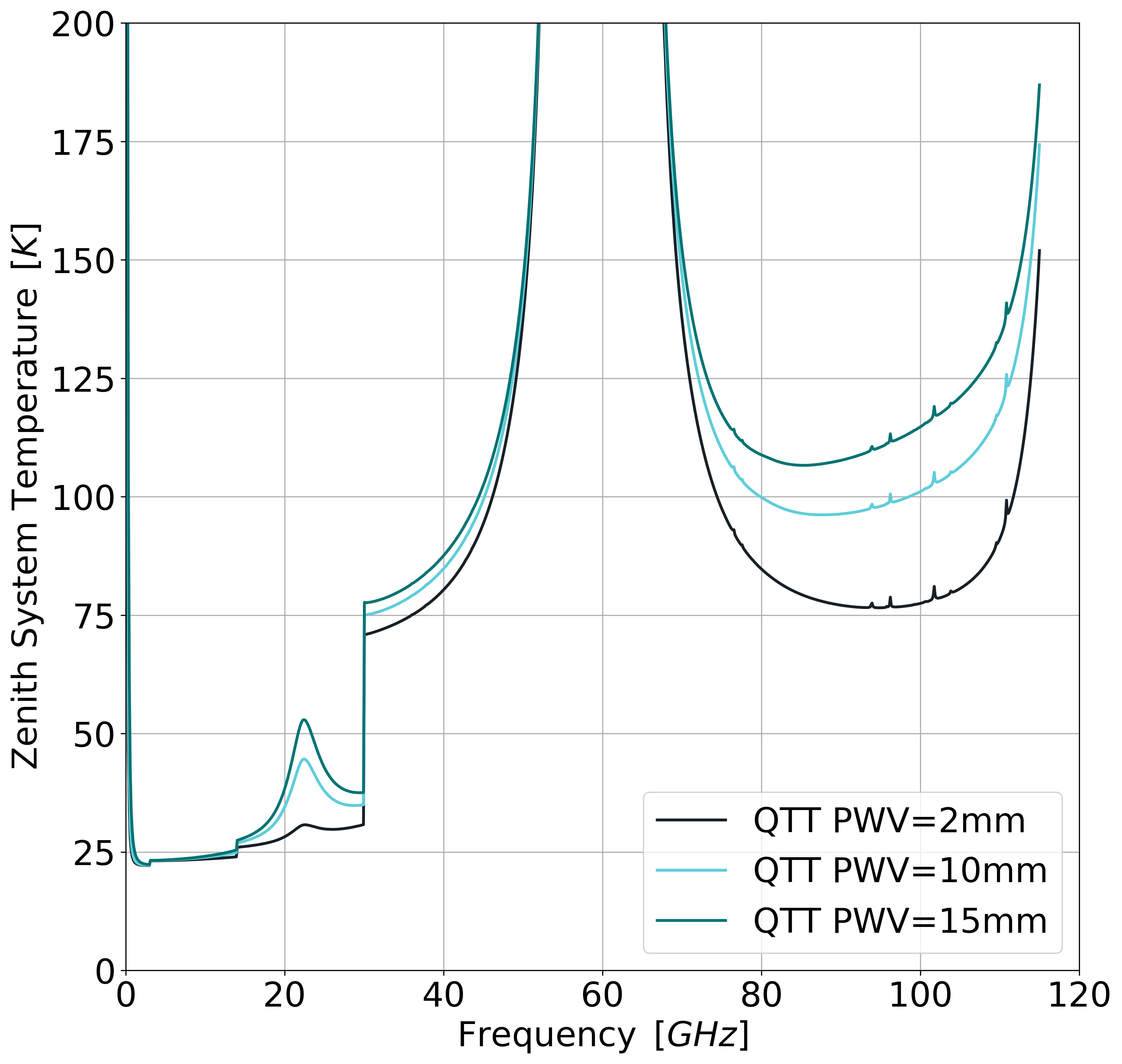}
	\caption{The simulation results of the zenith system temperature with frequency change of six telescopes. The telescopes are: (top panel) FAST, TMRT, and SKA1-MID; (bottom panel) SKA2-MID, GBT, and QTT, from left to right in order. The black, wathet blue, and dark blue lines correspond to the $10^{th}$, $50^{th}$, and $90^{th}$ percentiles of those distributions, respectively.}
	\label{fig:system-temperature}
\end{figure*}

Finally, we present the simulation results of the telescope's zenith atmosphere temperature with observation frequency in Fig. \ref{fig:system-temperature}. We use different colors to show the three different average annual precipitable water vapors, just like with zenith atmosphere opacity.  For our final results, we use the simulation results of the system temperature at the $50^{th}$ astrophysical background temperature and the average annual precipitable water vapor. These telescope parameters are summarized up in Table \ref{tab:telescope-parameters-2} at appendix \ref{sec:quantities-rewritten-in-the-nautural-lorentz-heaviside-untis} based on the size of the telescope.

\subsection{The signal-to-noise ratio.}
\label{sec:the-signal-to-noise-ratio}

As we will show in Section \ref{Results}, we use two types of radio telescopes to discuss the signals of two different timescales, so we need to calculate four signal-to-noise ratios (S/Rs). In this subsection, we discuss the four S/Rs separately.

For a single-dish telescope such as FAST, in the persistent event case, the S/R is equal to the flux $F$ of the radio sources divided by the error of the measurement $\Delta F$:
\begin{equation}
	\mathrm{S/R}=\frac{F}{\Delta F}=\frac{F\sqrt{n_\mathrm{pol}\Delta\nu\Delta t_{\mathrm{int}}}}{\mathrm{SEFD}},\label{snr-single}
\end{equation}
where $n_{\mathrm{pol}}$ denotes the number of polarization channels, and $\Delta\nu$ signifies the frequency bandwidth associated with each radio telescope's observations of neutron stars; and $\Delta t_{\mathrm{int}}$ indicates the integration time $\Delta t_{\mathrm{int}}=t_{\mathrm{sur}}\tau$, where $t_{\mathrm{sur}}$ represents the total time survey duration and $\tau=\lambda_{\gamma}/2\pi D$ is the fractional time that an object on the celestial equator transits the field of view of the telescope; and SEFD is the system-equivalent flux density mentioned earlier \citep{Yu:2013bia}. Therefore, it is evident that a longer observation time for the same target source will result in a higher S/R. However, during the actual observation, the observer will not have unlimited time. The location of the target source, its proper motion, and the radio telescope's performance all influence the accessible observational time.We can select the target source based on the actual situation to extend the observational time, but the telescope can not perform as expected. For example, for FAST, which is not a fully steerable telescope and has the largest target tracking time approximately 6 hours. Therefore, we estimate the observational time for all telescopes used to observe to be 6 hours \citep{2020RAA....20...64J}. For a transient event, the formula for calculating the energy flux becomes an integral: $F_\mathrm{peak}=\frac{1}{\omega_2-\omega_1} \int_{\omega_1}^{\omega_2} F(\omega) d \omega$ and the integral time changes from the observation time to the duration of the signal \citep{Lorimer:2013roa}. This limits the minimum sampling time of the telescope, and the approximate order of magnitude of the duration of the signal is mentioned in Sec. \ref{Results} and in the second subsection of Sec. \ref{methods}. For example, this means that observations of transient signals cannot be received by using FAST's SETI backend, which has a sampling time of 10 seconds \citep{2023AJ....165..132L,2023AJ....166..190T,2022AJ....164..160T,2023AJ....166..245H}.

For a radio interferometer like SKA in the persistent event case, we treat the signal as a collection of synthesized beams. Therefore, we start with the assumption that the observed signal is Gaussian, that each signal is independent of the other, and that there is only one kind of GW. And these assumptions are sufficient for discussion. Then, we can obtain the likelihood function $\tilde{\mathcal{L}}(\boldsymbol{\theta}_{set})$ of observational energy flux $F_{\mathrm{obs},i}$, accounting for the measurement error $\sigma_{i}$ and theoretical model energy flux $F_{\mathrm{model},i}$
\begin{equation}
	\begin{aligned}
		\tilde{\mathcal{L}}(\boldsymbol{\theta}_{set}) & =P\left(F_{\mathrm{obs},i}\mid \boldsymbol{\theta}_{set}\right) \\
		&=\left(\prod_i^{N_{\mathrm{syn}}} \frac{1}{\sqrt{2 \pi \sigma_{i}^2}}\right) \exp \left(-\frac{\chi^2}{2}\right),\label{likelihood}
	\end{aligned}
\end{equation}
where $\boldsymbol{\theta}_{set}$ is an idiographic set of parameters such as the one-sided power spectral density of GWs $S_h(\omega)$, the amplitude of GWs $\Omega_{\mathrm{GW}}$, the characteristic strain of GWs $h_c$, and so on. And the $\chi^2$ statistic is
\begin{equation}
	\chi^2=\sum_i^{N_{\mathrm{syn}}} \frac{\left[F_{\mathrm{model},i}\left( \boldsymbol{\theta}_{set}\right)-F_{\mathrm{sys},i}-F_{\mathrm{obs},i}\right]^2}{\sigma_{i}^2},
\end{equation}
where $F_{\mathrm{sys},i}$ is the predicted background flux density for the single synthesized, which includes the system-equivalent flux density and other known astrophysical processes flux density. The single-dish error $\Delta F$ can be used to estimate the measurement error $\sigma_{i}$. Therefore, we can obtain the S/R from the $\chi^2$ statistic \citep{Hook:2018iia}
\begin{equation}
	\mathrm{S/R}=\sqrt{n_{\mathrm{pol}}\mathrm{Gain}_{\mathrm{array}}^2 \sum_{i=1}^{N_{\mathrm{syn}}} \frac{F_{\mathrm{model},i}^2\Delta\nu_i \Delta t_{\mathrm{int},i}}{T_{\mathrm{sys},i}^2}},
\end{equation}
where $\mathrm{Gain}_{\mathrm{array}}=\mathrm{Gain}\sqrt{N(N-1)}$  represents the array gain for that particular synthesized beam, based on the individual antennas and the number of array elements $N$. For each synthesized beam, there is a single signal flux $F_{\mathrm{model},i}$, a single integrated time $\Delta t_{\mathrm{int},i}$, a single beam's frequency resolution $\Delta\nu_{i}$, and a single system temperature $T_{\mathrm{sys},i}$. We are aware that another possibility is the simultaneous observation of mixed multiple GW signals, a situation we will discuss in our future research, which will require significant effort. Because statisticians have discussed the construction of a likelihood as the product of sub-likelihoods under the term composite likelihood inference, they discover that the statistical foundation is not solid. And there does not seem to exist a general strategy for constructing a combination of composite likelihoods that is both computationally convenient and statistically appealing. One can refer to these papers \citep{Hong:2023ieq,article-Composite-Likelihood,7ba3ad00-48e3-3e1a-bb81-112e59ea3834,ae96fe4b-cd0b-3a2b-9a88-5bd548187d36,cfad4d52-869f-35f3-8652-8534d93d28ad,23c7e5f7-f5a5-3aad-91bb-4641d35779f8}. Finally, we modify the single signal flux $F_{\mathrm{model},i}$ and the single integrate time $\Delta t_{\mathrm{int},i}$ for radio interferometers in transient events, respectively, to the single peak density flux $F_{\mathrm{peak},i}$ and the single signal duration $\Delta t_{\mathrm{sam},i}$.

\section{Results}
\label{Results}
\subsection{Anticipated observational signal.}
With these preconditions, we can simulate the expected signal shapes for both timeframes. We use FAST as our telescope to generate simulation signals. The telescope's and neutron star's parameters for generating the simulation signal with FAST are as follows: The center frequency of the telescope is $1420\mathrm{~MHz}$, the bandwidth is $500\mathrm{~MHz}$, the number of sampling channels is 4000, and the data sampling time is $98.304\mathrm{~us}$. For a magnetar or pulsar, we set its magnetic field as $B=1.85\times 10^{14}\mathrm{~Gauss}$, its spin period as $P=5.76\mathrm{~Sec}$, and distance $d=2\mathrm{~kpc}$ exactly as PSR J0501+4516. Since the energy of the final signal requires studying the system-equivalent flux density of the telescope, we will discuss the parameters of the GW in Sec. \ref{Results}. As discussed earlier, for the duration of the signal, the transient signal depends on the conversion probability with the distance the GW travels through the magnetic field, while the persistent signal is present throughout the entire observation process. We inject the $1\sigma$ Gaussian noise in the simulation process, and the simulated signal energy flux is at least $5\sigma$ relative to the FAST minimum detectable energy flux. The results of the simulation are shown in Fig. \ref{fig:anticipated-signal} of the Sec. \ref{Results}. After determining the shape of the radio signal that we have observed, we can proceed to analyze the specific spectral line properties of the source responsible for generating the signal. 

Reviewing the discussion in Section \ref{subsec:gw-photon mixing and photon specific intensity}, the distance at which this signal is generated is much greater than the radius of the optical column surface of the neutron star. And this distance decreases as the frequency of the GW increases and the spin period of the magnetar increases. resulting in a signal that looks like a fast radio burst, but its signal duration is much longer than the fast radio burst \citep{Zhang:2022uzl}. Therefore, we refer to this signal as a ``novel rapid optical variation signal".

For this ``novel rapid optical variation signal", from the conversion probability of GWs to photons, as long as the GW enters the magnetic field range of the neutron star, photons can be generated, so the overall signal can theoretically be an integral result. However, if the space-accumulation effect of the GZ effect is insufficient, it leads to a small number of produced photons, low energy flux, and a faint signal. With the increase of GW frequency, the height of electromagnetic wave emission decreases, the magnetic field increases, and the conversion probability increases. Depending on the behavior of GWs, signals can be divided into transient signals and persistent signals, which are going to be discussed in Section \ref{Results}. Therefore, without considering the dispersion in the propagation process, for a transient signal, after the observed frequency is fixed, the complete signal should present a likely bell-shaped signal on a two-dimensional graph where the horizontal axis is time and the vertical axis is energy flux. The bell-shaped signal exhibits asymmetry in the distribution of peak energy flux between its left and right sides. The slope of the signal curve to the left of the peak is significantly greater than the slope of the signal curve to the right of the peak, which is estimated by the variational trend of the conversion probability. And the timeline shows that the signal to the left of the peak lasts much less than the signal to the right.  In the whole observation frequency, on the two-dimensional heat map, the horizontal axis is the time, the vertical axis is the observation frequency, and the color bar of the heat map represents the signal intensity. The heat map should display a trapezoid signal, with the intensity increasing as the frequency increases. Similarly, when neglecting the effects of dispersion during signal propagation, for a persistent signal, after the observational frequency is fixed, the complete signal presents an overall integrated profile composed of transient signals on a two-dimensional graph where the horizontal axis is time and the vertical axis is energy flux. After carefully deducting the effects of background noise, it should be an almost straight line, with only some slight intensity perturbation. In the whole observation frequency range, a trapezoidal signal should be presented on the two-dimensional heat map with time as the horizontal axis and observation frequency as the vertical axis, and the color bar of the heat map represents the signal intensity, and the intensity increases with the increase of frequency.

The primary difference among various radio telescopes regarding signal duration is in the transient signal. Due to the different sensitivity of the telescope, the time of the transient signal detected by the high-sensitivity telescope is longer. Since there is a signal whenever a GW enters the neutron star's magnetic field, it can only be observed when the signal strength is strong, so the duration of the signal is calculated based on the conversion probability in Figs. \ref{figapp:conversion-probability-times} and \ref{figapp:conversion-probability-plus}, the energy density of the GW, and the neutron star's own properties such as rotation period, magnetic field strength, and other characteristics. Based on the current calculation results, a fixed characteristic amplitude $h_c$ of the GW leads to an increase in the transient signal's duration as the neutron star's magnetic field intensity increases.

Furthermore, for the spectral index, we can use $S\propto\alpha\nu^{\beta}$ to model the change of the energy flux $S$ with the frequency $\nu$, where $\beta$ is the spectral index, approximately equal to 1, $\alpha$ corresponds to the GW energy density $h_0^2\Omega_{\mathrm{GW}}(\nu)$ and the neutron star conversion probability $\left\langle P_{g \rightarrow \gamma}^{\mathrm{coherence}}(\omega,L,\theta)\right\rangle$ is a fitting constant. We can theoretically determine the conversion probability of the observational neutron star first, which enables us to derive the change in GW energy density with respect to frequency.

\begin{figure}
	\centering
	\includegraphics[width=0.9\linewidth]{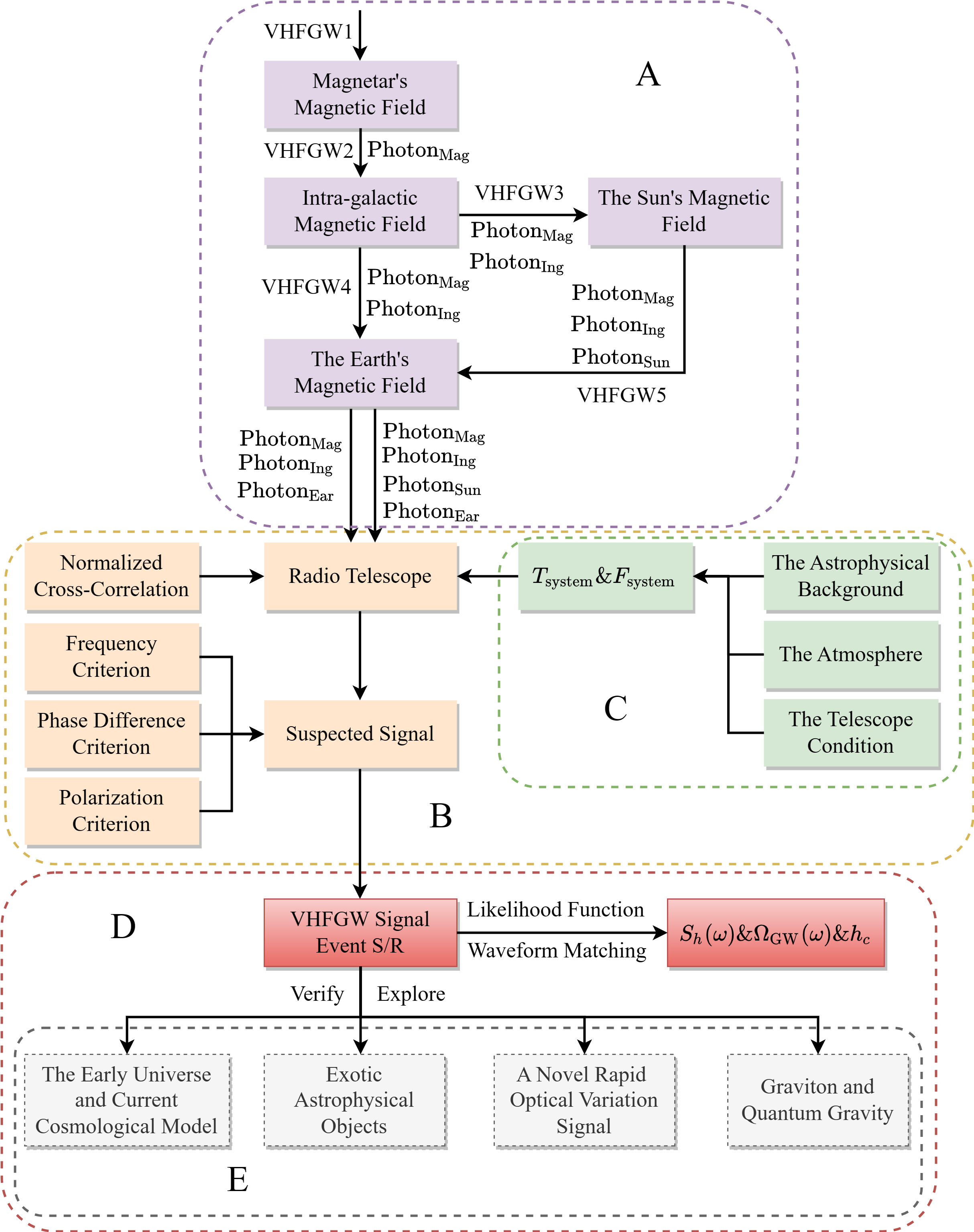}
	\caption{The diagram illustrates the entire process from signal generation to analysis. The process can be divided into several parts: (A) photon generation and propagation path; (B) radio signal processing pipeline; (C) simulation of partial telescope parameters; (D) very high-frequency GW signal and parameter constraint; and (E) motivation to observe very high-frequency GWs with radio telescopes.}
	\label{fig:overall-process}
\end{figure}

Combining the spatial-accumulation effects of the GZ effect, we endeavor to take full account of VHF GWs passing through magnetic fields around neutron stars. By doing so, we can increase the detectable field of view and then improve our detection capabilities along the trajectory along which GWs travel. The photons undergo an extensive and complex radiation process before being captured by radio telescopes. The process of very high-frequency GWs and photons traversing the magnetic field, distinguishing the signals, and the motivation for detection are illustrated in Fig. \ref{fig:overall-process}. The entire process is as follows: (A) The intense magnetic field of the neutron star creates an effective setting for the GZ effect to converse gravitational waves into electromagnetic waves. The spatial-accumulation effect results in coherent superposition of the conversion probability of gravitational waves as they traverse the magnetic field along the magnetospheric path. The impact of the residual magnetic fields along the observation path of neutron stars can be categorized into interstellar magnetic fields \citep{Domcke:2020yzq} and planetary magnetic fields \citep{Liu:2023mll}. In comparison, the intensity of electromagnetic wave signals converted from gravitational waves in these magnetic fields is relatively low. The signal intensity produced by the interstellar magnetic field is $10^2$ times lower than that conversed by the neutron star magnetic field, while the signal intensity from the planetary atmospheric magnetic field is also $10^4$ times lower than that of the neutron star magnetic field. The effect on our observations is minimal. The propagation path of the generated electromagnetic wave is influenced by the density distribution of the magnetospheric plasma. Free electrons induce a frequency shift in electromagnetic waves via Compton scattering, whereas the Faraday rotation effect alters the polarization angle, leading to variations in Stokes parameters associated with the magnetic field direction. The signals obtained from radio telescopes. (B) and (C) require multi-level processing to delineate the characteristics of gravitational waves. The performance variations among various radio telescopes directly influence the detection thresholds. This paper investigates the use of the equivalent flux density of the telescope system as a reference for background noise. It employs RM synthesis for polarization decomposition and a time-frequency heat map to identify the bell-shaped profile of transient signals and the broadband continuous spectrum of these signals. (D) By aligning the signal template with the observed data, the essential parameters of gravitational waves can be determined: energy density and characteristic strain. The mass distribution of primordial black holes and the predicted parameters of quantum gravity theory can be constrained through multi-band observations. (E) Scientific motivation. The exploration of VHF GWs, particularly in the MHz-GHz range, has opened a new avenue for probing physics beyond the standard cosmological model. Such frequency bands are largely inaccessible to current ground-based interferometers but may carry imprints of early-universe phenomena and fundamental interactions at energy scales close to the Planck scale. Here are some inspiring scientific goals. One compelling motivation arises from the inflationary paradigm. In canonical models of slow-roll inflation, the primordial tensor spectrum is predicted to be red-tilted, leading to an exponential suppression of gravitational wave amplitudes at high frequencies. Consequently, any detectable relic signal in the GHz regime would challenge the standard inflationary framework. As discussed by \cite{Vagnozzi:2022qmc}, the presence of a cosmic graviton background (CGB) at such frequencies would lend support to alternative scenarios, including bouncing and emergent cosmologies, which generically predict enhanced high-frequency tensor power. Detection of such a signal would provide an empirical handle on trans-Planckian physics and the microphysical origin of the early universe. VHF GWs also emerge naturally from the dynamics of PBHs. PBHs can form from enhanced scalar perturbations in the early universe and may subsequently merge to generate a stochastic gravitational wave background (SGWB) peaking in the MHz–GHz band. As shown by \cite{Kohri:2024qpd}, the shape and amplitude of this SGWB are sensitive to the initial PBH mass distribution and formation history. Hence, constraints or detection of such a background would offer crucial insights into the physics of PBH formation and early-universe density fluctuations. In addition, topological defects such as cosmic strings, predicted by various high-energy symmetry-breaking mechanisms, provide another robust source of high-frequency gravitational radiation. Oscillating string loops can emit bursts of gravitational waves through the formation of cusps and kinks, contributing to a distinct SGWB spectrum. The recent analysis by \cite{Aggarwal:2025noe} emphasizes that GHz-band detectors could potentially identify the unique non-thermal spectral features associated with cosmic strings, thereby offering an observational probe of beyond-Standard-Model phase transitions. Finally, HFGWs provide a potential observational window into quantum gravity. In particular, loop quantum cosmology predicts a strongly blue-tilted tensor spectrum due to a superinflationary phase preceding standard expansion. \cite{Copeland:2008kz} demonstrate that this leads to an enhancement of tensor modes at high frequencies while suppressing large-scale signals. Such a spectral signature, if observed, would offer direct evidence for quantum gravitational corrections to classical spacetime dynamics.

Based on the behavior of VHF GWs and the width of the magnetic field, the radio signals can be categorized into two principal classes: (a) transient and (b) persistent signals. Transient radio signals we anticipated observing are more closely associated with GW transients generated by low-mass binary system mergers, whereas persistent radio signals align more closely with a primordial stochastic GW background from the early universe. A photon signal is unequivocally transient when one of several GW signals traverses the magnetic fields of spheroidal celestial bodies, and the temporal gap between the current and the most recent GW signals exceeds the typical time scales of $\sim 100\mathrm{~seconds}$ for magnetars or pulsars, $10\mathrm{~seconds}$ for the Sun, and $10^{-2}\mathrm{~seconds}$ for the Earth, characterized by the typical length scale $t=L/c$ at $1\mathrm{~GHz}$. This characterization is elucidated in Section \ref{methods}. The duration of the transient signal is theoretically determined by three major factors: First, the neutron star's properties, including its magnetic field strength, rotation period, radius, and the density of surrounding charged matter, determine the conversion probability of GWs to electromagnetic waves. Second, the source of GWs determines the energy density of itself. Third, the temperature of the system, which includes the astrophysical background temperature, the atmosphere temperature, and the telescope electronics temperature, determines the sensitivity of the telescope. Therefore, a distinctly continuous signal emerges when a cohort of very high-frequency GWs passes through the magnetic field at intervals shorter than the typical time scales or when several very high-frequency GW signals sporadically traverse the interstellar magnetic fields.

\subsection{Detection sensitivity.} 
\begin{figure*}
	\centering
	\includegraphics[width=0.42\linewidth]{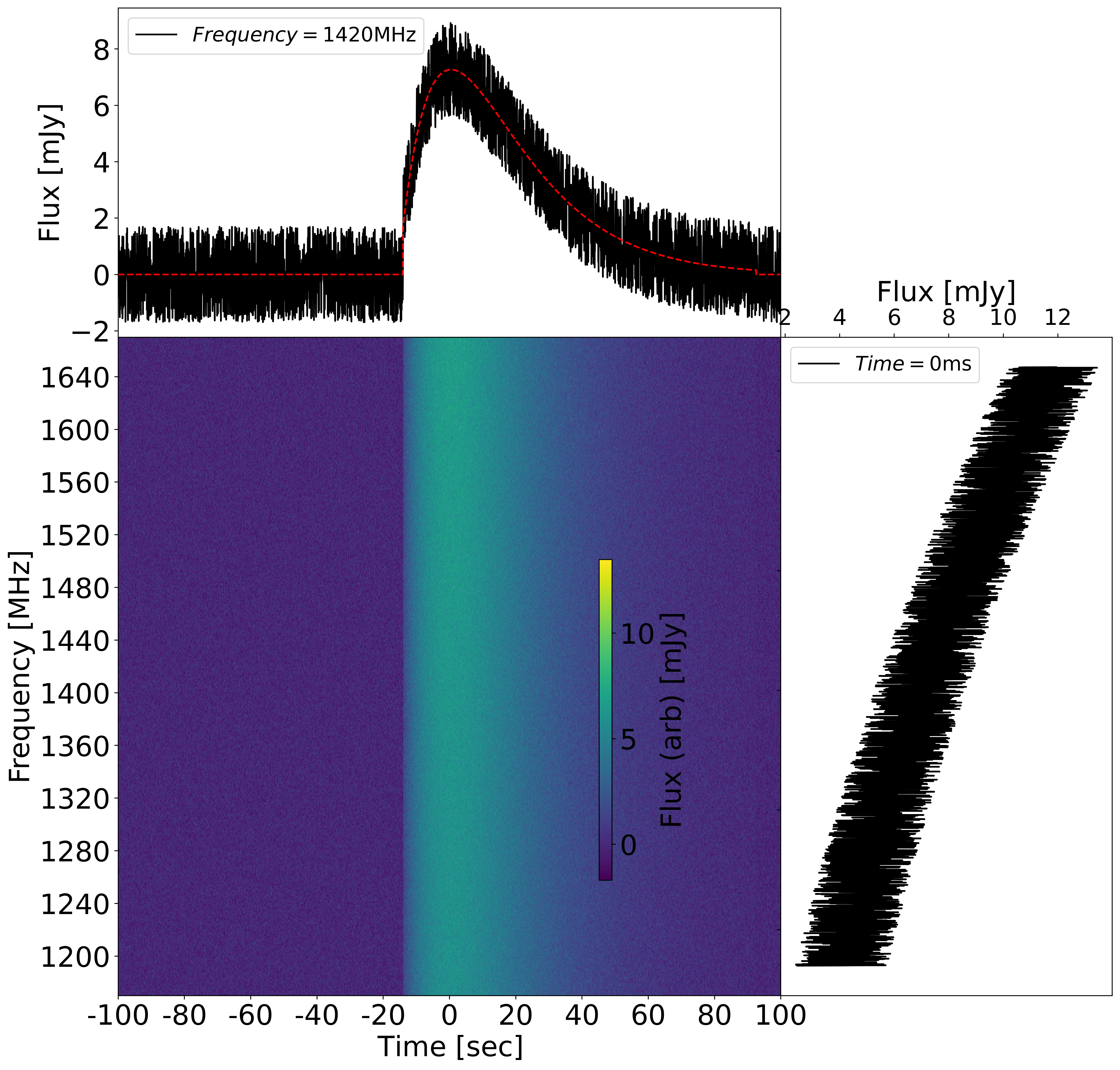}
	\includegraphics[width=0.42\linewidth]{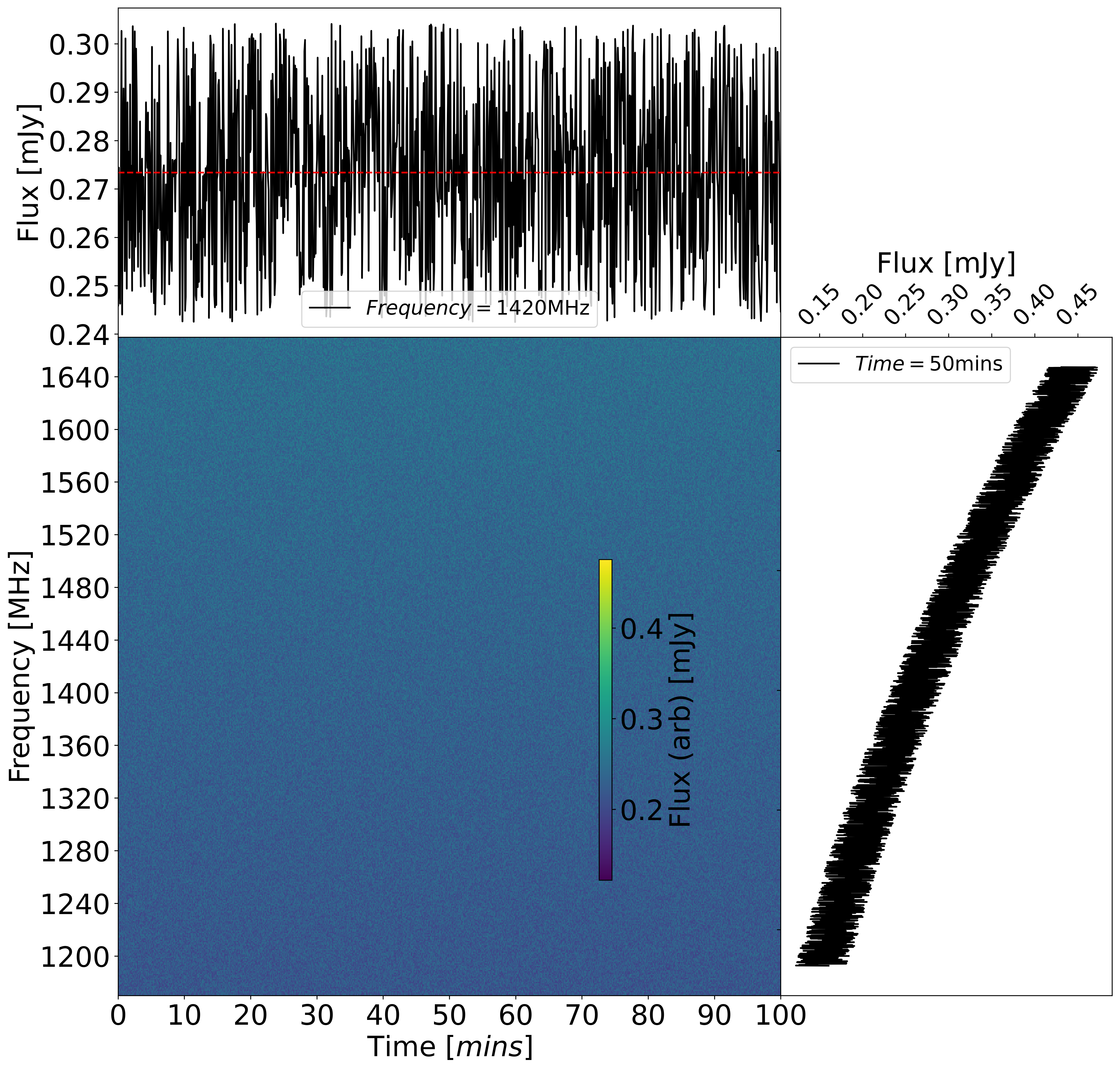}
	\caption{The anticipated observational signals of (left panel) transient and (right panel) persistent events. We show the change of energy flux with observation time at the center frequency of FAST $1420\mathrm{~MHz}$, the change of energy flux with observation frequency in the bandwidth range of $500\mathrm{~MHz}$ at the center time, and the waterfall diagram of energy flux with time and frequency.}
	\label{fig:anticipated-signal}
\end{figure*}

In this research, we consider six radio telescopes: (a) single-dish telescopes: Five-hundred-meter Aperture Spherical Radio Telescope (FAST) \citep{Jiang:2019rnj,2020RAA....20...64J,2020Innov...100053Q,fastref}, TianMa Radio Telescope (TMRT) \citep{tmrtref,tmrtsr,tmrtintr,tmrtpar}, QiTai Radio Telescope (QTT) \citep{Wang:2023occ,qttref}, and Green Bank Telescope (GBT) \citep{gbtref,GBT_Proposer_Guide}; and (b) the radio interferometers Square Kilometre Array Phase 1 MID-197 (SKA1-MID) \citep{skaref,2019arXiv191212699B} and Square Kilometre Array Phase 2 MID-2000 (SKA2-MID) \citep{skaref}, utilized for capturing radio signals transmuted by very high-frequency GWs. By integrating the aforementioned three criteria, we simulate and depict the anticipated dual types of radio signals expected to be received by FAST in Fig. \ref{fig:anticipated-signal}. From the previous discussion, we can get that the system temperatures of FAST are $27.389\mathrm{~K}$ from $1170\mathrm{~MHz}$ and $26.829\mathrm{~K}$ to $1670\mathrm{~MHz}$. Therefore, we can obtain the minimum detectable energy fluxes for persistent events from $3.093\times 10^{-5}\mathrm{~Jy}$ to $3.158\times 10^{-5}\mathrm{~Jy}$ that FAST can receive for six hours of observational time at a $500\mathrm{~MHz}$ bandwidth with $1420\mathrm{~MHz}$ as the center; these are the energy fluxes of $1\sigma$. Obviously, the energy fluxes of $5\sigma$ are $1.547\times 10^{-4}\mathrm{~Jy}$ and $1.579\times 10^{-4}\mathrm{~Jy}$. Correspondingly, outside $2\mathrm{~kpc}$, the energy fluxes of $5\sigma$ generated by the conversion of GWs into photons in the neutron star's magnetic field are $1.875\times10^{11}\mathrm{~Jy}$ and $1.914\times10^{11}\mathrm{~Jy}$. In combination with Figs. \ref{figapp:conversion-probability-times}, \ref{figapp:conversion-probability-plus}, \ref{figapp:conversion-probability-total}, and Appendix \ref{sec:quantities-rewritten-in-the-nautural-lorentz-heaviside-untis}, we can calculate that $h^{2}_{0}\Omega_{\mathrm{GW}}$ must be at least 0.05 for the overall observed energy to have a S/R higher than 5. Similarly, the minimum detectable energy fluxes for transient events from $1170\mathrm{~MHz}$ to $1670\mathrm{~MHz}$ are $1.695\times 10^{-3}\mathrm{~Jy}$ and $2.025\times 10^{-3}\mathrm{~Jy}$. Clearly, the energy fluxes of $5\sigma$ are $8.073\times 10^{-3}\mathrm{~Jy}$ and $1.012\times 10^{-2}\mathrm{~Jy}$. Accordingly, outside $2\mathrm{~kpc}$, the energy fluxes of $5\sigma$ generated by the conversion of GWs into photons in the neutron star's magnetic field are $1.027\times10^{13}\mathrm{~Jy}$ and $1.227\times10^{13}\mathrm{~Jy}$. As with the persistent events, we can calculate that $h^{2}_{0}\Omega_{\mathrm{GW}}$ must be at least 8 for the overall observed energy to have a S/R higher than 5. To make this picture, we assume the energy of the GW $h^{2}_{0}\Omega_{\mathrm{GW}}\approx 3.142\times 10^{9}$ stays the same over the bandwidth $500\mathrm{~MHz}$. And we also assume the characteristic strains of the GW from $h_c\approx 7.079\times10^{-23}$ to $h_c\approx 1.011\times10^{-22}$ are varying with frequency for persistent events. And, we assume the energy of the GW $h^{2}_{0}\Omega_{\mathrm{GW}}\approx 5.027\times 10^{11}$ stays the same over the bandwidth $500\mathrm{~MHz}$. And we also assume the characteristic strains of the GW from $h_c\approx 1.278\times10^{-21}$ to $h_c\approx 8.955\times10^{-22}$ are varying with frequency for transient events. It needs to be emphasized that the signal intensities shown in Fig. \ref{fig:anticipated-signal} do not describe a comprehensive representation of our incoming observations. Instead, they solely illustrate the dynamic nature of the signal. Fig. \ref{fig:anticipated-signal} (left panel) showcases the transient signal converted from PSR J0501+4516. The sudden, bright peak represents this signal, delineated by the red dashed line encompassing the entire signal. The strength of the signal depends on $\Omega_{\mathrm{GW}}(\omega)$, while the temporal width correlates with the magnetic field width of the magnetar and the frequency of the GW. In a rough estimation of a bell-like signal at the frequency of $1420\mathrm{~MHz}$, the duration of the signal to the left of the peak is $\sim\frac{3\times 10^6\mathrm{~km}-3\times 10^5\mathrm{~km}}{3\times 10^5 \mathrm{~km/s}}\times 2=18\mathrm{~s}$, and the duration of the signal to the right of the peak is $\sim\frac{1.5\times10^7\mathrm{~km}-3\times 10^6\mathrm{~km}}{3\times 10^5 \mathrm{~km/s}}\times 2=80.0\mathrm{~s}$. Fig. \ref{fig:anticipated-signal} (right panel) illustrates the persistent signal from PSR J0501+4516. The red dashed line also delineates the entire signal. Unlike the transient signal, this persistent signal is expected to persist throughout the entirety of the observation process and be detectable across the entire observation frequency spectrum.

Integrating the aforementioned discourse on the two signal typologies, we contemplate the signal-to-noise ratio for single-dish telescopes in persistent event scenarios, which equals the flux energy $F$ of the radio sources divided by the measurement error $\Delta F$, $\mathrm{S/R}=\frac{F}{\Delta F}=\frac{F\sqrt{n_\mathrm{pol}\Delta\nu\Delta t_{\mathrm{int}}}}{\mathrm{SEFD}}$, where $n_{\mathrm{pol}}$ denotes the number of polarization channels, and $\Delta\nu$ stands for the bandwidth, which we regard as the bandwidth for each radio telescope's observation of magnetars. $\Delta t_{\mathrm{int}}=t_{\mathrm{sur}}\tau$ represents the integration time, where $t_{\mathrm{sur}}$ signifies the total observation time, and $\tau=\lambda_{\gamma}/2\pi D$ is the fractional time an object on the celestial equator traverses the telescope's field of view. The SEFD denotes the system equivalent flux density, $\mathrm{SEFD}=\frac{T_\mathrm{sys}(\omega)}{\mathrm{Gain}}$, where $T_\mathrm{sys}(\omega)=T_{\mathrm{AST}}(\omega)+T_{\mathrm{ATM}}(\omega)+T_{\mathrm{RT}}(\omega)$ is the frequency-dependent system temperature comprising (a) astrophysical backgrounds $T_{\mathrm{AST}}(\omega)$, (b) atmospheric contributions $T_{\mathrm{ATM}}(\omega)$, and (c) telescope electronics $T_{\mathrm{RT}}(\omega)$. Subsequently, in transient event scenarios for single-dish telescopes, we should respectively adjust the energy density flux \citep{Lorimer:2013roa} $F$ and integration time $\Delta t_{\mathrm{int}}$ to the peak density flux $F_\mathrm{peak}=\frac{1}{\omega_2-\omega_1} \int_{\omega_1}^{\omega_2} F(\omega) d \omega$ and the sampling time $\Delta t_{\mathrm{sam}}$.

\begin{figure*}
	\centering
	\includegraphics[width=0.47\linewidth]{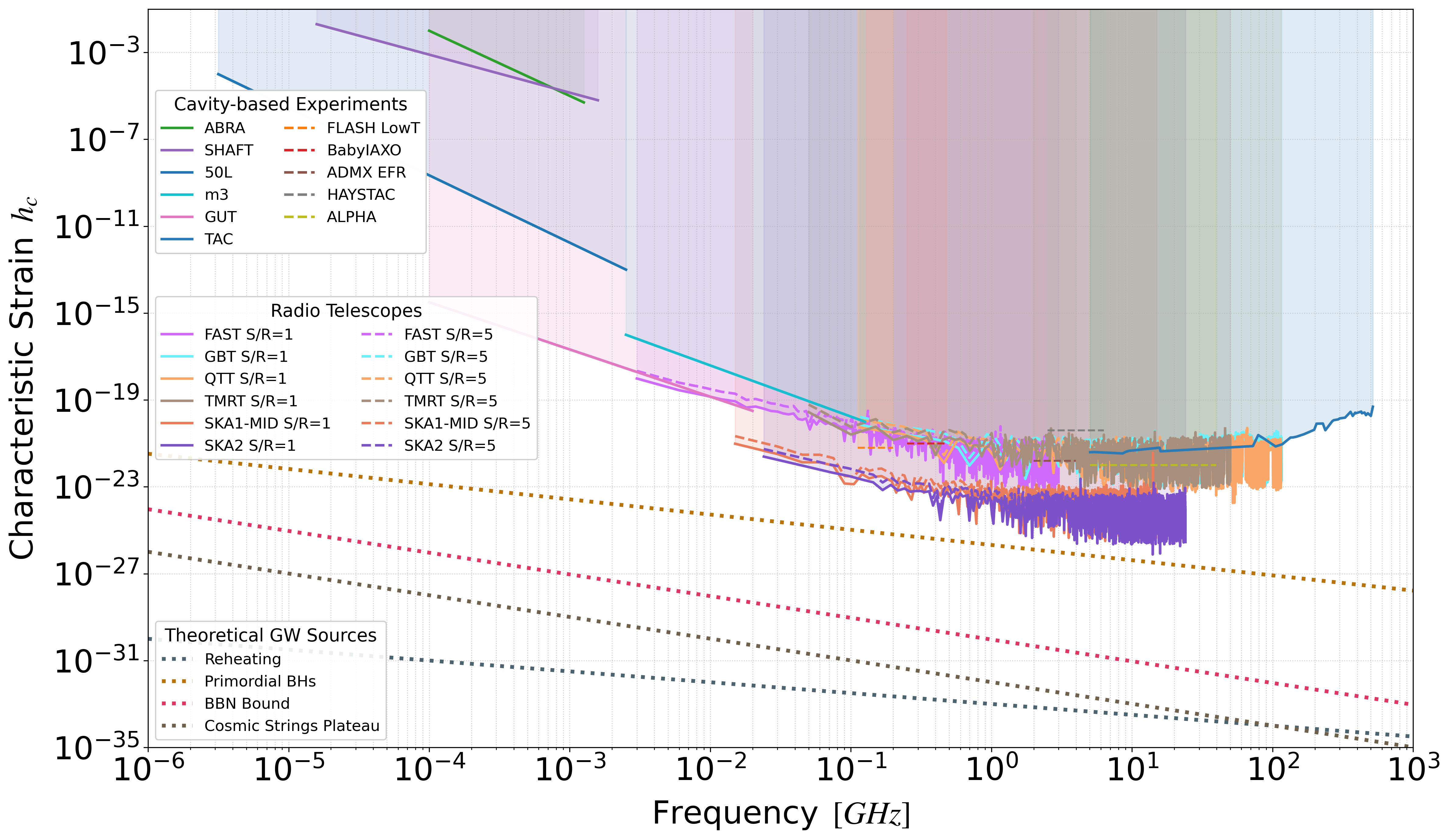}
	\includegraphics[width=0.47\linewidth]{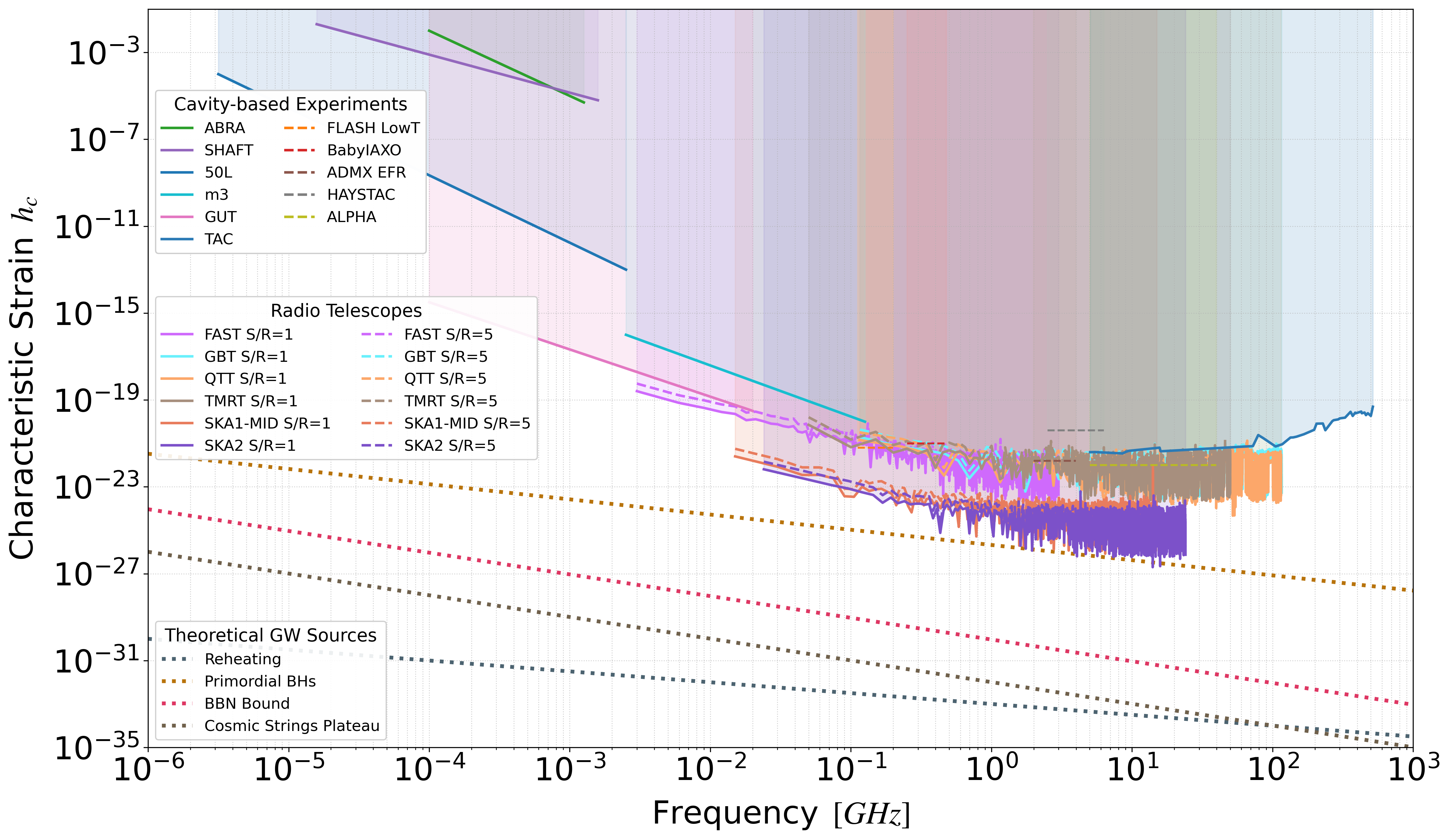}
	\caption{Upper bounds on the very high-frequency GWs derived from six radio telescopes for (left panel) transient and (right panel) persistent events. When the S/R is 1, the colored solid line indicates the telescope detection limit, and when the S/R is 5, the colored dashed line indicates the telescope detection limit. The corresponding color also indicates the scope of the telescope's detection capability. The colored dots and lines show how the characteristic $h_c$ of different GW sources varies with frequency. \added{And various cavity-based experiments indicate the sensitivities of GW, as reported in these studies: \citep{Domcke:2022rgu,Salemi:2021gck,Gramolin:2020ict,Gatti:2024mde,ADMX:2018gho,ADMX:2019uok,ADMX:2021nhd,Alesini:2023qed,Lawson:2019brd,ALPHA:2022rxj,Brubaker:2017ohw,IAXO:2020wwp,https://doi.org/10.1002/andp.202300326,Goryachev:2014yra}.}}
	\label{fig:signal-sensitivity}
\end{figure*}

Additionally, in the case of persistent events with radio interferometers, we have the total auto-correlation $\mathrm{S/R}=\sqrt{n_{\mathrm{pol}}\mathrm{Gain}_{\mathrm{array}}^2 \sum_{i=1}^{N_{\mathrm{syn}}} \frac{F_{\mathrm{model},i}^2\Delta\nu_i \Delta t_{\mathrm{int},i}}{T_{\mathrm{sys},i}^2}}$ \citep{2017isra.book.....T,1989ASIC..283..163W,Hook:2018iia,Safdi:2018oeu}, where $\mathrm{Gain}_{\mathrm{array}}=\mathrm{Gain}\sqrt{N(N-1)}$ represents the array gain for a specific synthesized beam, and $N$ denotes the number of array elements. Each term of the synthesized beam includes a single signal flux $F{\mathrm{model},i}$, a single integration time $\Delta t_{\mathrm{int},i}$, a single beam bandwidth $\Delta\nu$, and a single system temperature $T_{\mathrm{sys},i}$. Similarly to single-dish telescopes, in the case of transient events with radio interferometers, we respectively adjust the single signal flux $F_i$ and the single integration time $\Delta t_{\mathrm{int},i}$ to the single peak density flux $F_{\mathrm{peak},i}$ and the single signal duration $\Delta t_{\mathrm{sam},i}$. We demonstrate the sensitivity of these two anticipated signals with PSR J0501+4516 in Fig. \ref{fig:signal-sensitivity}, combined with some typical sources of very high-frequency GWs \citep{Figueroa:2017vfa,Barman:2023ymn,2016RvMP...88a5004C,Figueroa:2012kw,Auclair:2019wcv}. For persistent signals, we set the observation time to the maximum tracking time of a single FAST observation $t_{\mathrm{obs}}=6\mathrm{~hours}$. Therefore, a detectable $h_0^2\Omega_{\mathrm{GW}}(\omega)$ can be obtained by combining Eqs. \ref{specific-intensity} and \ref{total-energy-flux} and the signal-to-noise ratio $S/R$
\begin{equation}
	\begin{aligned}
		h_0^2\Omega_{\mathrm{GW}}(\omega)=&\frac{4d^2\omega F_i}{30000R_{tot}^2M_{\mathrm{planck}}^2\left\langle P_{g \rightarrow \gamma}(\Omega,\omega,L)\right\rangle}\\
		=&11.4\times\left(\frac{d}{1\mathrm{~kpc}} \right)^2\left(\frac{10^9\mathrm{~km}}{R_{tot}}\right)^2\left(\frac{\omega}{10^6\mathrm{~Hz}}\right)\\
		&\times \left(\frac{10^{-5}}{\left\langle P_{g \rightarrow \gamma}(\Omega,\omega,L)\right\rangle}\right)\left(\frac{F_i}{1\mathrm{~Jy}}\right),
	\end{aligned}
\end{equation}
and the characteristic strain $h_c(\omega)$
	\begin{equation}
		\begin{aligned}
			h_c(\omega)=&\sqrt{\frac{3H_0^2\Omega_{\mathrm{GW}}(\omega)}{2\pi^2\omega^2}}\\
			=&2.26 \times 10^{-24}\left(\frac{d}{1\mathrm{~kpc}} \right)\left(\frac{10^9\mathrm{~km}}{R_{tot}}\right)\left(\frac{10^6\mathrm{~Hz}}{\omega}\right)^{1/2}\\
			&\times \left(\frac{10^{-5}}{\left\langle P_{g \rightarrow \gamma}(\Omega,\omega,L)\right\rangle}\right)^{1/2}\left(\frac{F_i}{1\mathrm{~Jy}}\right)^{1/2},
		\end{aligned}
	\end{equation}
where $F_i$ is the measured energy flux with $i=1$ in $1\sigma$ of S/R and with $i=5$ in $5\sigma$ of S/R. And we summarize these parameters of $h_0^2\Omega_{\mathrm{GW}}(\omega)$, $h_c(\omega)$, and $F_i$ at $1\mathrm{~GHz}$ under 6 hours of observation time in Table. \ref{tab:sample-for-detectable}.
\begin{table*}
	\caption{\label{tab:sample-for-detectable}The list of detection capability of radio telescopes at $1\mathrm{~GHz}$ with 6 hours of observational time.}
	\resizebox{1.0\linewidth}{!}{
		\begin{threeparttable}
			\begin{tabular}{cccccc}
				\toprule
				Telescope&Observation&$1\sigma$ Detectable&$5\sigma$ Detectable&Minimum Detectable Normalized&Minimum Detectable\\
				Name&bandwidth&Energy Flux&Energy Flux&Energy Density $h_0^2\Omega_{\mathrm{GW}}(\omega)$&Characteristic Strain $h_c(\omega)$\\
				&(MHz)&(Jy)\tnote{a}&(Jy)\tnote{b}&(Dimensionless)\tnote{c}&(Dimensionless)\tnote{d}\\ 
				\midrule
				FAST&500&$2.98\times 10^{-5}$&$1.49\times 10^{-4}$&($3.25\times 10^{7}$, $1.62\times 10^{8}$)&($7.20\times 10^{-24}$, $1.61\times 10^{-23}$)\\
				TMRT&1500&$1.12\times 10^{-4}$&$5.64\times 10^{-4}$&($7.55\times 10^{10}$, $6.58\times 10^{12}$)&($3.47\times 10^{-22}$, $3.24\times 10^{-21}$)\\
				SKA1-MID&300&$1.17\times 10^{-9}$&$5.85\times 10^{-9}$&($3.82\times 10^{4}$, $1.92\times 10^{5}$)&($2.47\times 10^{-25}$, $5.53\times 10^{-25}$)\\
				SKA2-MID&300&$1.15\times 10^{-10}$&$5.75\times 10^{-10}$&($1.14\times 10^{4}$, $5.72\times 10^{4}$)&($1.35\times 10^{-25}$, $3.02\times 10^{-25}$)\\
				GBT&800&$1.66\times 10^{-4}$&$8.30\times 10^{-4}$&($3.89\times 10^{10}$, $1.94\times 10^{11}$)&($2.49\times 10^{-22}$, $5.57\times 10^{-22}$)\\
				QTT&1500&$5.91\times 10^{-5}$&$2.96\times 10^{-4}$&($3.21\times 10^{9}$, $1.60\times 10^{10}$)&($7.16\times 10^{-23}$, $1.60\times 10^{-22}$)\\
				\bottomrule
			\end{tabular}
			\begin{tablenotes}
				\footnotesize
				\item[a] The minimum detectable energy flux under the $1\sigma$ condition is derived from the $\Delta F$ estimation in the Eq. \ref{snr-single}.				
				\item[b] The minimum detectable energy flux under the $5\sigma$ condition is derived from the $\Delta F$ estimation in the Eq. \ref{snr-single}.
				\item[c] The energy density in parentheses, from left to right, is the set of $1\sigma$ and $5\sigma$, separated by commas.
				\item[d] The characteristic strain in parentheses, from left to right, is the set of $1\sigma$ and $5\sigma$, separated by commas.
			\end{tablenotes}
		\end{threeparttable}
	}
\end{table*}

From Fig. \ref{fig:signal-sensitivity}, it is evident that the telescope exhibits greater sensitivity to persistent events than transient events, implying that persistent events can include a wider variety of GW signals, rendering the discovery of all such signals more challenging. Consequently, identifying a suspicious signal in the absence of a GW template is of paramount importance. We suggest employing signal cross-correlation techniques \citep{Shao:2023agv,1965ApJ...142..419P,2001CQGra..18.4081T,2001PhRvD..64f2002H} to screen observational data at various observation frequencies.

\begin{figure*}
	\centering
	\includegraphics[width=0.47\linewidth]{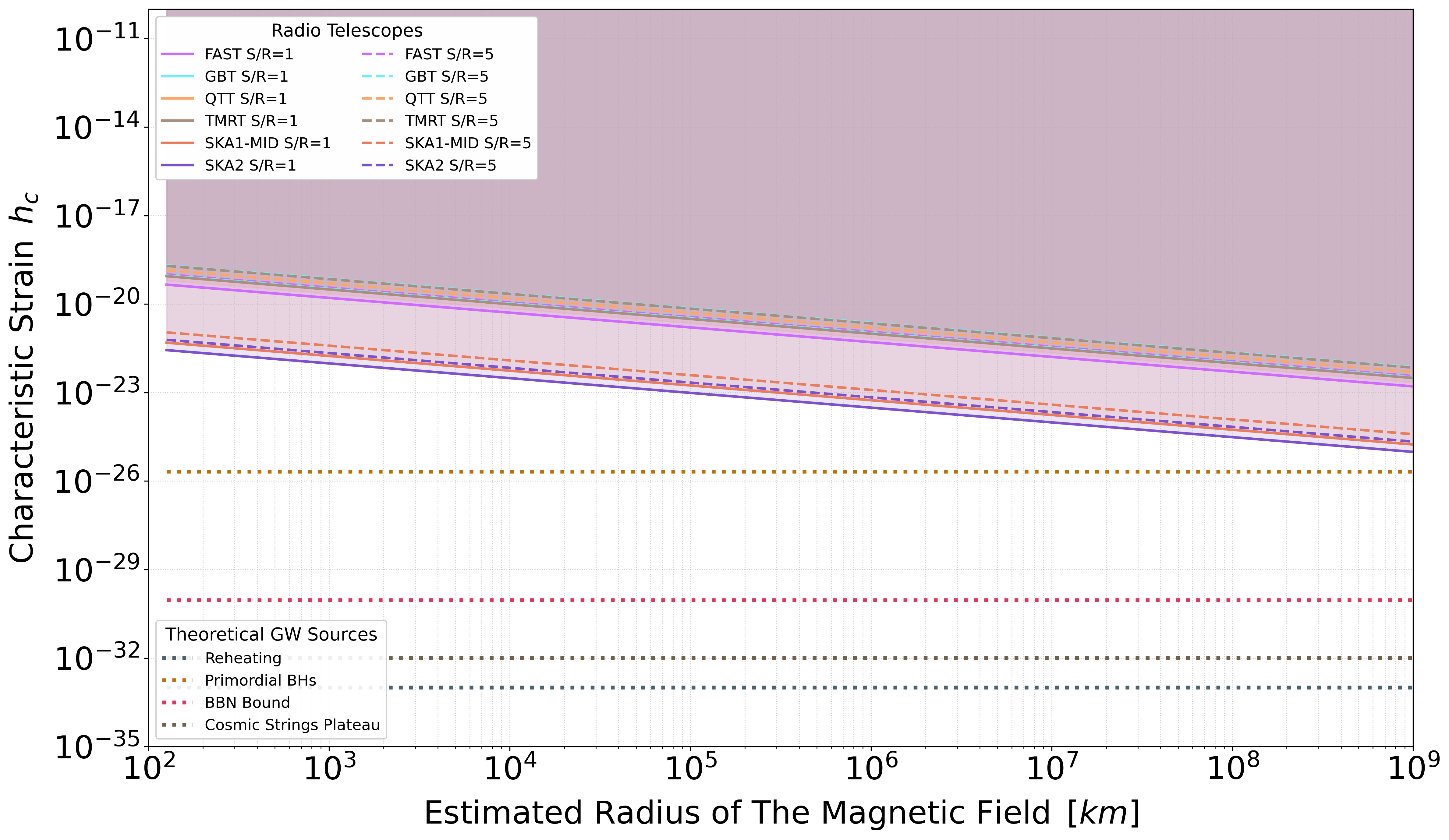}
	\includegraphics[width=0.47\linewidth]{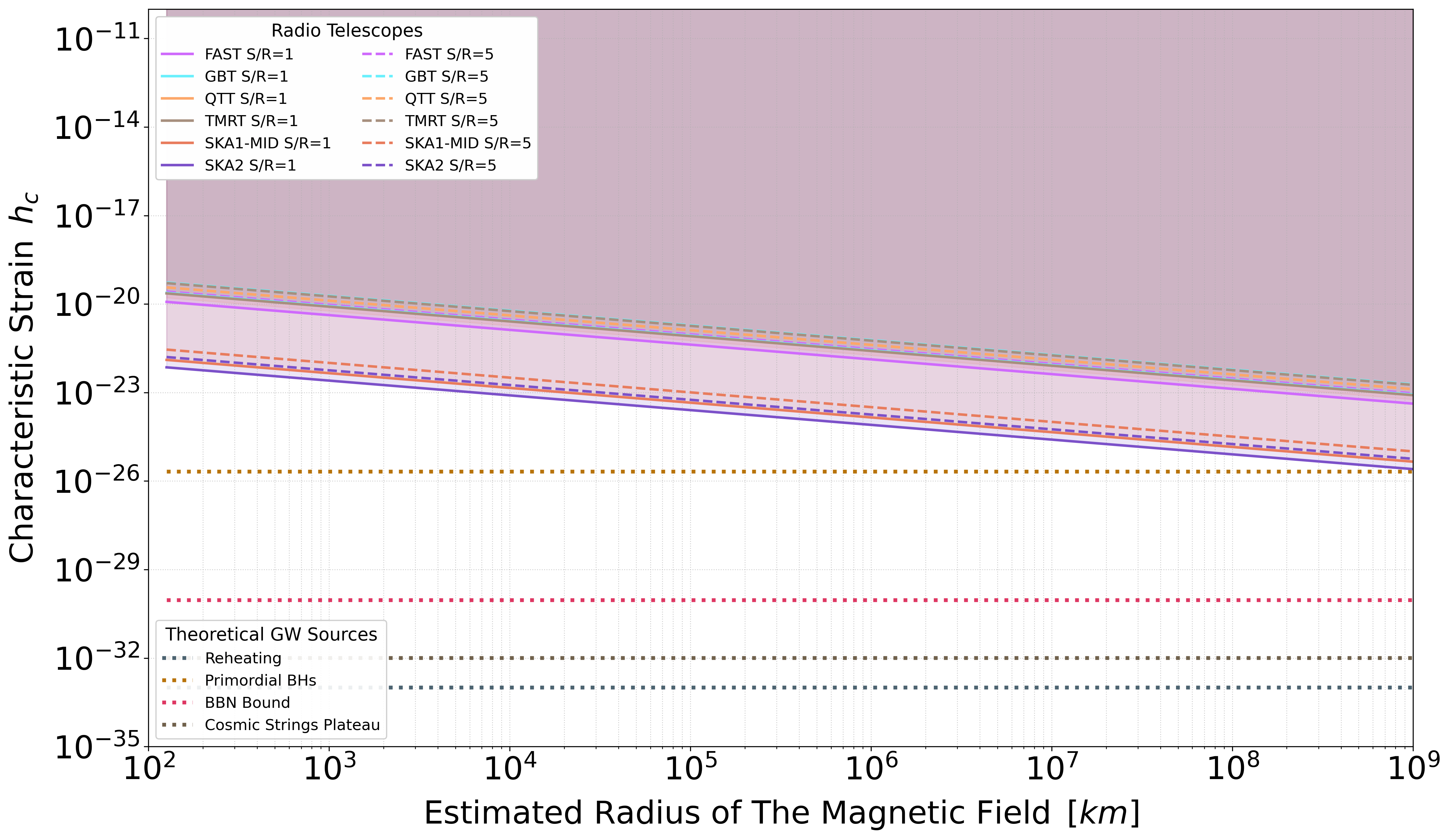}
	\caption{Upper limits on very high-frequency GWs derived from six radio telescopes are shown for (left panel) transient and (right panel) persistent events. The solid colored lines represent the detection thresholds for each telescope assuming a signal-to-noise ratio S/N of 1, while the dashed colored lines correspond to an S/N of 5. The color coding reflects the sensitivity range of each instrument. Colored dots and lines illustrate the dependence of the characteristic strain $h_c$ of various GW sources on the optimistic estimate of the magnetic field region's radius.}
	\label{fig:signal-sensitivity-km}
\end{figure*}

To further investigate how the telescope's detectability varies under different physical conditions, Fig. \ref{fig:signal-sensitivity-km} presents the upper bounds on very high-frequency gravitational wave signals derived from six representative radio telescopes, for both transient (left panel) and persistent (right panel) scenarios. The solid and dashed colored lines correspond to signal-to-noise ratios of 1 and 5, respectively, with the color scheme indicating the sensitivity range of each instrument. The evolution of the characteristic strain $h_c$ for a range of gravitational wave sources is shown as a function of the magnetic field region's radius, assuming the same neutron star magnetic parameters used in Fig. \ref{fig:signal-sensitivity} at 1 GHz.

This analysis demonstrates that the predicted sensitivity is strongly influenced by the spatial extent of the coherent magnetic field. The optimistic cases correspond to magnetic regions bounded by the telescope's full-beam field of view and the typical size of a pulsar's magnetosphere, under the condition that the field strength remains above the interstellar background. In more conservative scenarios, the gravitational wave may only interact with a limited portion of the magnetosphere, or the effective signal may be reduced due to geometrical factors such as finite source distance or beam coverage. These constraints reduce the conversion probability and lower the expected signal amplitude. For instance, one possibility is that the gravitational wave traverses only a small fraction of the pulsar's magnetosphere, thereby significantly reducing the effective interaction volume over which gravitational wave–photon conversion can occur. In such cases, the interaction path length becomes comparable to or even shorter than the coherence scale of the background magnetic field, leading to a marked suppression in the conversion amplitude. Another scenario involves the regime in which only a single photon induced by the GZ effect may be generated and potentially detected, representing the quantum limit of detectability. This situation is especially relevant when the gravitational wave signal is weak, and the magnetic field strength or interaction region is limited. Moreover, some more geometrically constrained situations that cannot be adequately modeled without reference to specific observational conditions. For example, the standard assumption that the gravitational wave source lies effectively at infinity may not hold in all cases. If the source is a compact object, such as a nearby stellar-mass binary, the angular extent of the incoming gravitational wavefront may become non-negligible. In such cases, only the component of the converted electromagnetic signal that is aligned with the observer’s line of sight will be detectable, thereby decreasing the effective conversion probability. Similarly, for gravitational wave sources with limited spatial extent or anisotropic emission, the telescope may intercept only a subset of the generated electromagnetic signal—potentially through just one or a few beams in a multi-beam system—which further suppresses the expected signal strength. Accordingly, Fig. \ref{fig:signal-sensitivity-km} provides a more comprehensive perspective on observational feasibility by bracketing both favorable and constrained conditions for conversed photon in neutron star environments.

\section{Conclusion and Discussion}
\label{conclusion}
In this paper, we propose two signals and detection sensitivities based on the resonance of radio and very high-frequency GWs. All the telescopes we use to estimate the sensitivity of GW detection meet and exceed the upper limit of detection of primordial black holes' GWs, of which FAST has the highest sensitivity for detection of GWs $h_c<10^{-23}$ near $h_c=10^{-24}$ in the single-dish telescopes at the Low Frequency Class, and QTT has the highest sensitivity for detection of GWs $h_c<10^{-23}$ in the single-dish telescopes at the Very High Frequency Class. In contrast, SKA2-MID offers the potential to detect stochastic background GWs from the late universe, as its detection sensitivity surpasses the critical upper limit for primordial black holes. In order to detect weaker GW signals, it is necessary to use telescopes that have larger fields of view, enhanced gain, and multiple beams. This will lead to the development of the FASTA and SKA. Our observations of VHF GWs in radio will provide ground-based laboratories with more refined measurements, allowing for better prediction. With the assistance of high-precision measurements from ground-based laboratories, GW sources can be better separated. Additionally, this will provide insignificant upper limits on the mass of gravitons for large particle colliders.

Our encapsulation of the findings is as follows: (1) Predicated upon the motion of GWs, temporally, two categories of radio signals exist: (a) transient signals and (b) persistent signals. The time scale of the transient signal is derived from the distance traveled by the GW at the point where the conversion probability is greatest, divided by the velocity of the GW. The morphology of transient events is similar to that of fast radio bursts, but the physical process of very high-frequency GWs converting into photons within the magnetic field of a magnetar or pulsar does not constitute the physical origin of fast radio bursts, as the typical distance $L$ is larger than the radius of the light cylinder of a neutron star $R_{\mathrm{LC}}=c P / 2 \pi \simeq 5 \times 10^9 \mathrm{~cm} P_0$. This implies a prospective novel paradigm for rapid optical variation signals.

(2) Although our analysis focuses on converted photons within the radio band, our methodology can be extended to include the entire X-ray spectrum and portion of the gamma-ray spectrum, potentially extending up to $10^{20}\mathrm{~Hz}$. This extrapolation relies on a comparison of the converted photon's wavelength with the electron's Compton wavelength, which means that the electromagnetic field should be almost constant over the scale of the Compton wavelength. When the deviation of the former being shorter than the latter necessitates an amendment termed ``Beyond the Heisenberg-Euler Effective Action", which renders isolated neutron stars as optimal candidates for observing vacuum birefringence, thereby enhancing our comprehension of the magnetic structure of neutron stars or magnetars. However, numerous articles studying very high-frequency GW detection push the frequency upper limit to $10^{25}\mathrm{~Hz}$ when using Heisenberg-Euler effective action, which is inappropriate. 

(3) Our theoretical framework postulates a minimum detection frequency, stipulating that the GW's frequency must surpass that of the plasma it traverses to enable electromagnetic wave penetration through plasma. This minimum frequency is necessary for the detection of GWs. This is also the reason for the lower conversion probability obtained at low frequencies. In order for photons at low frequencies from $10^{6}\mathrm{~Hz}$ to $10^{7}\mathrm{~Hz}$ to pass through the magnetic field of the neutron star, we require a lower electron density surrounding the neutron star, implying that GWs are farther away from the center of the neutron star, where the magnetic field strength is lower. This is also why our detection sensitivity in low-frequency bands is lower than that in high-frequency bands.

(4) Leveraging the properties of gravitons to delineate the spectral line shapes. Once we have established that the observed radio signal is from the GWs converted into photons in the magnetic field, and after carefully deducting all the noise, the spectral lines show clear, distinguishable spikes in emission or absorption, we can demarcate constraints on graviton properties and examine the fundamental principles of quantum gravity theory. Moreover, we can give the test results of graviton mass from $10^{-9}\mathrm{~eV}$ to $10^{-4}\mathrm{~eV}$.

(5) From a perspective of sensitivity, we confirm the detectability of GW-generating cosmological events transcending extant limits on primordial black hole detection. By analyzing the transient events and persistent events detected by FAST, we can give a lower mass limit of $\sim 3.4\times10^{-2}M_{\bigodot}$ and $\sim1.1\times10^{-2}M_{\bigodot}$ for the primordial black hole with $5\sigma$ confidence, respectively. Thus, we can explore potential exotic astrophysical bodies. Moreover, we ascertain the detectability of GW-generating cosmological events near the epoch of Big Bang nucleosynthesis, though current observations are unable to capture very high frequency GW events from the early universe. This reason allows FAST to probe aspects of early-universe cosmology that extend beyond the standard cosmological model. Once we observe the very high-frequency background GWs produced by the early universe with certainty, this means that the extra amount of radiation parametrized by extra neutrino species $\Delta N_{\mathrm{eff}}$ is greater than $0.2$, which will completely change our understanding of the present cosmology. Due to the higher detection sensitivity of SKA2-MID, the observations of SKA2-MID can be used to verify the results of FAST and better explore GW events in the late universe.

In the upcoming paper, we will conduct the overall numerical simulation and examine two cross-correlation methods: (a) the system noise of various radio telescopes, and (b) the signal and the temperature of the matter surrounding the neutron star. And two methods when the cross-correlation methods are invalid: (a) single-detector excess power statistic \citep{1965ApJ...142..419P}, and (b) null channel method \citep{2001CQGra..18.4081T,2001PhRvD..64f2002H}. Finding the radio signal generated by the GZ effect from the neutron star baseband data is not an easy task. When we use the characteristic GW template and meticulously subtract all known radio signals from celestial bodies, this can treat the radio signal generated by the currently theoretically unknown GW through the GZ effect as a noise deduction. Based on this, we plan to propose using the cross-correlation technique of finding stochastic background GWs to assist us in searching for VHF GW signals. Cross-correlation quantifies the degree of resemblance between two different signals. Since radio observational data usually persists for a certain duration, we can define the integral form of signal cross-correlation \citep{Bracewell:1966}
\begin{equation}
	\left(s_{i}\otimes s_{j} \right)\left(t,\tau\right)=\int_{t}^{t+\Delta t_{\mathrm{int}}}s_{i}(t)s_{j}(t+\tau)dt,
\end{equation}
where $s_i$ and $s_j$ are two signals, $\Delta t_{\mathrm{int}}$ is the integration time of observation, and $\tau$ is the displacement of time. A zero integral means the signals are orthogonal and uncorrelated, while a non-zero integral means they are correlated. 

\begin{figure}
	\centering
	\includegraphics[width = 0.45\textwidth]{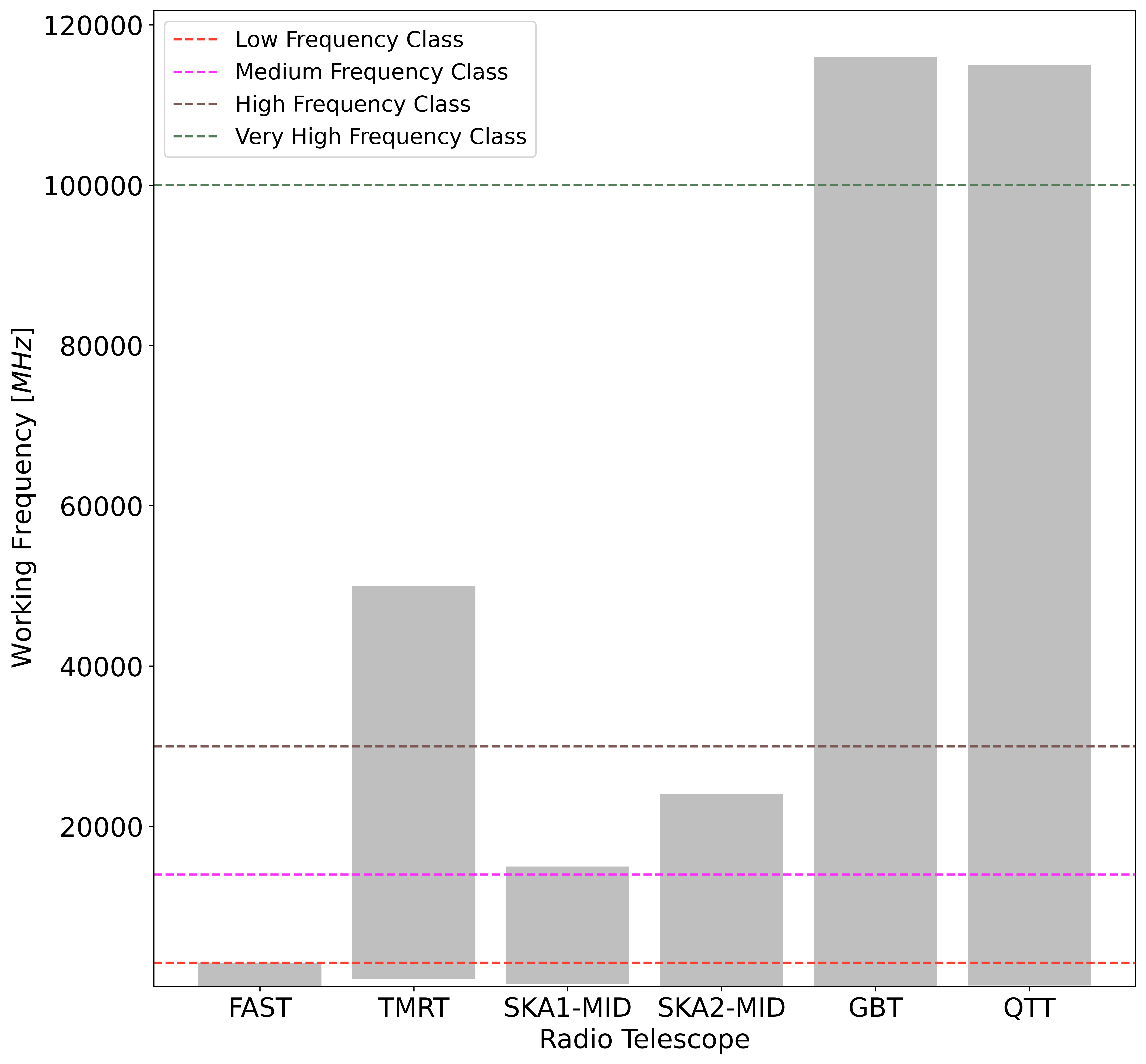}
	\caption{The frequency allocation of radio telescopes for the upcoming paper. The red dashed line is low frequency. The pink dashed line is medium-frequency. The brown dashed line is high-frequency. The green dashed line represents a high-frequency class.}
	\label{fig:telescope-class}
\end{figure}

Since our ultimate goal is to perform cross-correlation with observations from different telescopes, and the sources of very high-frequency GWs at different frequencies can be inconsistent, we need to consider the common frequency interval between telescopes as much as possible. For the convenience of subsequent discussion, the radio telescopes considered are divided into ``Low Frequency Class" (LFC), ``Medium Frequency Class" (MFC), ``High Frequency Class" (HFC), and ``Very High Frequency Class" (VHFC) according to their working frequencies, as shown in Fig. \ref{fig:telescope-class}. In Fig. \ref{fig:telescope-class}, we use dashed lines of different colors to classify the telescopes. Based on this classification, we also identify the telescopes used by the cross-correlation at different frequencies. At the Low Frequency Class, the telescopes involved are FAST, TMRT, SKA1-MID, SKA2-MID, GBT, and QTT. At the Medium Frequency Class, the telescopes involved are TMRT, SKA1-MID, SKA2-MID, GBT, and QTT. At the High Frequency Class, the telescopes involved are TMRT, SKA2-MID, GBT, and QTT. At the Very High Frequency Class, the telescopes involved are TMRT, GBT, and QTT.

Utilizing cross-correlation in conjunction with the properties of the astrophysical magnetic field allows for the implementation of specific observation modes to cross-verify the gravitational wave signal. (a) distinct neutron stars observed by various telescopes, (b) identical neutron stars observed by various telescopes, (c) distinct neutron stars observed by a single telescope, and (d) the same neutron stars observed by a single telescope on different dates, considering the proper motion of the neutron stars, with a sufficiently large interval between observations. The objective of these observational methods is to optimize the utilization of varying noise distributions across distinct observational paths. We will address the specific issues related to these observation patterns in our upcoming paper.

\begin{acknowledgments}
We thank Pei Wang, Fang-Yu Li, Xianfei Zhang, Fan Zhang, Zhoujian Cao, Kang Jiao and Jian-Kang Li for useful discussions. This work was supported by National SKA Program of China, No.2022SKA0110202 and the China Manned Space Program with grant No. CMS-CSST-2025-A01. This research used Pulsar data from the ATNF Pulsar Catalogue and McGill Online Magnetar Catalog.
\end{acknowledgments}

\begin{contribution}
	T.-J.Z. and W. H developed the idea of Very High-Frequency Gravitational Waves; Z.-Z.T, and P. H contributed to the scientific analysis and discussion of Very High-Frequency Gravitational Waves observation on FAST; W. H drafted the paper; All authors contributed to the final version of the manuscript.
\end{contribution}


\appendix
\label{appendix}
\section{Hurwitz zeta function}
\label{sec:hurwtiz-zeta-function}
This appendix is intended to describe the Hurwitz zeta function required to calculate the QED effect when the magnetic field strength exceeds the critical magnetic field strength $B_\mathrm{critical}=m_e^2 c^3/\hbar|q_f|$.

The generalized Hurwitz zeta function $\zeta(z, a)$ is defined by $\zeta(z, a)=\sum_{n=0}^{\infty}(n+a)^{-z}, \quad \operatorname{Re} z>1, a \neq 0,-1,-2, \ldots$ \citep{Elizalde:1986}. For the particular values $\operatorname{Re} z<0$, this function reduces to $\zeta(z, a)=-\frac{B_{1-z}(a)}{1-z}$ under the Bernoulli polynomials representation. Straightforwardly, we can donate that $\zeta^{\prime}(z, a) \equiv \frac{\partial}{\partial z} \zeta(z, a)$. When dealing with the magnetic field of a magnetar, as stated in the Section Result, it is necessary to take into account that $\operatorname{Re} z<0$. And as $0<a=\rho/2=B_\mathrm{critical}/2|B|<1$, we can rewrite the Hurwitz zeta function under the Fourier series representation
\begin{equation}
	\zeta(z, a)=\frac{2 \Gamma(1-z)}{(2 \pi)^{1-z}}\left[\sin \left( \frac{\pi z}{2}\right)  \mathcal{C}_{1-z}(a)+\cos \left( \frac{\pi z}{2}\right)\mathcal{S}_{1-z}(a)\right],
\end{equation}
where $\Gamma(z)$ is the gamma function. Therefore, we can easily express the $\zeta^{\prime}(z, a)$
\begin{equation}
	\begin{aligned}
		\zeta^{\prime}(z, a)& =\frac{\pi \Gamma(1-z)}{(2 \pi)^{1-z}}\left[\cos\left(\frac{\pi z}{2}\right) \mathcal{C}_{1-z}(a)-\sin\left(\frac{\pi z}{2}\right)\mathcal{S}_{1-z}(a)\right]\\
		&+ \frac{2 \Gamma(1-z)}{(2 \pi)^{1-z}}\left[\sin\left(\frac{\pi z}{2}\right) \mathcal{C}_{\mathrm{fc},1-z}^{(1)}(a)+\cos\left(\frac{\pi z}{2}\right) \mathcal{S}_{\mathrm{fc},1-z}^{(1)}(a)\right]\\
		&+\left[\ln 2 \pi-\psi_0(1-z)\right] \zeta(z, a),
	\end{aligned}
\end{equation}
where $\mathcal{C}_{1-z}(a)=\sum_{k=1}^{\infty} \frac{\cos \left( 2 \pi k a\right) }{k^{1-z}}$, $\mathcal{S}_{1-z}(a)=\sum_{k=1}^{\infty} \frac{\sin \left( 2 \pi k a\right)}{k^{1-z}}$, $\mathcal{C}_{\mathrm{fc},1-z}^{(n)}(a)=\sum_{k=1}^{\infty} \frac{\left(\ln k\right)^n}{k^{1-z}} \cos \left(2 \pi k a\right)$, $\mathcal{S}_{\mathrm{fc},1-z}^{(n)}(a)=\sum_{k=1}^{\infty} \frac{\left(\ln k\right)^n}{k^{1-z}} \sin\left(2 \pi k a\right)$, and $\psi_0(z)=\frac{d}{d z} \ln \Gamma(z)=-\gamma+\sum_{k=1}^{\infty}\left(\frac{1}{k}-\frac{1}{k+z-1}\right)$, where $\gamma$ is the Euler-Mascheroni constant. Similarly, we can obtain the $\zeta''(z, a)$ and $\zeta'''(z, a)$.

\section{Quantities rewritten in the natural Lorentz-Heaviside units and some parameters of the telescopes.}
\label{sec:quantities-rewritten-in-the-nautural-lorentz-heaviside-untis}
It is beneficial to rewrite some useful quantities in the natural Lorentz-Heaviside units, as it can facilitate our ability to carry out numerical calculations and allow for verification by others. In the $10^{6}-10^{11} \mathrm{~Hz}$ frequency, the photons' wavelengths are $\lambda_{\gamma}\approx 0.003-300\mathrm{~m}$ and their energy are $E_{\gamma}\approx10^{-10}-10^{-5}\mathrm{~eV}$. Therefore, we use the principle of ``lowest value" to determine unit conversion in our calculation of the frequency range and magnetic field strength range as $\omega\sim 10^{-10}\mathrm{~eV}\sim 10^{6}\mathrm{~Hz}$, $n_e \sim 10^{-7}\mathrm{~cm}^{-3}$, and $B_0 \sim 10^{-9}\mathrm{~Gauss}$. And we utilize the facts that $1 \mathrm{~eV} \simeq 5 \times 10^4 \mathrm{~cm}^{-1}=5 \times 10^9 \mathrm{~km}^{-1}$, $1 \mathrm{~Gauss} \simeq 1.95 \times 10^{-2} \mathrm{~eV}^2$, and fine structure constant $\alpha=e^2 / 4 \pi=1 / 137$. For example, the energy density of gravitational waves is used to calculate the sensitivity
\begin{widetext}
	\begin{equation}
			\frac{3 H_0^2M_{\mathrm{planck}}^2}{4 \pi\omega}\Omega_{\mathrm{GW}} (\omega)=2.66\times 10^{18}\left(\frac{10^6\mathrm{~Hz}}{\omega}\right)h_0^2\Omega_{\mathrm{GW}}(\omega)\mathrm{~Jy}\mathrm{~sr}^{-1},
	\end{equation}
\end{widetext}
where $H_0=100h_0\mathrm{~km}\mathrm{~s}^{-1}\mathrm{~Mpc}^{-1}=3.24h_0\times 10^{-18}\mathrm{~s}^{-1}=1.44h_0\times 10^{-42}\mathrm{~GeV}$ and $M_{\mathrm{planck}}=\sqrt{\frac{1}{8\pi G}}=\frac{m_e^2}{8\pi\alpha_{G}}=2.44\times 10^{18}\mathrm{~GeV}$ with $m_e=0.511\mathrm{~MeV}$ is the electron mass and $\alpha_{G}=1.75\times 10^{-45}$ is the gravitational coupling constant. It is worth noting that the $\mathrm{~sr}^{-1}$ in the $\frac{3 H_0^2M_{\mathrm{planck}}^2}{4 \pi\omega}\Omega_{\mathrm{GW}} (\omega)$ unit conversion's result is due to the $4\pi$ square degrees of sky that we have to take into account (a) the size of the radio telescope's and signal's field of view defined by the beamwidths of the antennas and (b) the per-solid angle of the spectral energy density of GWs as the energy flux $F=\int I d\Omega$. We can estimate the telescope's field of view from the observational wavelength $\lambda$, the area of a station $A$ and the multi-beam numbers $n_{mb}$ $\mathrm{FOV}=\lambda^2 n_{mb}/A\mathrm{~sr}=3850.37\lambda^2 n_{mb}/D^2\mathrm{~Deg^2}$ \citep{2017isra.book.....T,1989ASIC..283..163W}. The numbers of multi-beams for different telescopes are listed below: (a) FAST: $n_{mb}=19$ at frequency $\omega\in[0.07,3]\mathrm{~GHz}$ \citep{Jiang:2019rnj,2020RAA....20...64J,2020Innov...100053Q,fastref,2021AJ....162..151H}, (b) TMRT: $n_{mb}=1$ at frequency $\omega\in[1,18]\cup[30,34]\mathrm{~GHz}$ and $n_{mb}=2$ at frequency $\omega\in[18,30]\cup[35,50]\mathrm{~GHz}$ \citep{tmrtref,tmrtsr,tmrtintr,tmrtpar}, (c) SKA1-MID: $n_{mb}=1$ at frequency $\omega\in[0.35,15]\mathrm{~GHz}$ \citep{skaref,2019arXiv191212699B,2021AJ....162..151H}, (c) SKA2-MID: $n_{mb}=15$ at frequency $\omega\in[0.05,24]\mathrm{~GHz}$ \citep{skaref,2021AJ....162..151H}, (d) GBT: $n_{mb}=1$ at frequency $\omega\in[0.1,116]\mathrm{~GHz}$ \citep{gbtref,GBT_Proposer_Guide,2021AJ....162..151H}, and (e) QTT: $n_{mb}=1$ at frequency $\omega\in[0.15,30]\mathrm{~GHz}$, $n_{mb}=36$ at frequency $\omega\in[30,50]\mathrm{~GHz}$, $n_{mb}=1$ at frequency $\omega\in[50,80]\mathrm{~GHz}$, and $n_{mb}=37$ at frequency $\omega\in[80,115]\mathrm{~GHz}$ \citep{Wang:2023occ,qttref}. Fortunately, the telescope field of view is several orders of magnitude larger than the field of view of the signal $\pi \left(10^{9}\mathrm{~km}/2\mathrm{~kpc}\right)^2=8.25\times10^{-16}\mathrm{~sr}$.

Further, let us list the three temperatures of the telescope used in the calculation as follows:
\begin{table*}
	\caption{\label{tab:telescope-parameters-2}The list of the temperatures of radio telescopes used in this paper.}
	\renewcommand\arraystretch{0.5}
	\resizebox{1.0\linewidth}{!}{
		\begin{threeparttable}
			\begin{tabular}{cccccccc}
				\toprule
				Frequency&Telescope&Working&Zenith System&Astrophysical Background&Precipitable&Surface Air&Telescope\\Class&Name&Frequency&Temperature $T_\mathrm{sys}$&Temperature $T_\mathrm{ast}$&Water Vapour&Temperature $T_\mathrm{atm}$&Temperature $T_\mathrm{rt}$\\
				& &(GHz)&(K)\tnote{a}&(K)\tnote{b}&(mm)\tnote{c}&(K)\tnote{d}&(K)\tnote{e}\\ \midrule
				&FAST&0.07-3&$\sim$26.62-4750.98&$\sim$2.80-4577.21&$\sim$2-8&$\sim$297-300&$\sim$20\\
				LF&TMRT&1-3&$\sim$19.14-26.60&$\sim$2.80-22.94&$\sim$2-7&$\sim$282-308&$\sim$14-21\\
				Class&SKA1-MID&0.35-3&$\sim$15.14-38302.57&$\sim$2.80-83.25&$\sim$5.8-19.2&$\sim$290&$\sim$11-23\\
				&SKA2-MID&0.05-3&$\sim$15.14-11795.62&$\sim$2.80-10647.17&$\sim$5.8-19.2&$\sim$290&$\sim$11-23\\
				&GBT&0.1-3&$\sim$30.33-1310.84&$\sim$2.80-1871.42&$\sim$10-30&$\sim$250-270&$\sim$10-30\\
				&QTT&0.15-3&$\sim$22.14-1331.09&$\sim$2.80-678.112&$\sim$2-15&$\sim$246-30&$\sim$16-18\\ \midrule
				&TMRT&3-14&$\sim$17.04-27.77&$\sim$2.73-2.86&$\sim$2-7&$\sim$282-308&$\sim$12-22\\
				MF&SKA1-MID&3-14&$\sim$15.14-25.58&$\sim$2.73-2.86&$\sim$5.8-19.2&$\sim$290&$\sim$10-17\\
				Class&SKA2-MID&3-14&$\sim$15.14-24.23&$\sim$2.73-2.86&$\sim$5.8-19.2&$\sim$290&$\sim$10-17\\
				&GBT&3-14&$\sim$32.21-126.13&$\sim$2.73-2.86&$\sim$10-30&$\sim$250-270&$\sim$10-30\\
				&QTT&3-14&$\sim$23.11-27.48&$\sim$2.73-2.86&$\sim$2-15&$\sim$246-302&$\sim$18\\ \midrule
				&TMRT&14-30&$\sim$20.88-54.78&$\sim$2.730-2.731&$\sim$2-7&$\sim$282-308&$\sim$15-35\\
				HF&SKA2-MID&14-24&$\sim$22.73-67.11&$\sim$2.730-2.731&$\sim$5.8-19.2&$\sim$290&$\sim$17-24\\
				Class&GBT&14-30&$\sim$32.21-126.13&$\sim$2.730-2.731&$\sim$10-30&$\sim$250-270&$\sim$10-30\\
				&QTT&14-30&$\sim$26.01-77.68&$\sim$2.730-2.731&$\sim$2-15&$\sim$246-302&$\sim$20\\ \midrule
				&TMRT&30-50&$\sim$40.64-119.20&$\sim$2.73&$\sim$2-7&$\sim$282-308&$\sim$30-40\\
				VHF&GBT&30-116&$\sim$98.08-373.0&$\sim$2.73&$\sim$10-30&$\sim$250-270&$\sim$40-180\\
				Class&QTT&30-115&$\sim$70.94-340.0&$\sim$2.73&$\sim$2-15&$\sim$246-302&$\sim$35-40\\
				\bottomrule
			\end{tabular}
			\begin{tablenotes}
				\footnotesize
				\item[a] Since the system temperature of the telescope is related to the directionality of the telescope, we only show the zenith system temperature in this paper.
				\item[b] When the frequency exceeds 30 GHz, the average sky brightness temperature for galactic synchrotron radiation is already far below the temperature of the cosmic microwave background radiation.
				\item[c] The data from the telescope official website or telescope project book in this paper is only without loss of general predicted value, the actual water vapour is the real-time water vapour at the time of observation.
				\item[d] The data from the telescope official website or telescope project book in this paper is only without loss of general predicted value. The atmospheric temperature is closely related to the frequency of the actual observation, the altitude of the telescope station during the observation, and the weather at that time.
				\item[e] For unbuilt or unfinished radio telescopes, we use the planned indicators in their construction papers.
			\end{tablenotes}
		\end{threeparttable}
	}
\end{table*}

\section{Conversion probability calculation process}
\label{sec:function-detail}
Start with the simplest, we can now determine the conversion probability of neutron stars for a particular path at the location of $r_{\mathrm{occur}}$ at their equators $\theta=0$
\begin{equation}
	\begin{aligned}
		&P_{g \rightarrow \gamma}(L,\omega,\theta=0)  =\left|\left\langle\hat{A}_{\omega,\lambda}(L) \mid \hat{h}_{\omega,\lambda}(0)\right\rangle\right|^2 \\
		&=\left|\int_{-L/2}^{L/2} d l' \frac{\sqrt{2}B_\mathrm{eff}(l',\theta=0)}{2M_{\mathrm{planck}}}\exp \left(-i \int_{-L/2}^{l'} d l'' \frac{-\Delta_\omega^2(l'')}{ 2 \omega}\right)\right|^2.
	\end{aligned}\label{convertion-probability-two}
\end{equation}
We assume that the function after the integration of function $\frac{-\Delta_\omega^2(l'')}{ 2 \omega}$ is $\Delta \tilde{W}(l'')$. Using Lagrange's mean value theorem, we can obtain
\begin{equation}
	\frac{-\Delta_\omega^2(\xi_1)}{ 2 \omega}(l'+L/2)=\Delta \tilde{W}(l')-\Delta \tilde{W}(-L/2),
\end{equation}
where $\xi_1=-L/2+\zeta_1 l'$ and $0<\zeta_1<1$. Therefore,
\begin{equation}
	\begin{aligned}
		&\int_{-L/2}^{L/2} d l' i\frac{\sqrt{2}B_\mathrm{eff}(l',\theta=0)}{2M_{\mathrm{planck}}}\exp \left[-i \Delta \tilde{W}(l')+\Delta \tilde{W}(-L/2)\right]\\
		=&\frac{\sqrt{2}e^{-i \Delta \tilde{W}(-L/2)}e^{i\frac{\pi}{2}}}{2M_{\mathrm{planck}}}\int_{-L/2}^{L/2} d l' B_\mathrm{eff}(l',\theta=0)e^{-i \Delta \tilde{W}(l')}.\label{integrate-mid}
	\end{aligned}
\end{equation}
By integrating by parts, we get the further result of Eq. (\ref{integrate-mid})
\begin{equation}
	\begin{aligned}
		&\int_{-L/2}^{L/2} d l' B_\mathrm{eff}(l',\theta=0)e^{-i \Delta \tilde{W}(l')}\\
		&=F_B\left(\frac{L}{2},\theta=0\right)\left[ e^{i \Delta \tilde{W}(\frac{L}{2})}+e^{i \Delta \tilde{W}(-\frac{L}{2})}\right]\\
		&+e^{i\frac{\pi}{2}}F_B\left(\xi_2,\theta=0\right)\Delta \tilde{W}'(\xi_2) e^{i \Delta \tilde{W}(\xi_2)},
	\end{aligned}
\end{equation}
where $F_B\left(L,\theta=0\right)$ is the integral of $B_\mathrm{eff}(L,\theta=0)$, and $\xi_2=-L/2+\zeta_2L/2$ is the mean value parameter from the rest term of integrating by parts
\begin{equation}
\int_{-L/2}^{L/2} d l' i F_B\left(l',\theta=0\right)\Delta \tilde{W}'(l')e^{i\Delta \tilde{W}(l')},
\end{equation}
where $\Delta \tilde{W}'(l')=\frac{-\Delta_\omega^2(l')}{2\omega}$. Then, we can obtain the conversion probability $P_{g \rightarrow \gamma}(L,\omega,\theta=0)$
\begin{equation}
	\begin{aligned}
		&P_{g \rightarrow \gamma}(L,\omega,\theta=0)=\frac{1}{2M_{\mathrm{planck}}^2}\\
		&\times\left\lbrace 2+2\cos\left[2\Delta \tilde{W}\left(-\frac{L}{2}\right)\right]F_B^2\left(\frac{L}{2},\theta=0\right)\right\rbrace \\
		&+\frac{L^2 F_B^2\left(\xi_2,\theta=0\right)}{2M_{\mathrm{planck}}^2}\left[ \Delta \tilde{W}'(\xi_2)\right]^2\\
		&+\frac{L\Delta \tilde{W}'(\xi_2)}{M_{\mathrm{planck}}^2}F_B\left(\frac{L}{2},\theta=0\right)F_B\left(\xi_2,\theta=0\right)\\
		&\times\left\lbrace \sin\left[3\Delta \tilde{W}\left(-\frac{L}{2}\right)+\Delta \tilde{W}\left(\xi_2\right) \right] \right\rbrace\\
		&-\frac{L\Delta \tilde{W}'(\xi_2)}{M_{\mathrm{planck}}^2}F_B\left(\frac{L}{2},\theta=0\right)F_B\left(\xi_2,\theta=0\right)\\
		&\times\left\lbrace \sin\left[\Delta \tilde{W}\left(\frac{L}{2}\right)-\Delta \tilde{W}\left(\xi_2\right) \right] \right\rbrace,
	\end{aligned}
\end{equation}
where
\begin{equation}
F_B\left(L,\theta=0\right)=\frac{2B_0 L r_{0}^3}{r_{\mathrm{proj}}^2 \sqrt{L^2+4 r_{\mathrm{proj}}^2}}
\end{equation}
is an odd function where $r_{\mathrm{proj}}$ is the radius of a GW as it passes through the y-axis. Considering that the electromagnetic waves converted can escape the magnetic field of the magnetar, $r_{\mathrm{proj}}$ is larger than or equal to $r_{\mathrm{occur}}$. And, when $B_\mathrm{eff}(L,\theta=0)<B_\mathrm{critical}$
	\begin{equation}
		\begin{aligned}
		&\Delta \tilde{W}(L)=\\
		&B_0  r_0^3 \left[ \frac{4 \tilde{\eta}_{\mathrm{QED},\lambda}^{\mathrm{normal}} B_0  L r_0^3}{r_{\mathrm{proj}}^2 \left(L^2+4 r_{\mathrm{proj}}^2\right)^2}+\frac{3 \tilde{\eta}_{\mathrm{QED},\lambda}^{\mathrm{normal}} B_0  L r_0^3}{2 r_{\mathrm{proj}}^4 \left(L^2+4 r_{\mathrm{proj}}^2\right)}\right] \\
		&+B_0  r_0^3 \left[ \frac{3 \tilde{\eta}_{\mathrm{QED},\lambda}^{\mathrm{normal}} B_0  r_0^3 \arctan\left(\frac{L}{2 r_{\mathrm{proj}}}\right)}{4 r_{\mathrm{proj}}^5}-\frac{2 \tilde{\eta}_{\mathrm{plasma}}^{\mathrm{mag}} L}{r_{\mathrm{proj}}^2 \sqrt{L^2+4 r_{\mathrm{proj}}^2}}\right],
		\end{aligned}
	\end{equation}
where $\tilde{\eta}_{\mathrm{QED},\lambda}^{\mathrm{normal}}$ is the stability coefficient for $+$ and $\times$ states of the normal QED effect, which means that it is not affected by the integration variable $l''$ during the integration process. For example, $\tilde{\eta}_{\mathrm{QED},\times}^{\mathrm{normal}}=4.66 \times 10^{-54}\left(\frac{\omega}{10^{-9} \mathrm{~eV}}\right)\left(\frac{1}{10^{-9} \mathrm{~Gauss}}\right)^2\mathrm{~cm}^{-1}$. Similarly, $\tilde{\eta}_{\mathrm{plasma}}^{\mathrm{mag}}=2.51\times 10^{-18}\left(\frac{1}{10^{-9} \mathrm{Gauss}}\right)\left(\frac{1 \mathrm{sec}}{P}\right)\left(\frac{10^{-9} \mathrm{eV}}{\omega}\right)\mathrm{~cm}^{-1}$ is the stability coefficient for plasma near magnetar. Therefore, $\Delta \tilde{W}(L)$ is also an odd function in this case. For the case of $B_\mathrm{eff}(L)>B_\mathrm{critical}$, because the formulas for the beyond QED effect are too complicated, we do not show the specific analytic results and directly use the numerical method.

\begin{figure}
	\centering
	\includegraphics[width = 0.3\textwidth]{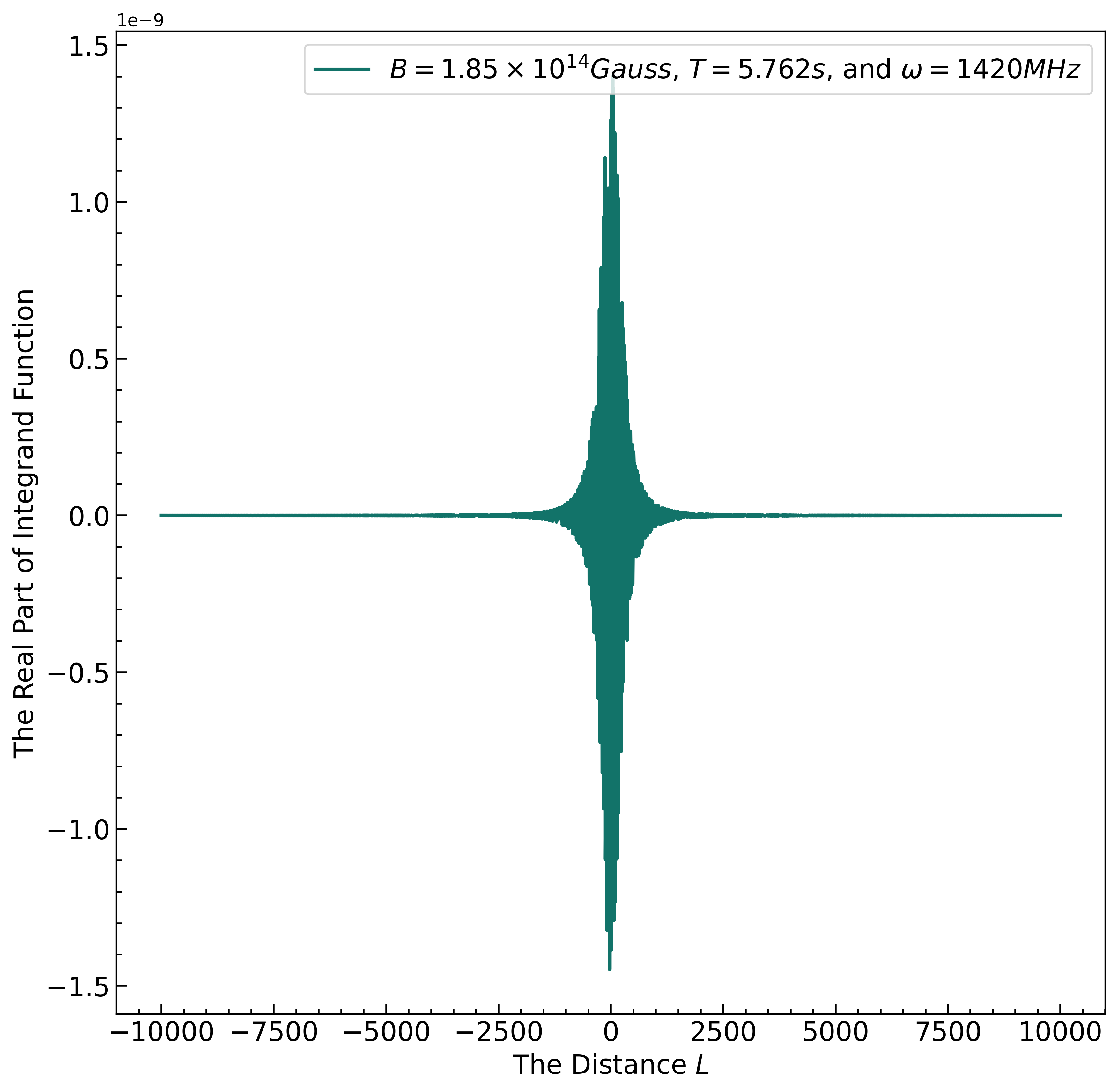}
	\includegraphics[width = 0.3\textwidth]{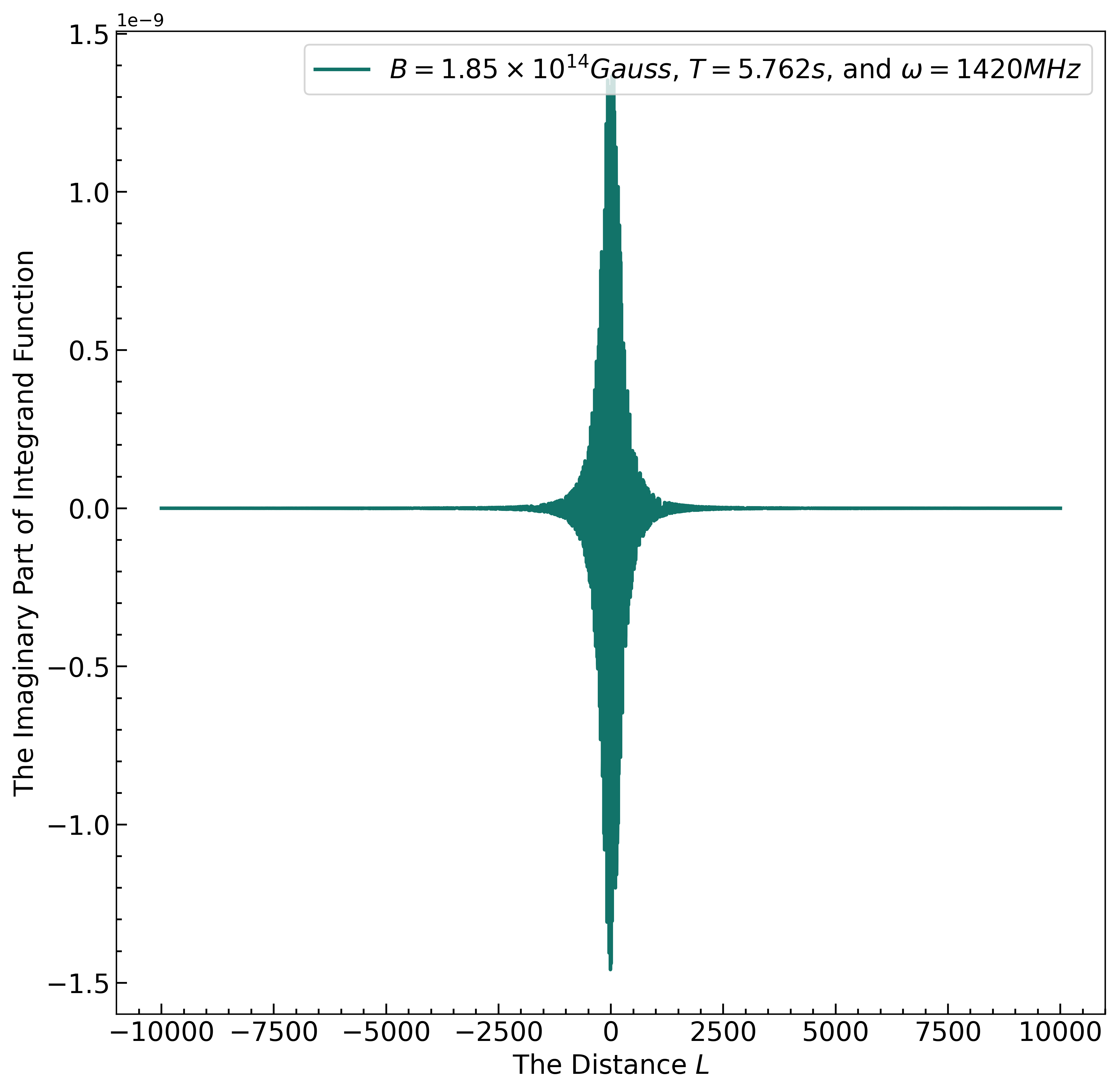}
	\caption{The real and imaginary parts of the integrand function from Eq. \ref{integrate-mid}. (Top panel) we set $B=1.85\times 10^{14}\mathrm{~Gauss}$, $T=5.762\mathrm{~sec}$, and $\omega=1420\mathrm{~MHz}$ to show the real parts of the integrand function in a blue solid line, which are the same as (bottom panel) with the distance that GWs travel from $0\mathrm{~km}$ to $10^{4}\mathrm{~km}$.}
	\label{fig:direction-dig-app}
\end{figure}

It should be noted that because our magnetic field distribution considers the radial $B(r,\theta)$, there is a square relationship between the distance $L$ of the GW in the magnetic field and the radius of the magnetic field where the GW is located $r^2=\frac{L^2}{4}+r_{\mathrm{proj}}^2$. And we only show the integral results at the magnetar equator, but in fact, for the conversion probabilities at different latitudes, our results only require small numerical corrections of $1/2\left[3 \cos \theta\mathbf{m}\cdot\mathbf{r}-\cos \theta_m\right]$ for the magnetic field at different latitudes. In Fig. \ref{fig:direction-dig-app}, we show the minimum radius $r_{\mathrm{occur}}$ at which GWs cross the magnetar equator and radio-observable converted photons.

As the GW travels out of the neutron star's magnetic field and into the interstellar magnetic field, the calculation of the conversion probability will become easier because we can approximate the magnetic field by simplifying it to a homogeneous one. In addition, it is important to note that the direction of the interstellar magnetic field needs to be averaged out since it has certain randomness, and the inhomogeneous magnetic field can be considered as a superposition of many local homogeneous magnetic field
\begin{equation}
	\begin{aligned}
		P_{g \rightarrow \gamma}(L,\omega)&=\left|\left\langle\hat{A}_{\omega,\lambda}(L) \mid \hat{h}_{\omega,\lambda}(0)\right\rangle\right|^2 \\
		&=\frac{64\pi GB_\mathrm{eff}^2 \omega ^2}{64\pi GB_\mathrm{eff}^2 \omega^2+\Delta_\omega ^4} \sin ^2\left(\frac{m_1-m_2}{2} L\right).\\
	\end{aligned}\label{convertion-probability-homo}
\end{equation}

\section{The figures with the complete calculation results.}
\label{sec:more-figures}
In this section, we show the figures of the complete calculation results that have been moved to the appendix for readers' readability. These figures show the conversion probabilities of GWs of the two polarization modes ($\times$ and $+$ modes) as they pass through the neutron star's magnetic field at its equator, with the neutron star's parameters changing. They also show the conversion probabilities of GWs changing when they pass through the neutron star's total solid angle integral magnetic field.
\begin{figure*}
	\centering
	\includegraphics[width=0.29\linewidth]{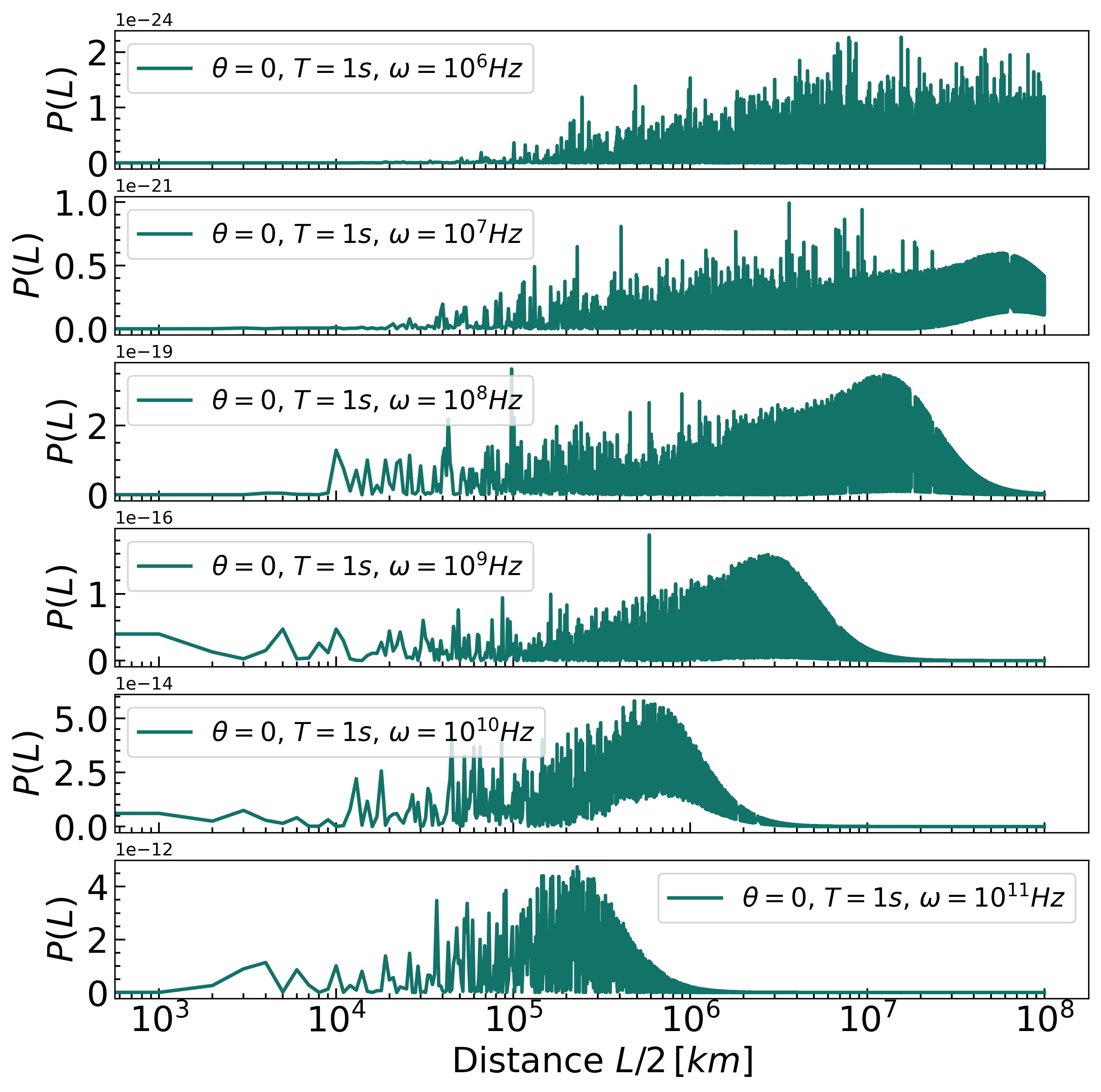}
	\includegraphics[width=0.29\linewidth]{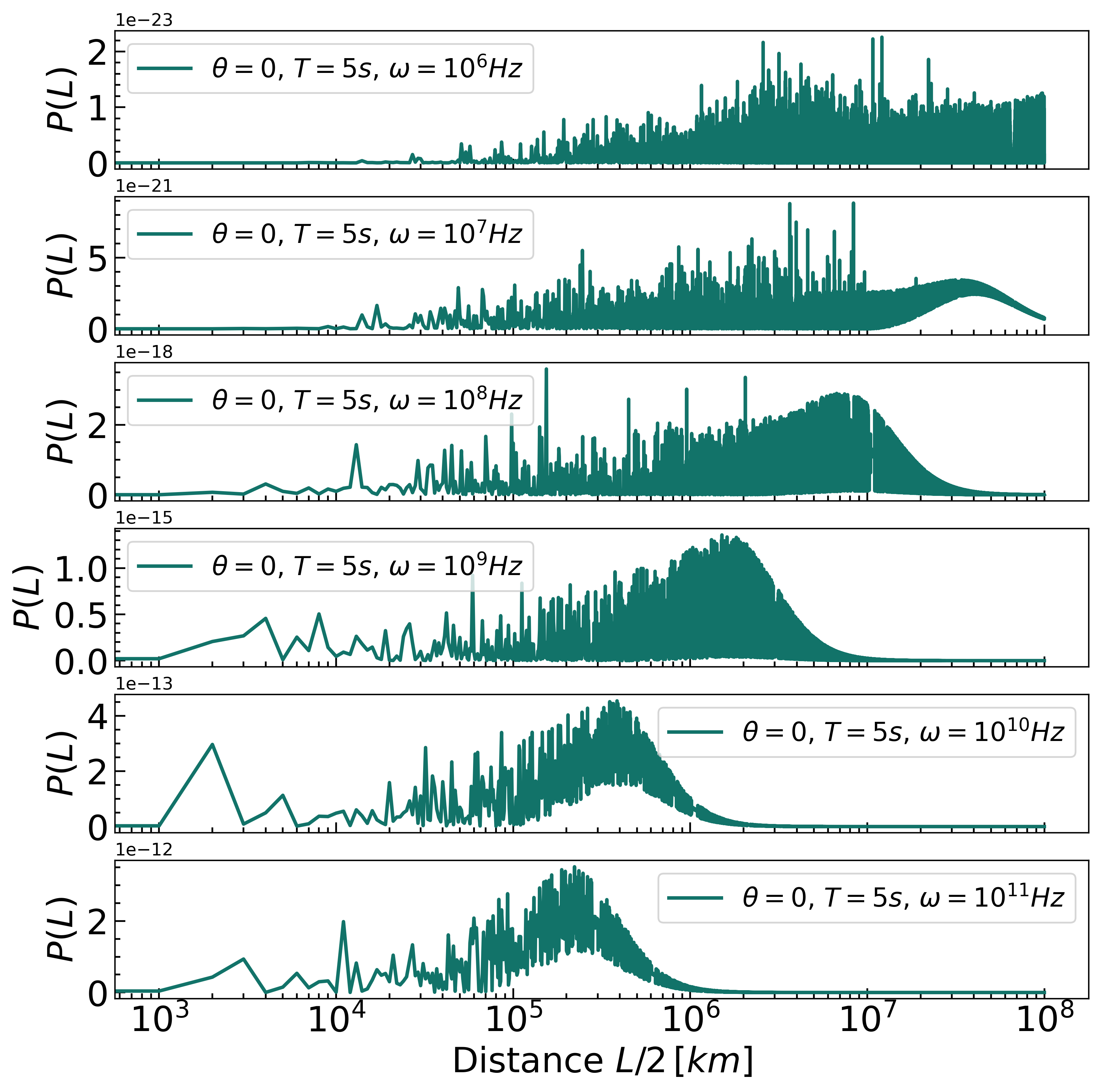}
	\includegraphics[width=0.29\linewidth]{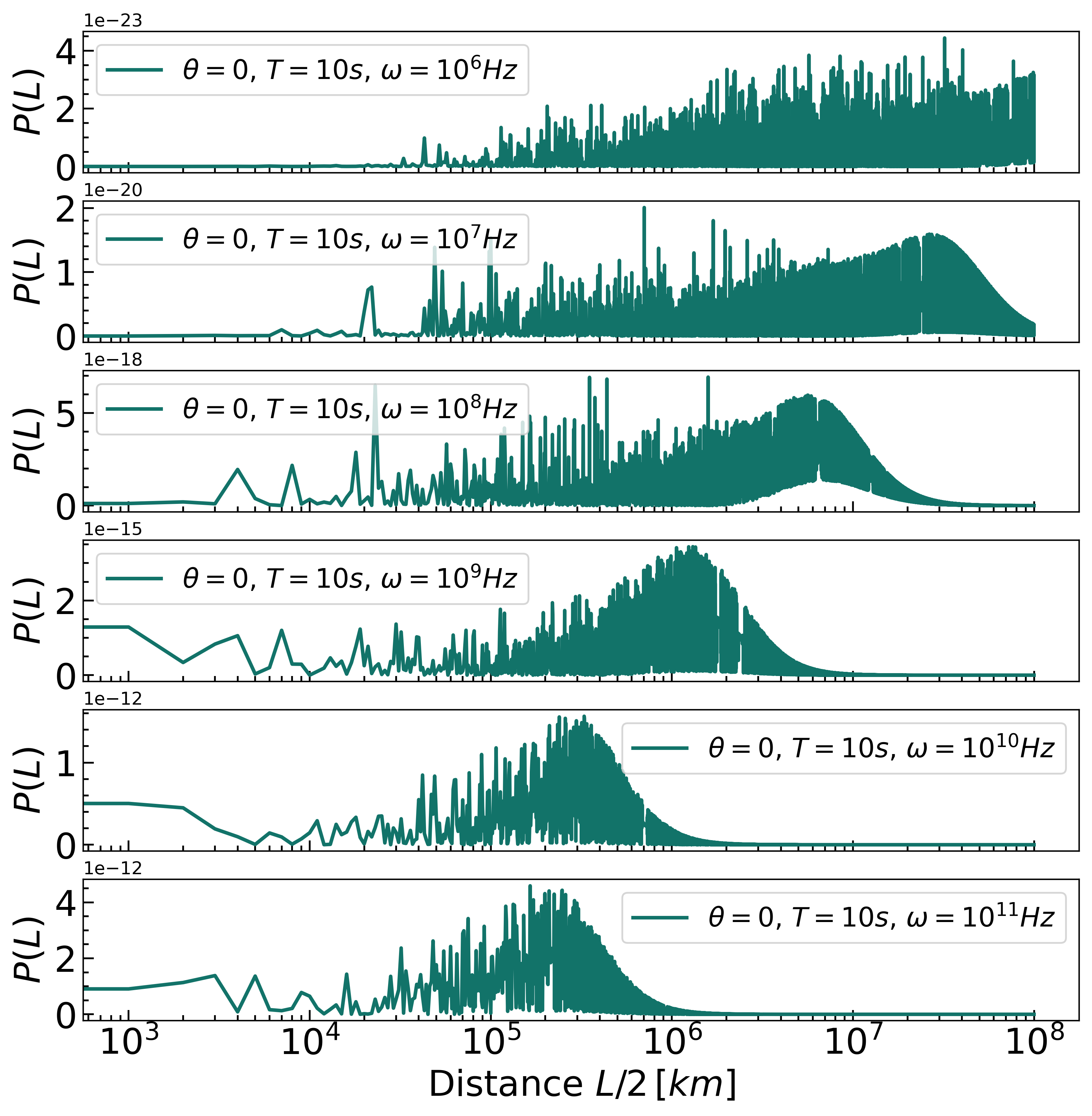}
	\includegraphics[width=0.29\linewidth]{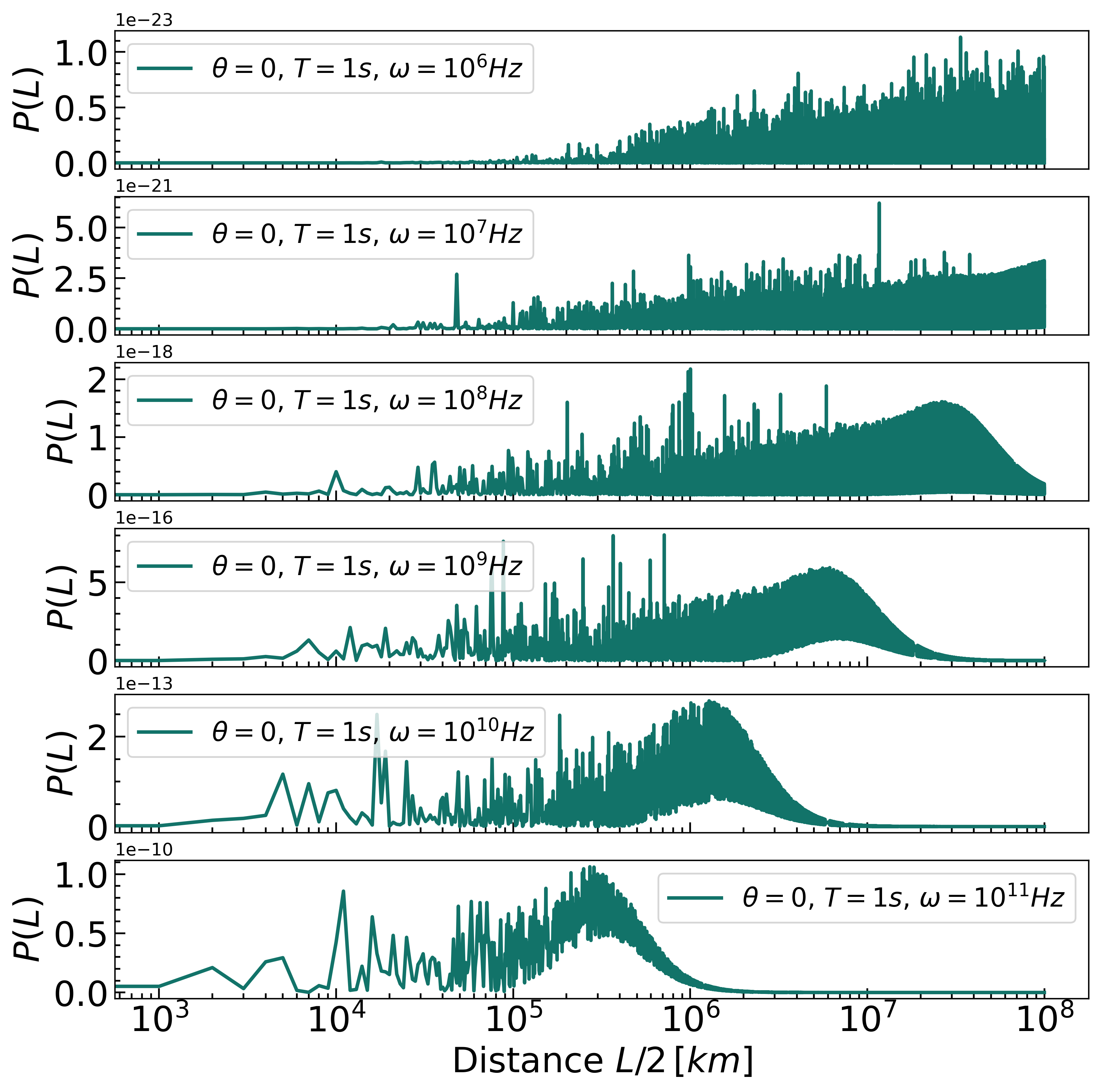}
	\includegraphics[width=0.29\linewidth]{pl-times-5.png}
	\includegraphics[width=0.29\linewidth]{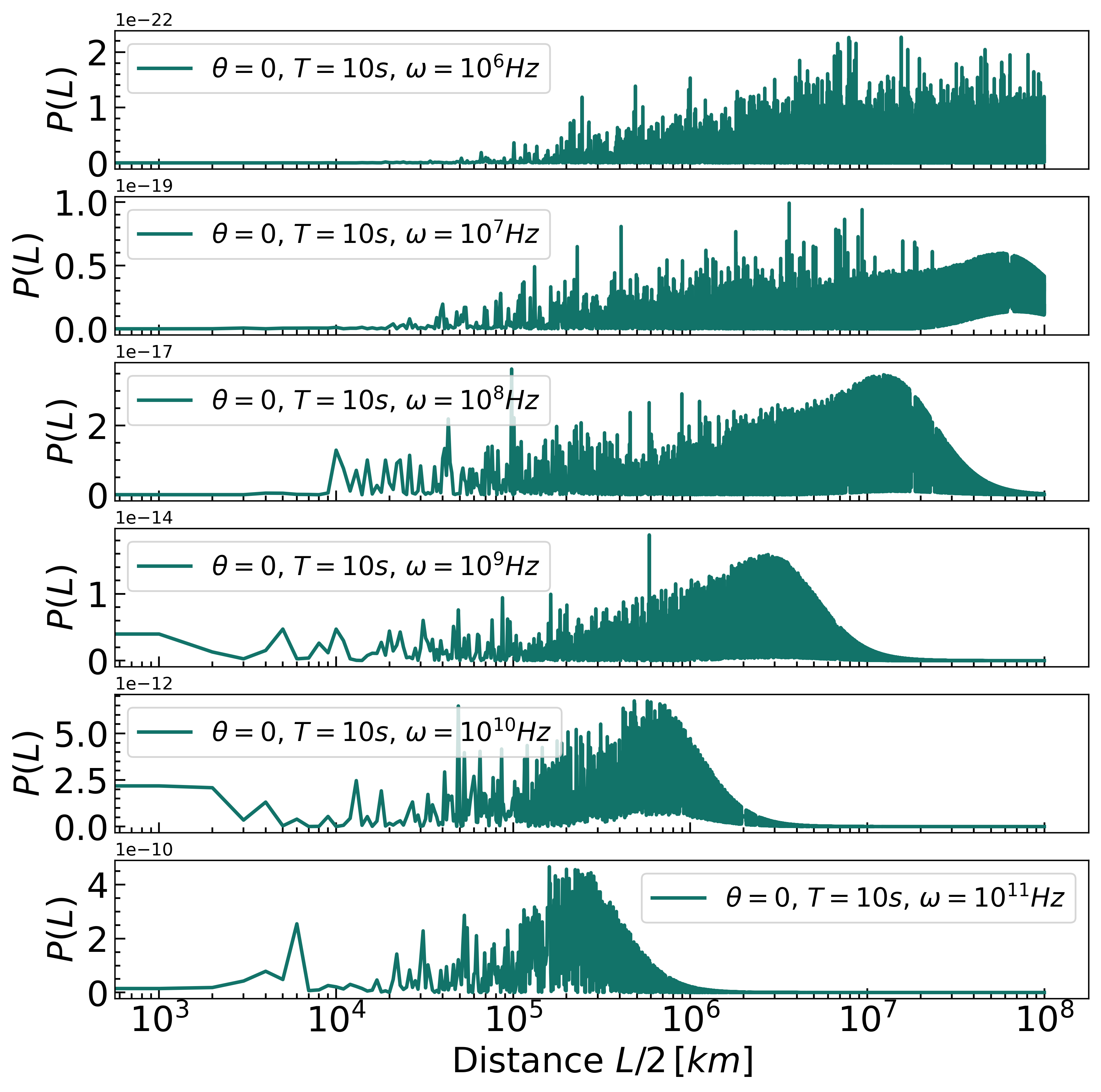}
	\includegraphics[width=0.29\linewidth]{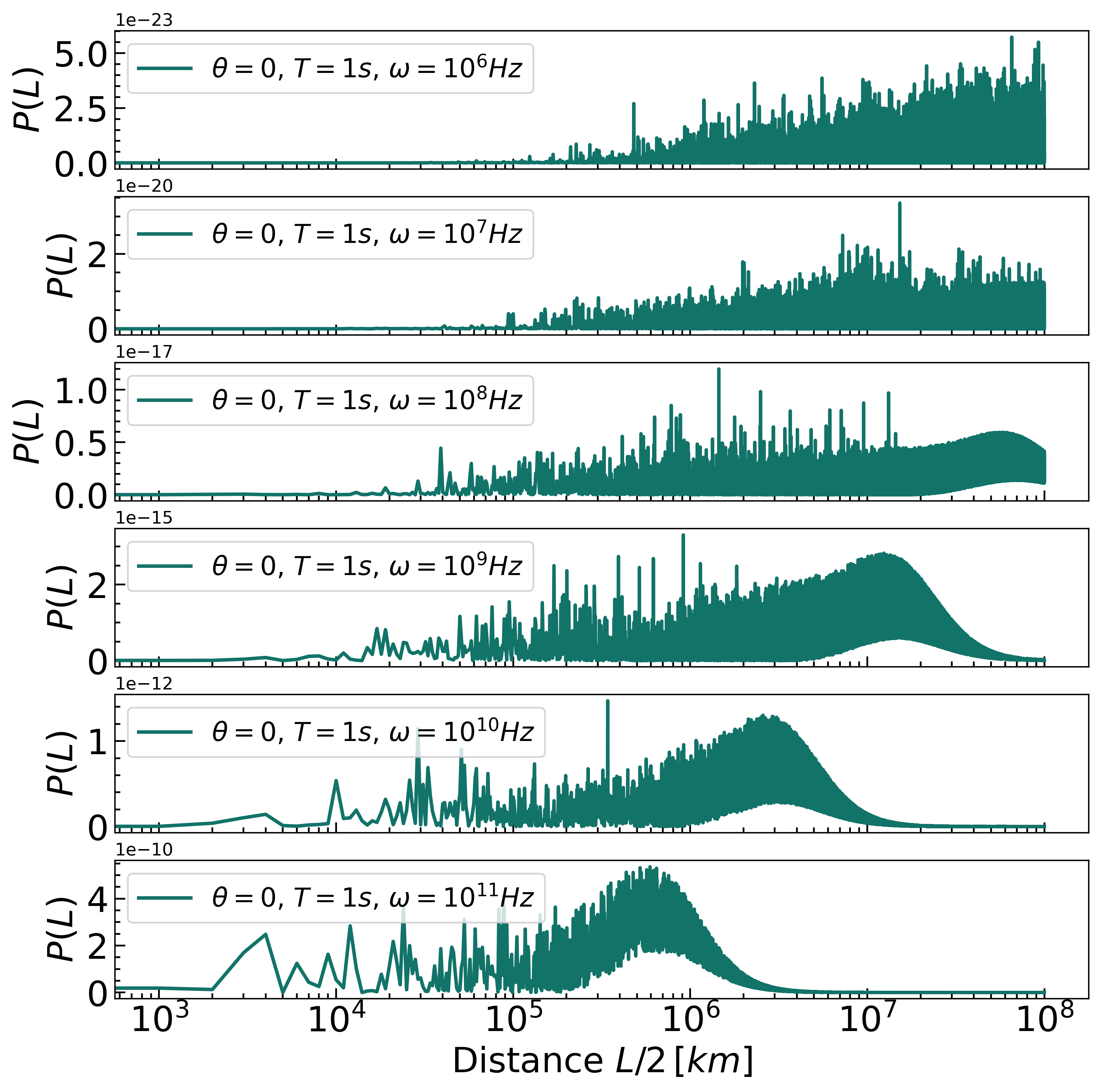}
	\includegraphics[width=0.29\linewidth]{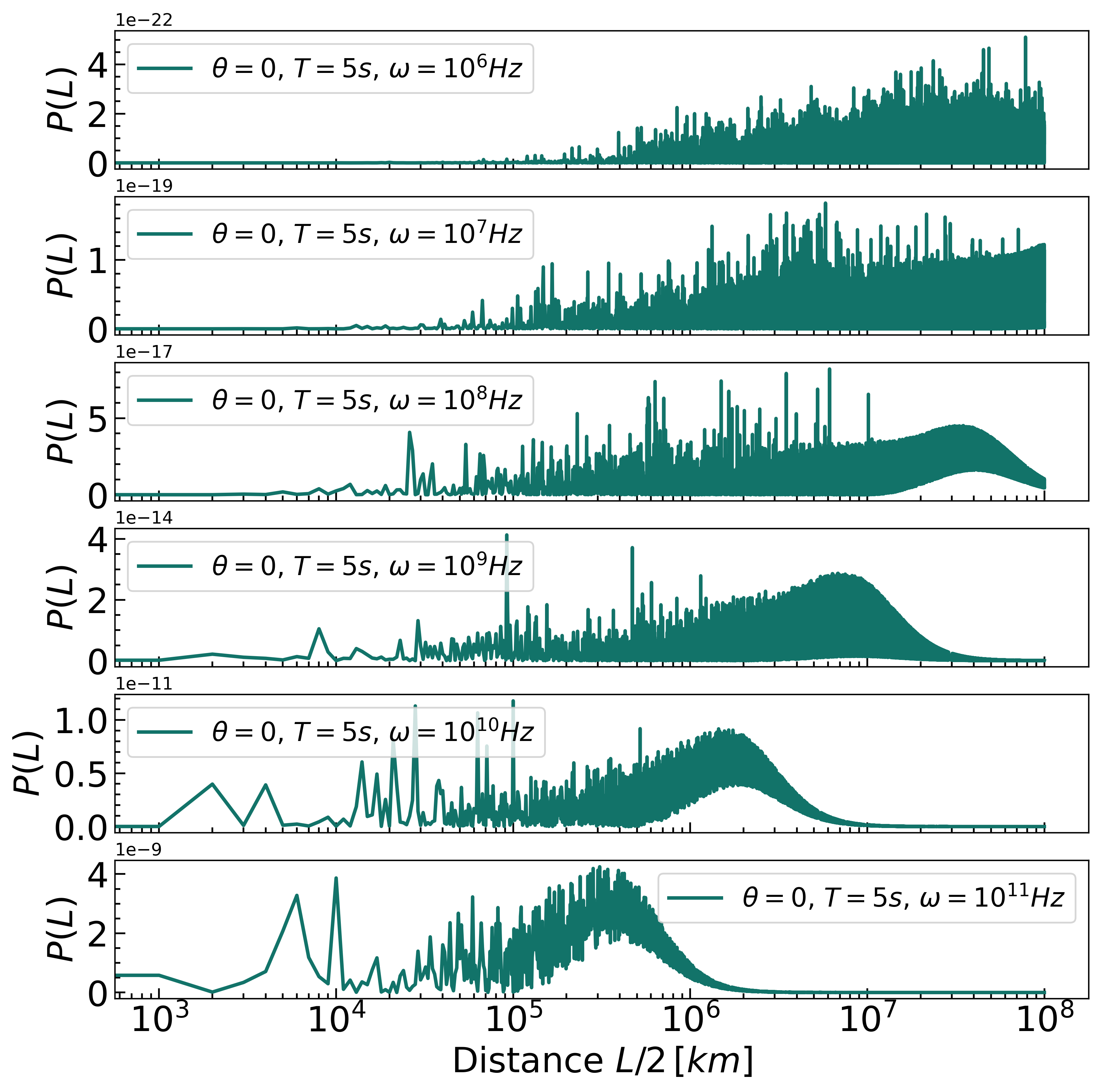}
	\includegraphics[width=0.29\linewidth]{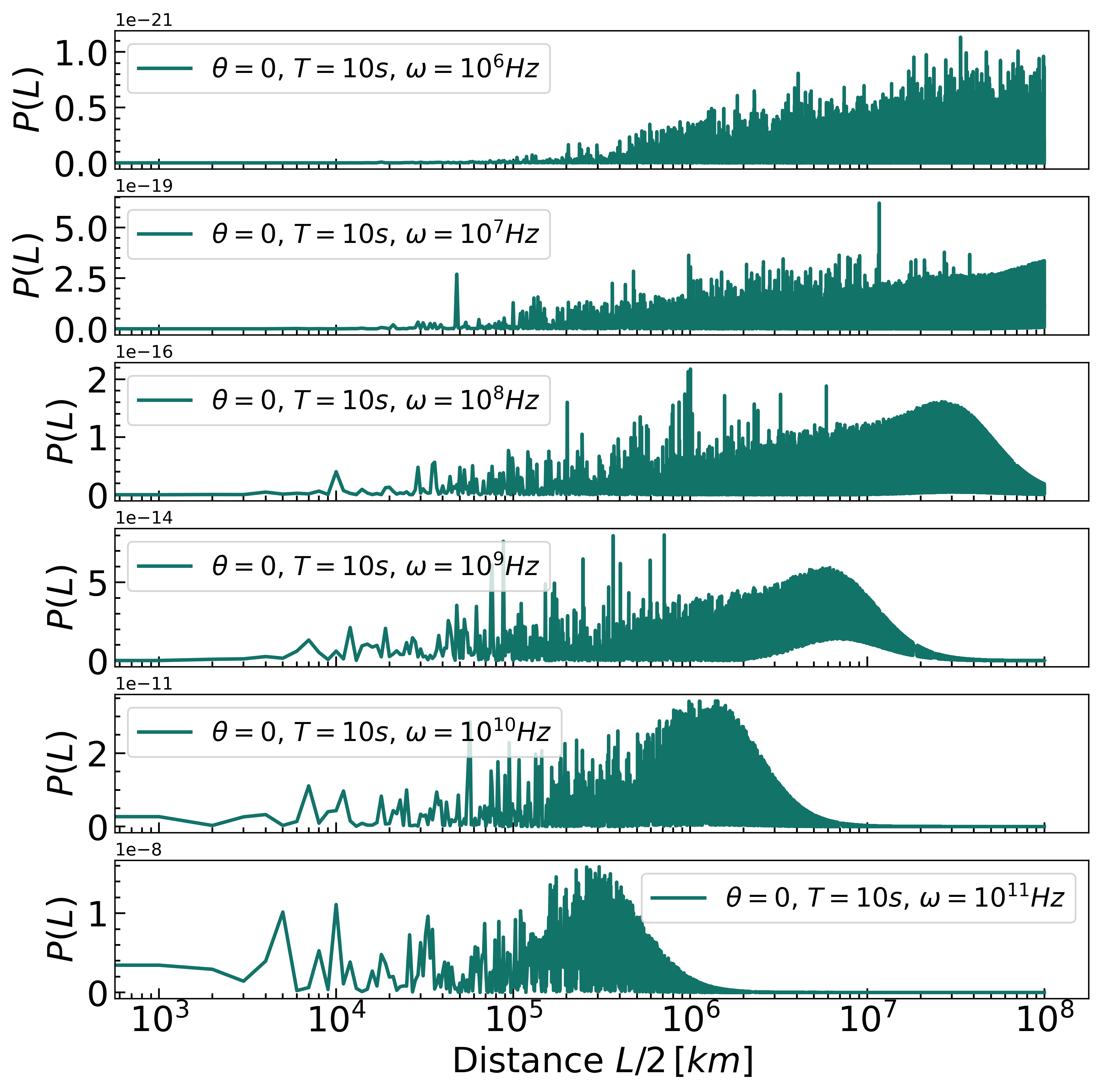}
	\caption{The conversion probability $P_{g \rightarrow \gamma}^{\mathrm{coherence}}(L,\omega,\theta=0)$ of cross-polarization (``$\times$"-polarization) of the GWs traveling different distances in magnetars and pulsars. We have assumed the following parameters for magnetars and pulsars: (top panel) from left to right: $B=10^{13}\mathrm{~Gauss}$, and $T=1\mathrm{~sec}$; $B=10^{13}\mathrm{~Gauss}$, and $T=5\mathrm{~sec}$; $B=10^{13}\mathrm{~Gauss}$, and $T=10\mathrm{~sec}$. (Middle panel) from left to right: $B=10^{14}\mathrm{~Gauss}$, and $T=1\mathrm{~sec}$; $B=10^{14}\mathrm{~Gauss}$, and $T=5\mathrm{~sec}$; $B=10^{14}\mathrm{~Gauss}$, and $T=10\mathrm{~sec}$. (Bottom panel) from left to right: $B=10^{15}\mathrm{~Gauss}$, and $T=1\mathrm{~sec}$; $B=10^{15}\mathrm{~Gauss}$, and $T=5\mathrm{~sec}$; $B=10^{15}\mathrm{~Gauss}$, and $T=10\mathrm{~sec}$.}
	\label{figapp:conversion-probability-times}
\end{figure*}

\begin{figure*}
	\centering
	\includegraphics[width=0.29\linewidth]{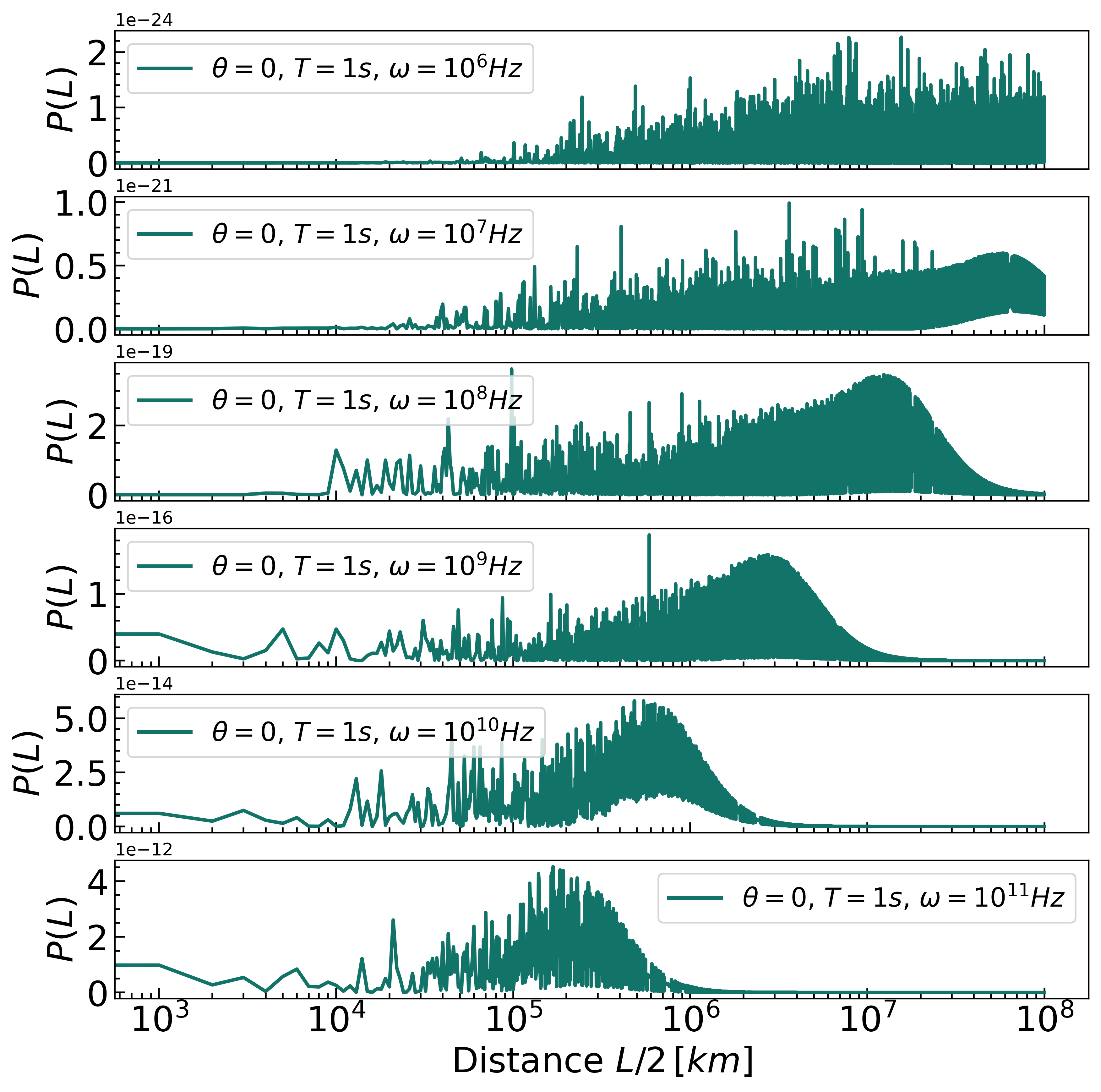}
	\includegraphics[width=0.29\linewidth]{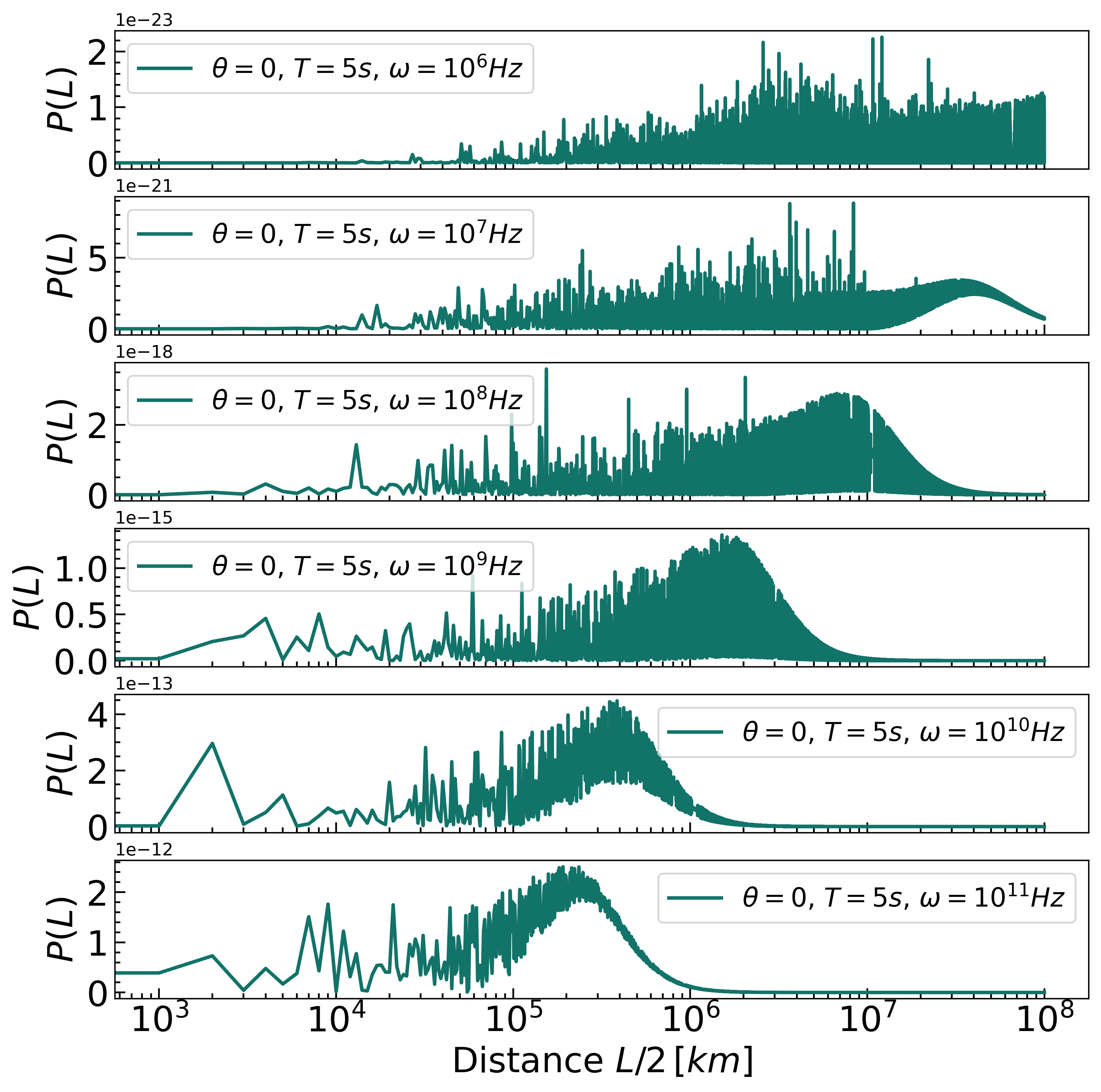}
	\includegraphics[width=0.29\linewidth]{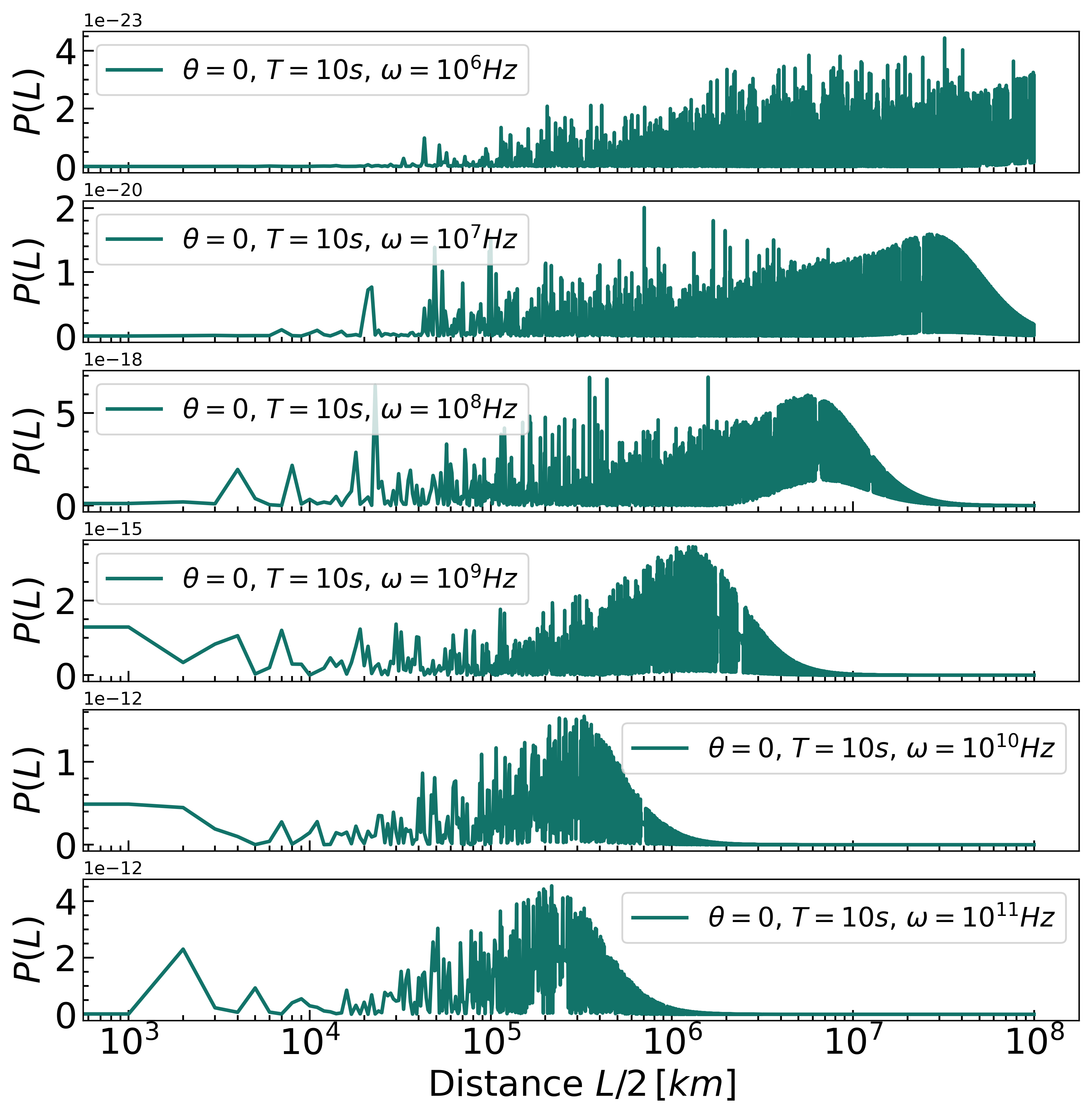}
	\includegraphics[width=0.29\linewidth]{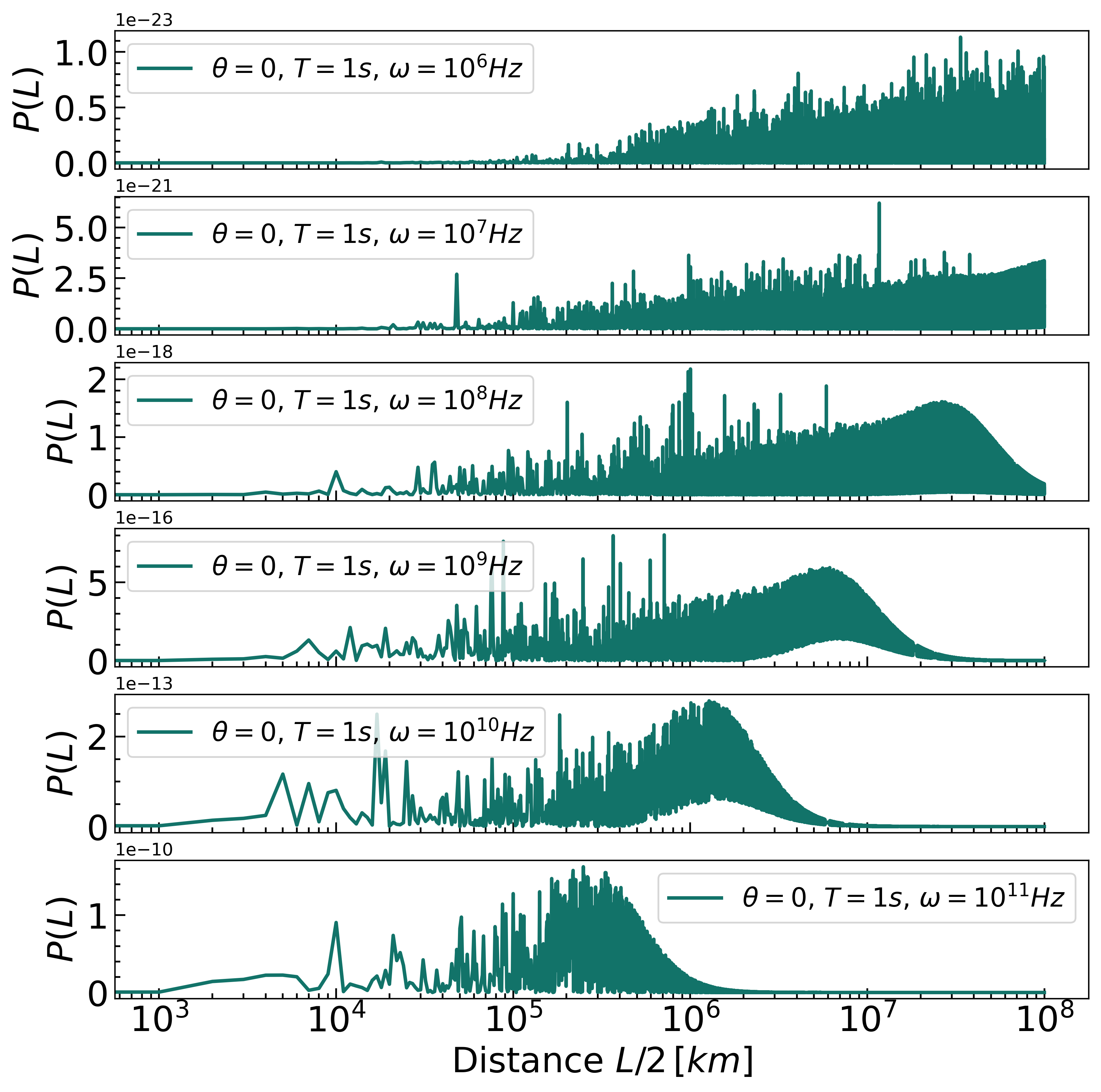}
	\includegraphics[width=0.29\linewidth]{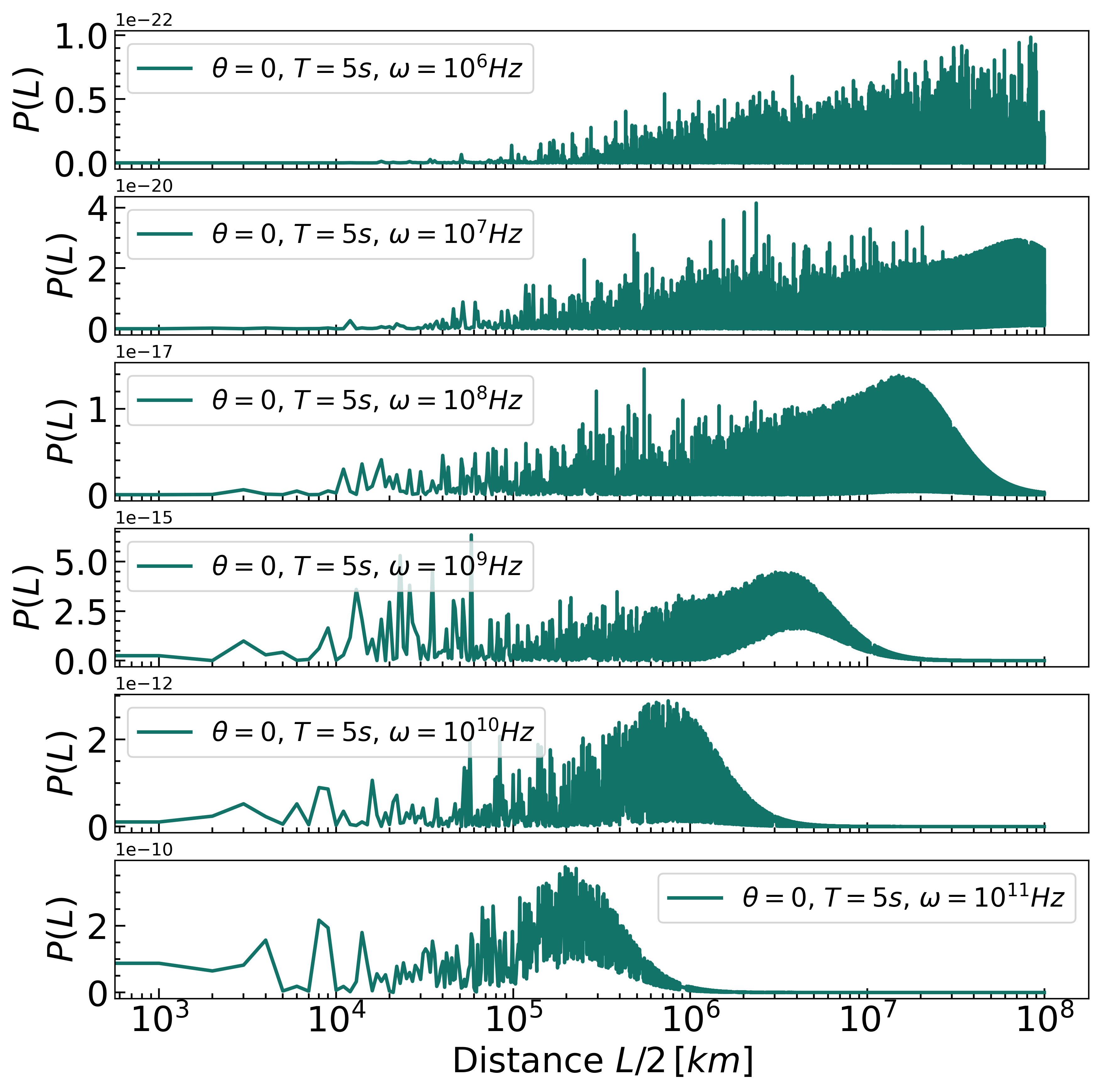}
	\includegraphics[width=0.29\linewidth]{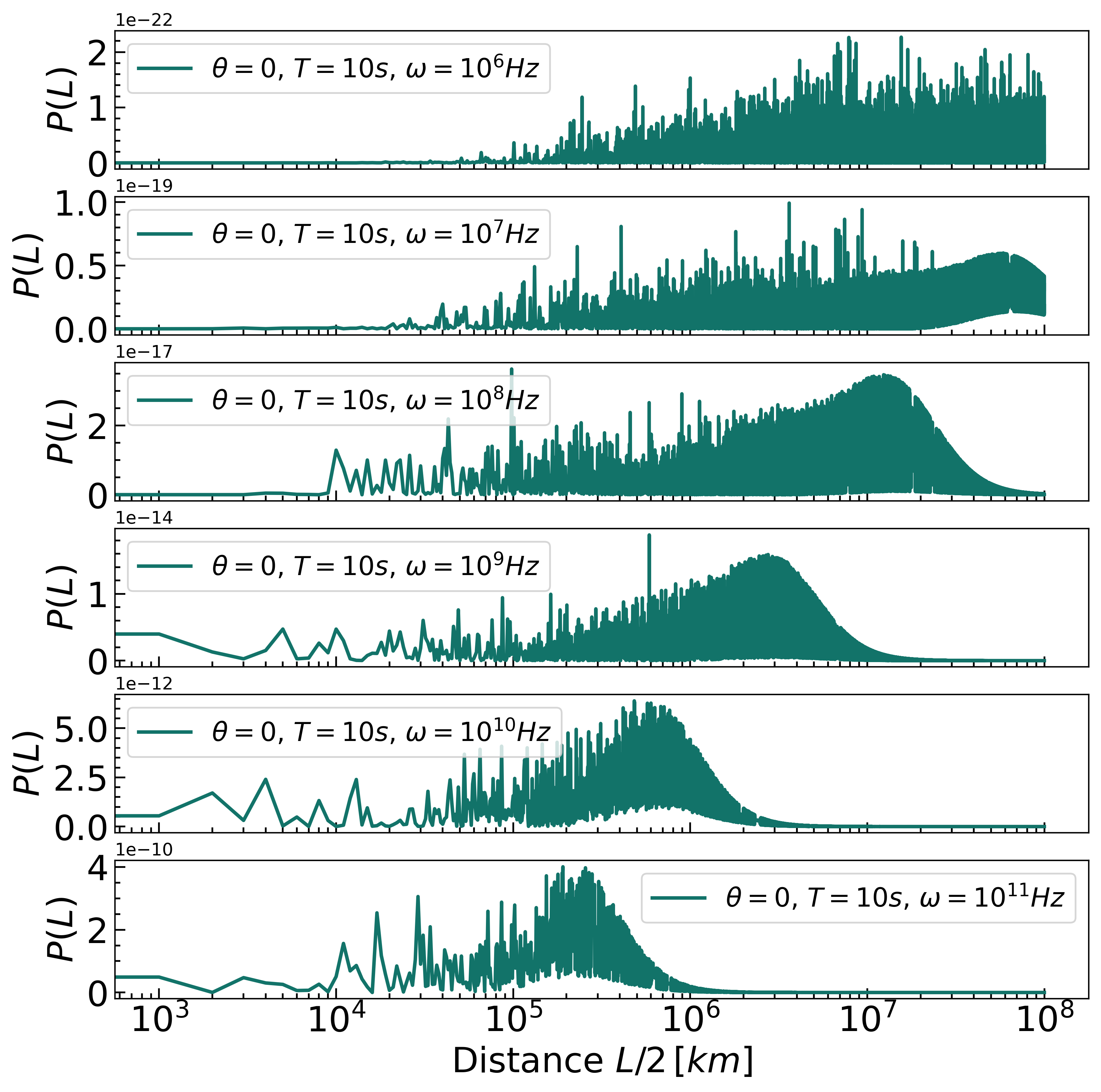}
	\includegraphics[width=0.29\linewidth]{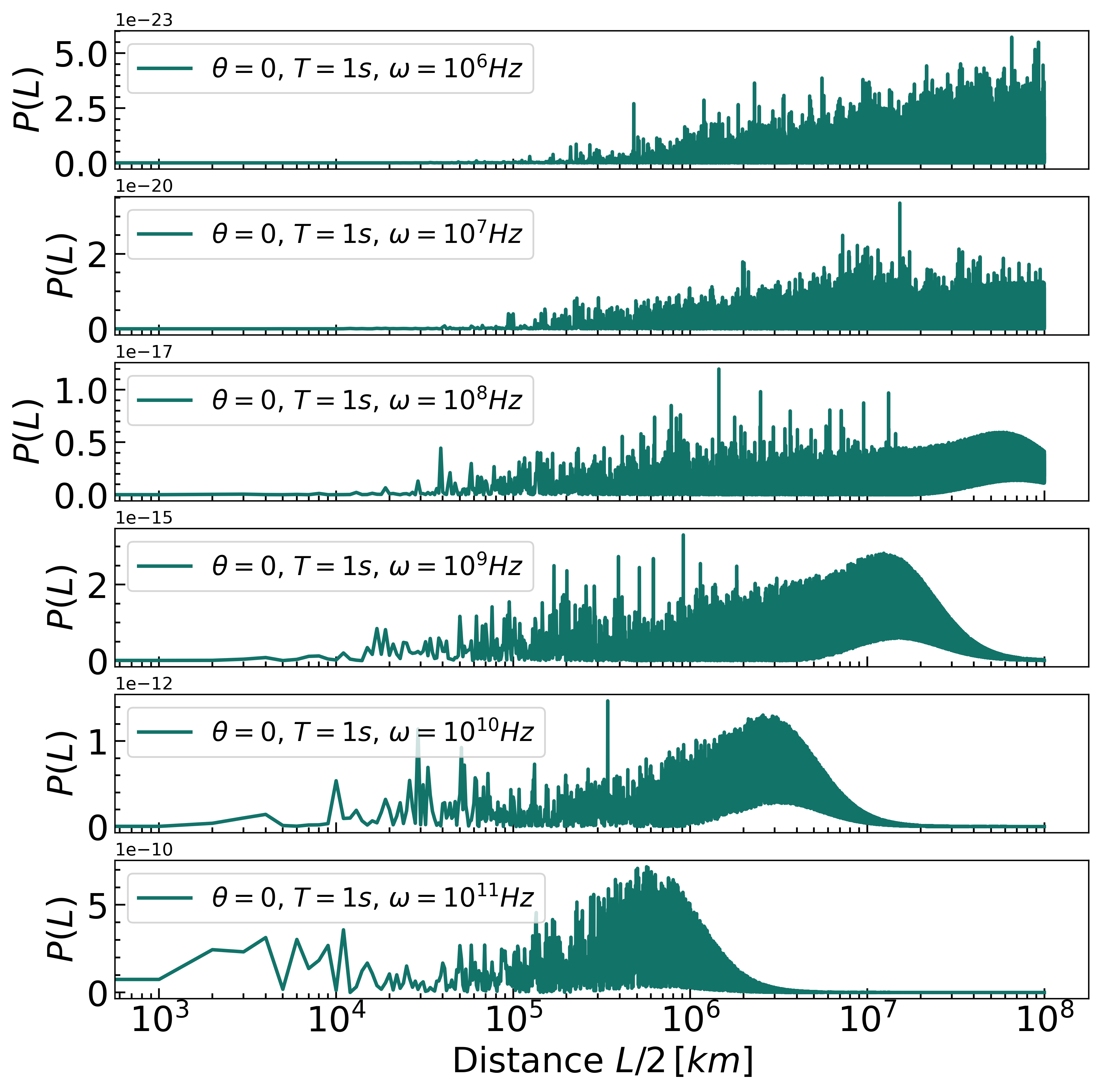}
	\includegraphics[width=0.29\linewidth]{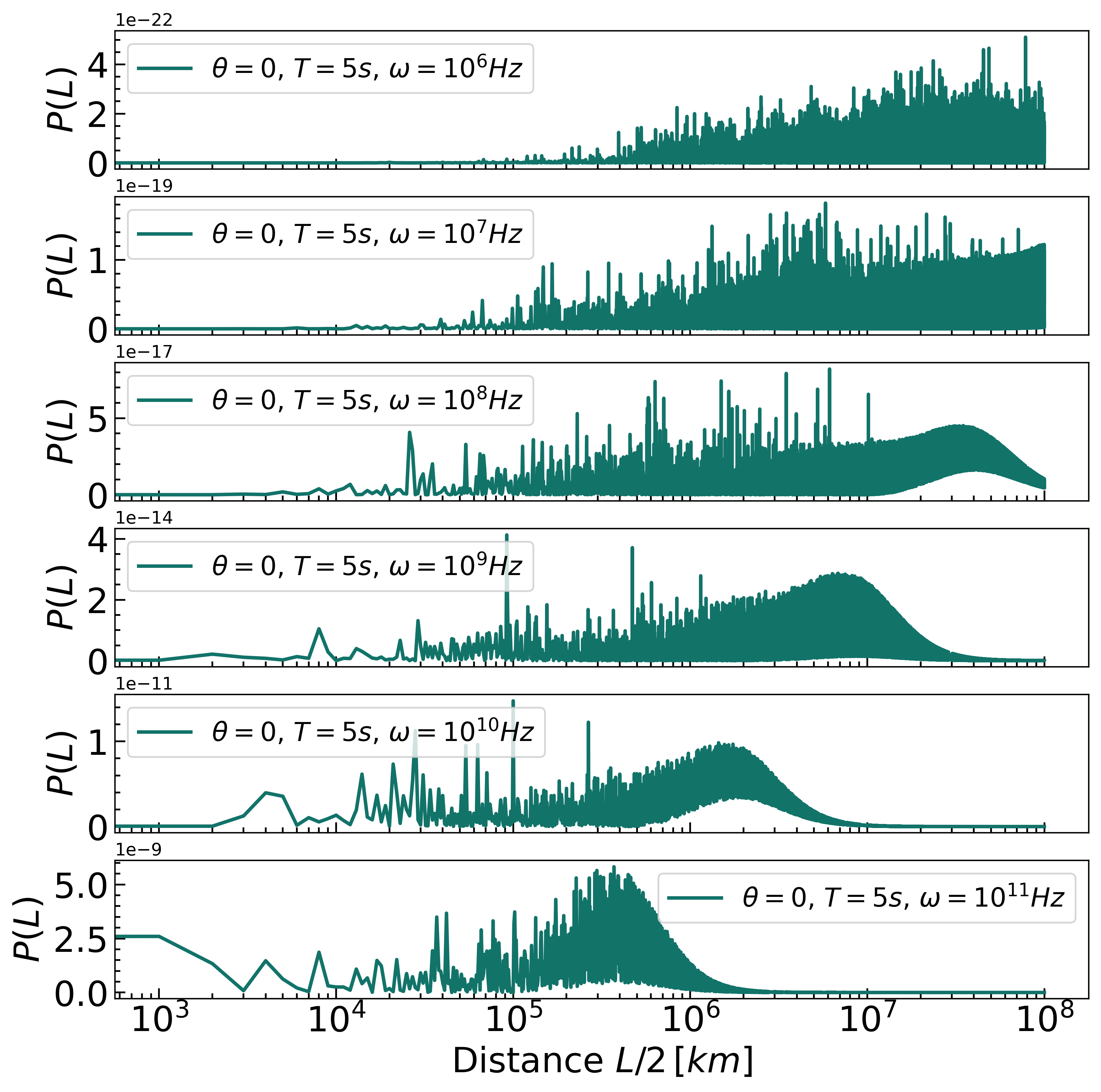}
	\includegraphics[width=0.29\linewidth]{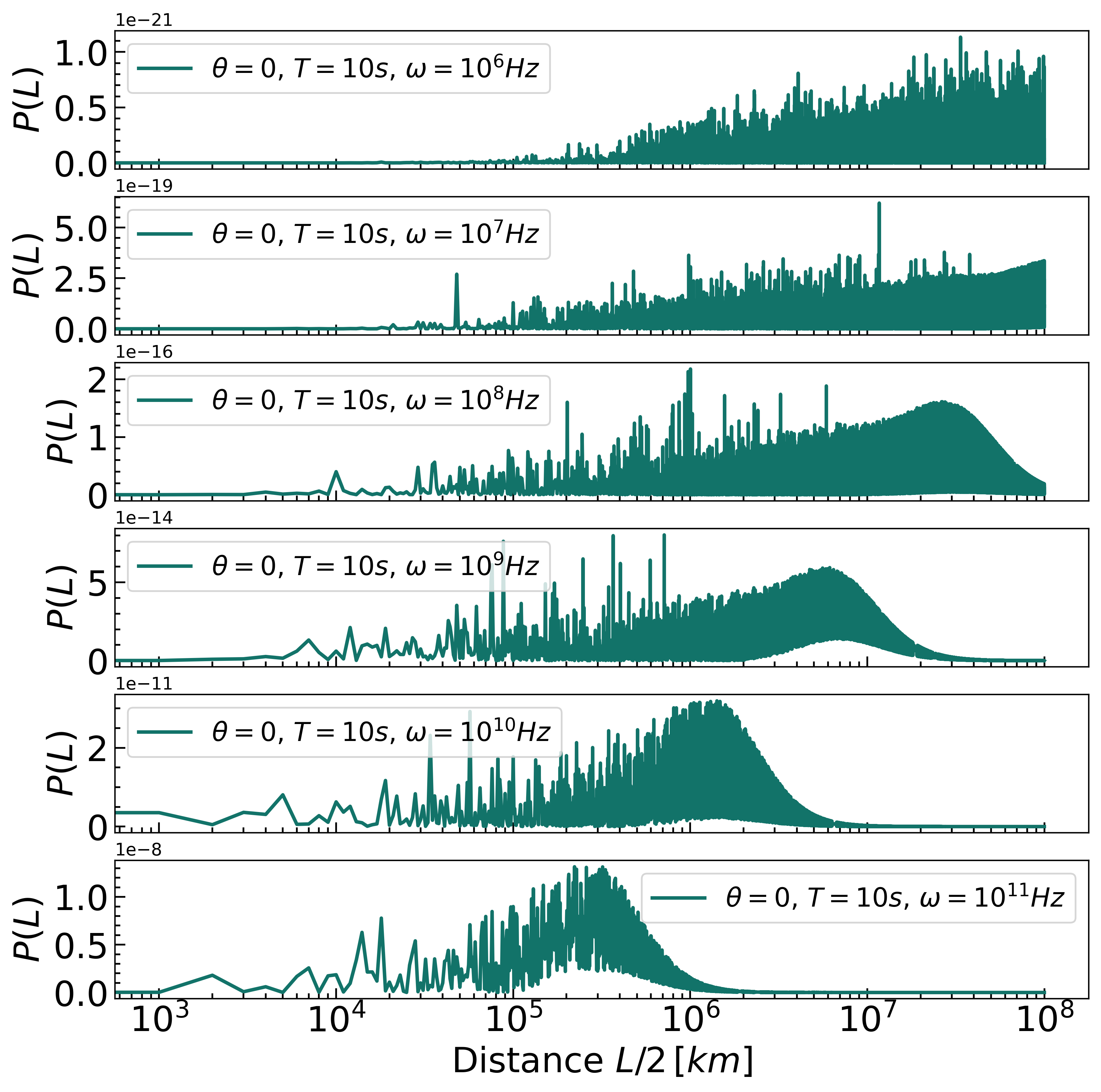}
	\caption{The conversion probability $P_{g \rightarrow \gamma}^{\mathrm{coherence}}(L,\omega,\theta=0)$ of plus-polarization (``$+$"-polarization) of the GWs traveling different distances in magnetars or pulsars. We have assumed the following parameters for magnetars and pulsars: (top panel) from left to right: $B=10^{13}\mathrm{~Gauss}$, and $T=1\mathrm{~sec}$; $B=10^{13}\mathrm{~Gauss}$, and $T=5\mathrm{~sec}$; $B=10^{13}\mathrm{~Gauss}$, and $T=10\mathrm{~sec}$. (Middle panel) from left to right: $B=10^{14}\mathrm{~Gauss}$, and $T=1\mathrm{~sec}$; $B=10^{14}\mathrm{~Gauss}$, and $T=5\mathrm{~sec}$; $B=10^{14}\mathrm{~Gauss}$, and $T=10\mathrm{~sec}$. (Bottom panel) from left to right: $B=10^{15}\mathrm{~Gauss}$, and $T=1\mathrm{~sec}$; $B=10^{15}\mathrm{~Gauss}$, and $T=5\mathrm{~sec}$; $B=10^{15}\mathrm{~Gauss}$, and $T=10\mathrm{~sec}$.}
	\label{figapp:conversion-probability-plus}
\end{figure*}

\begin{figure*}
	\centering
	\includegraphics[width=0.29\linewidth]{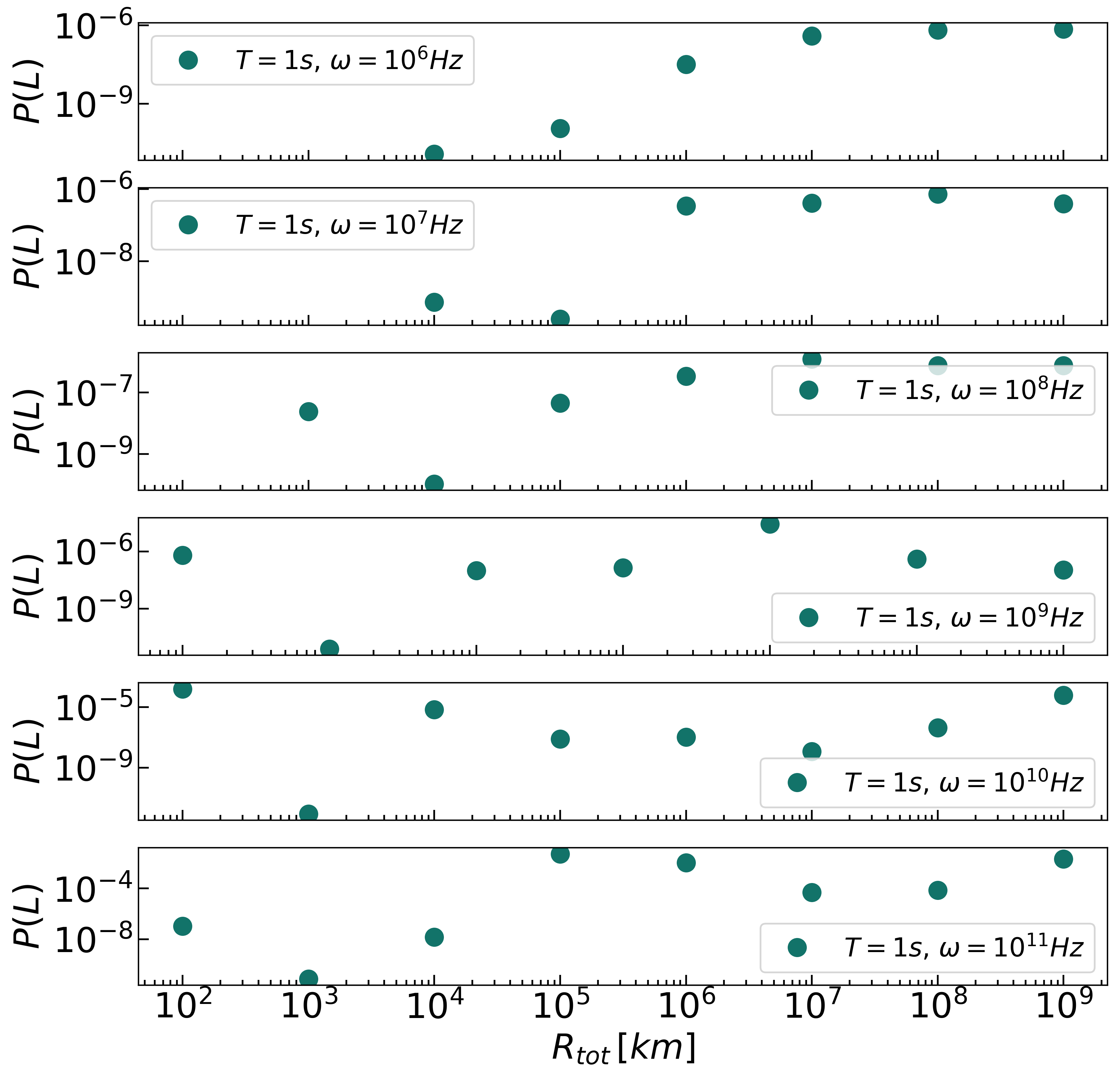}
	\includegraphics[width=0.29\linewidth]{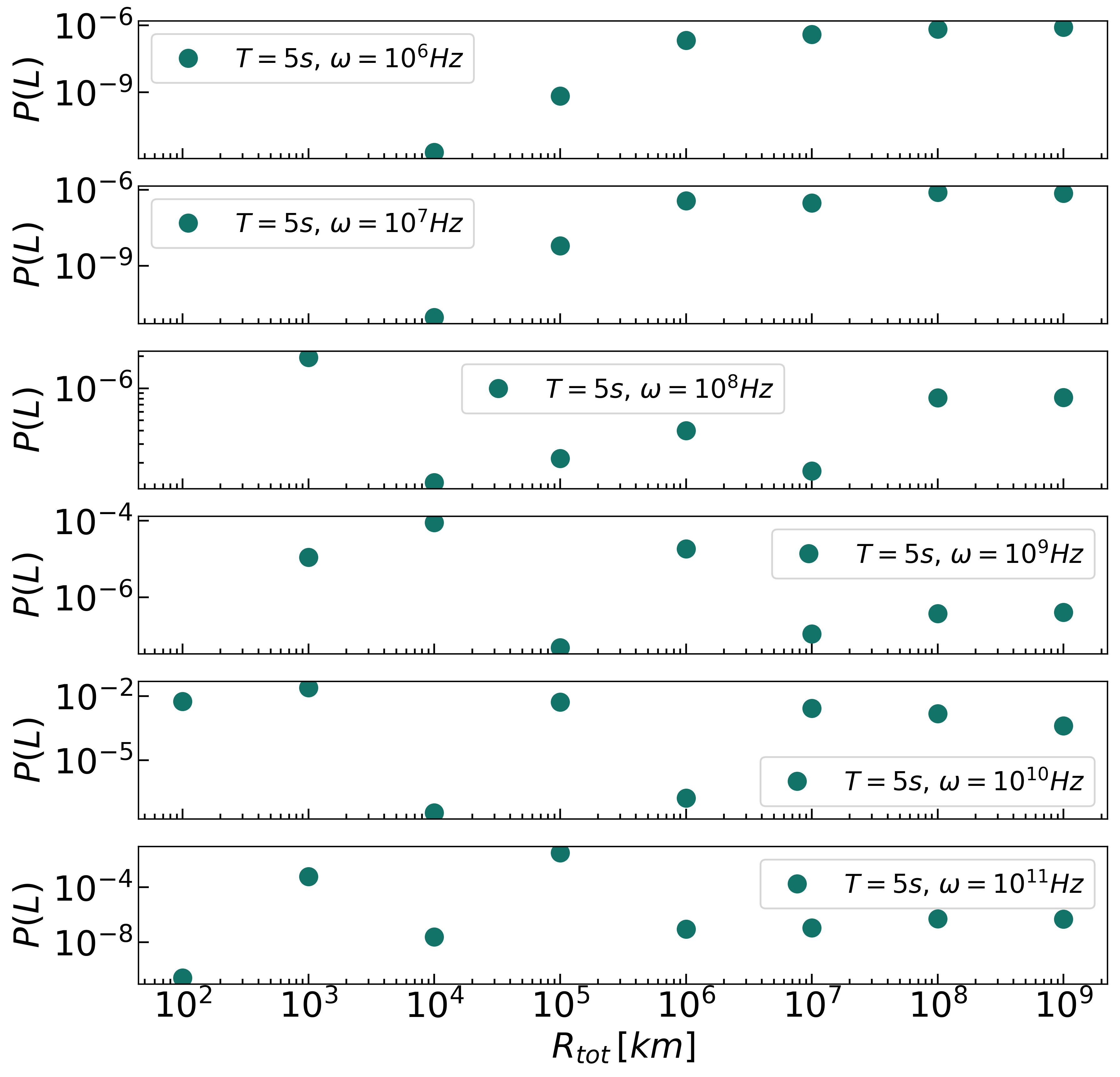}
	\includegraphics[width=0.29\linewidth]{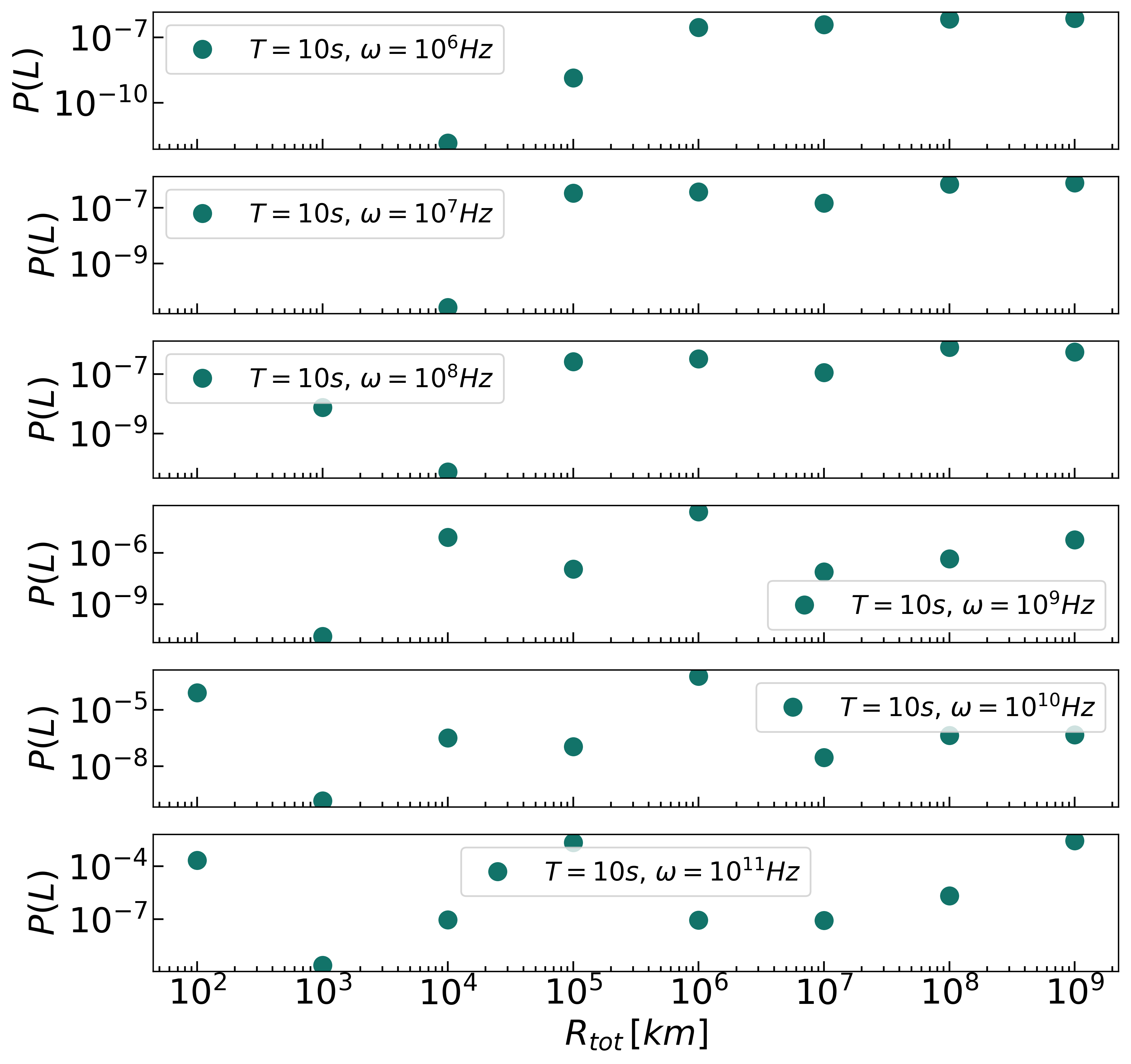}
	\includegraphics[width=0.29\linewidth]{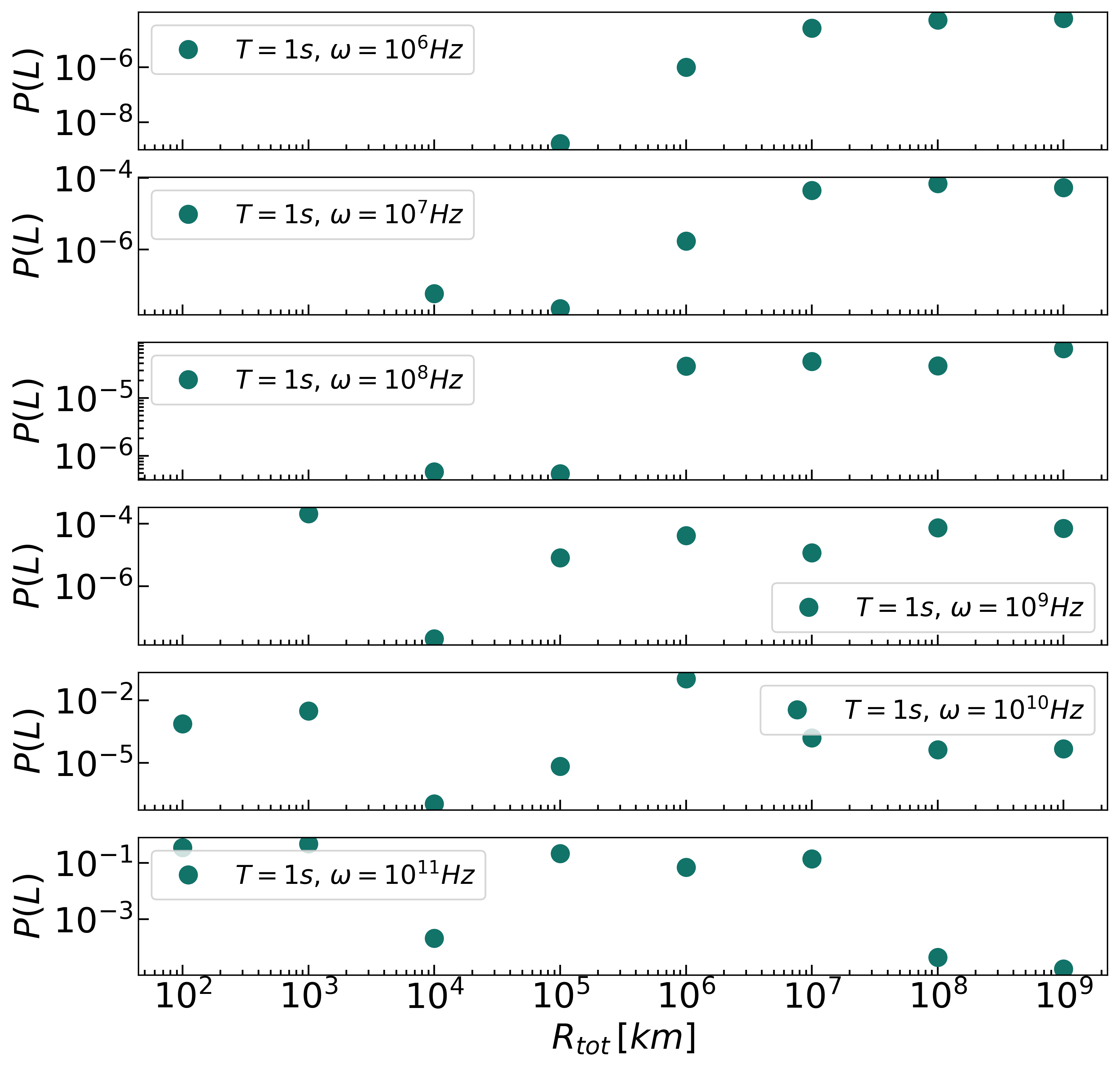}
	\includegraphics[width=0.29\linewidth]{total-5.png}
	\includegraphics[width=0.29\linewidth]{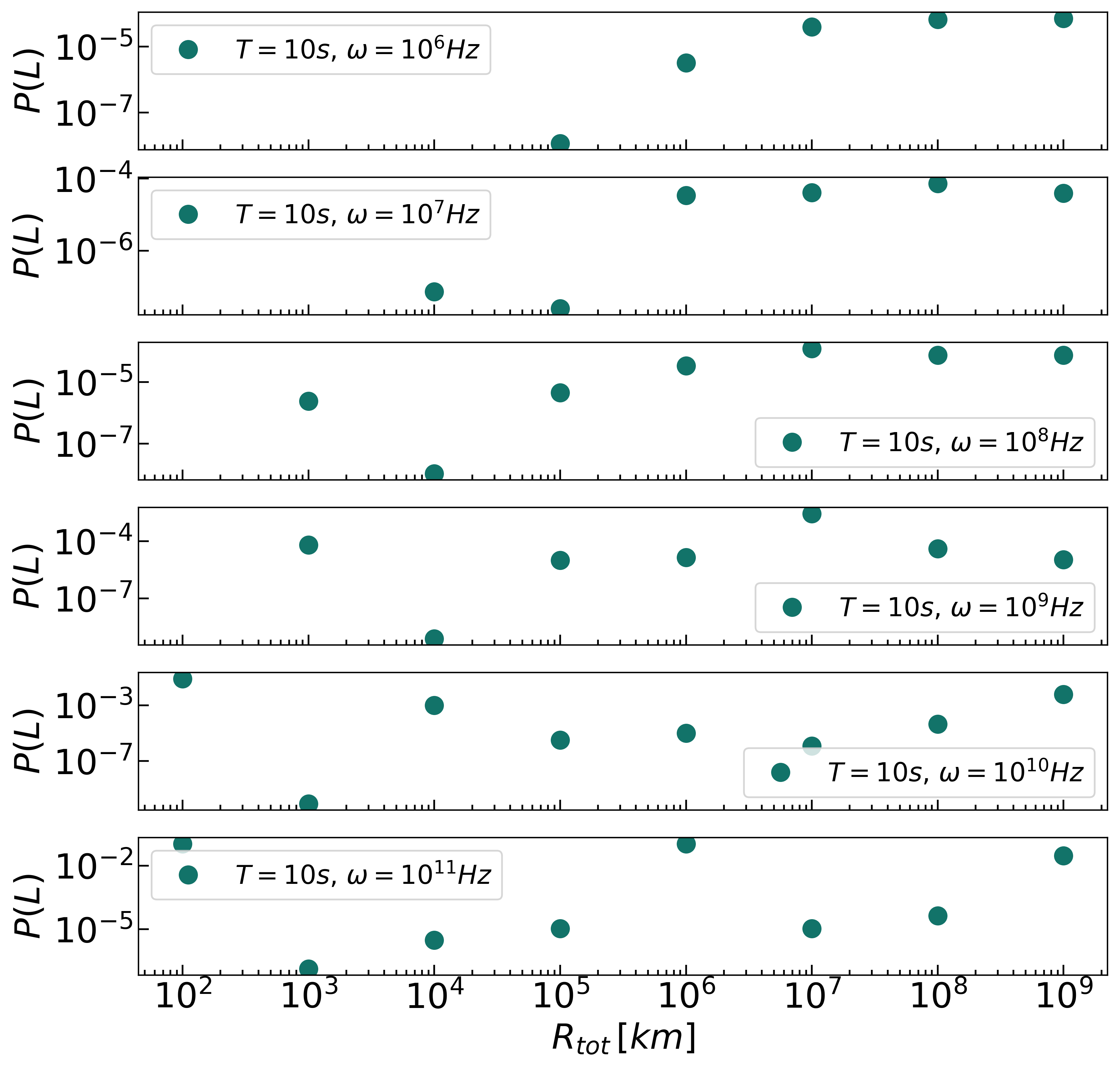}
	\includegraphics[width=0.29\linewidth]{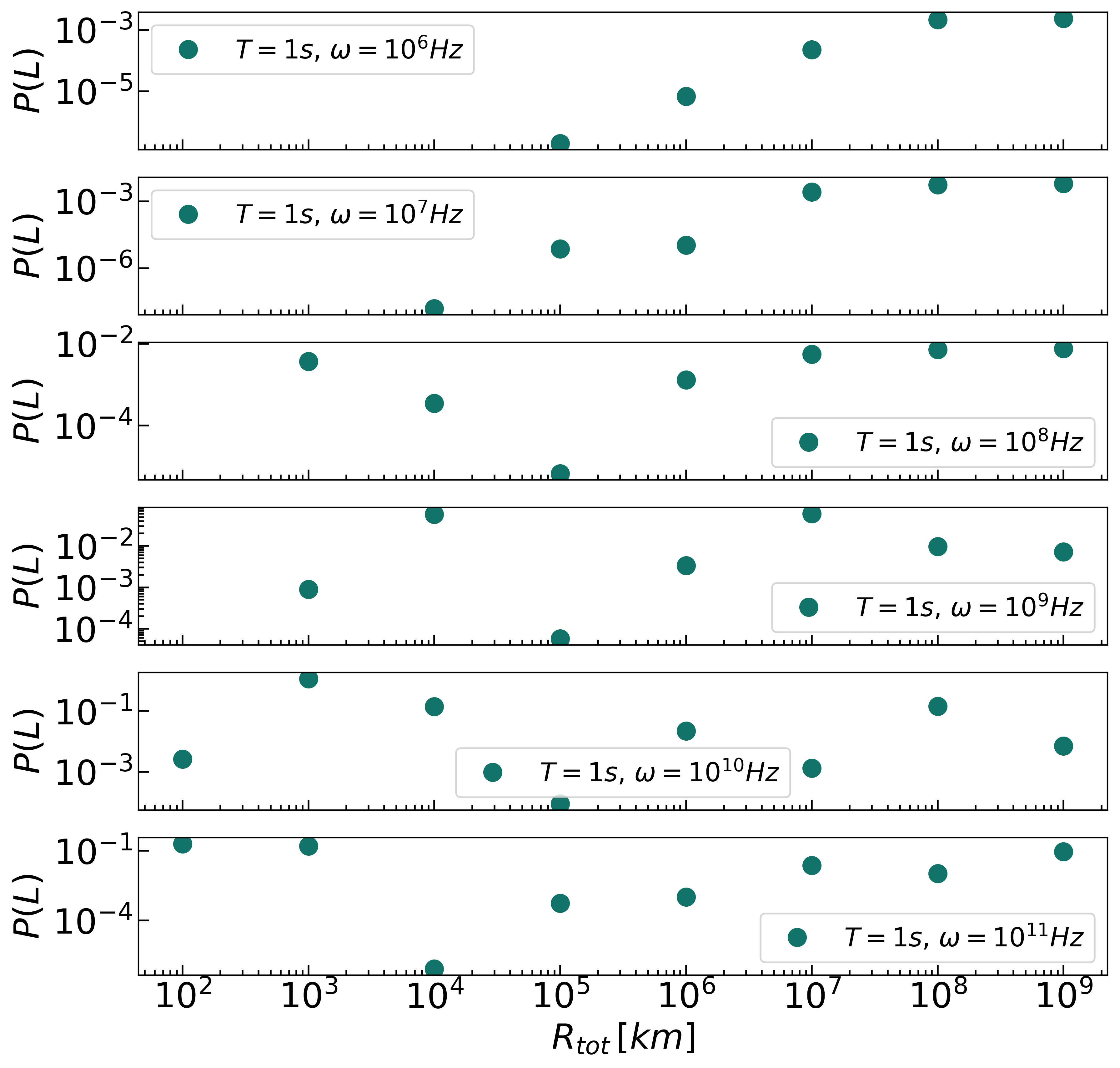}
	\includegraphics[width=0.29\linewidth]{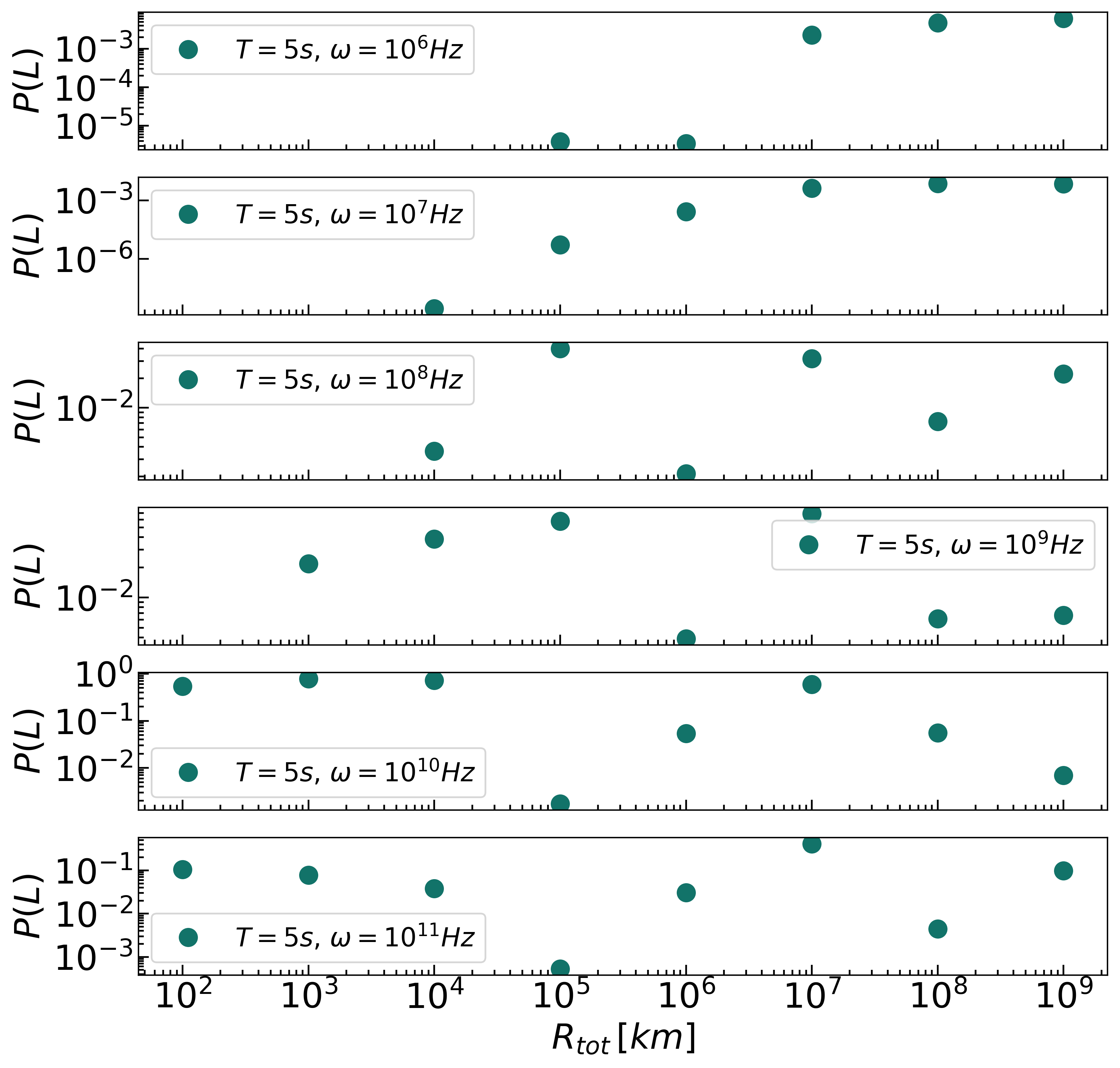}
	\includegraphics[width=0.29\linewidth]{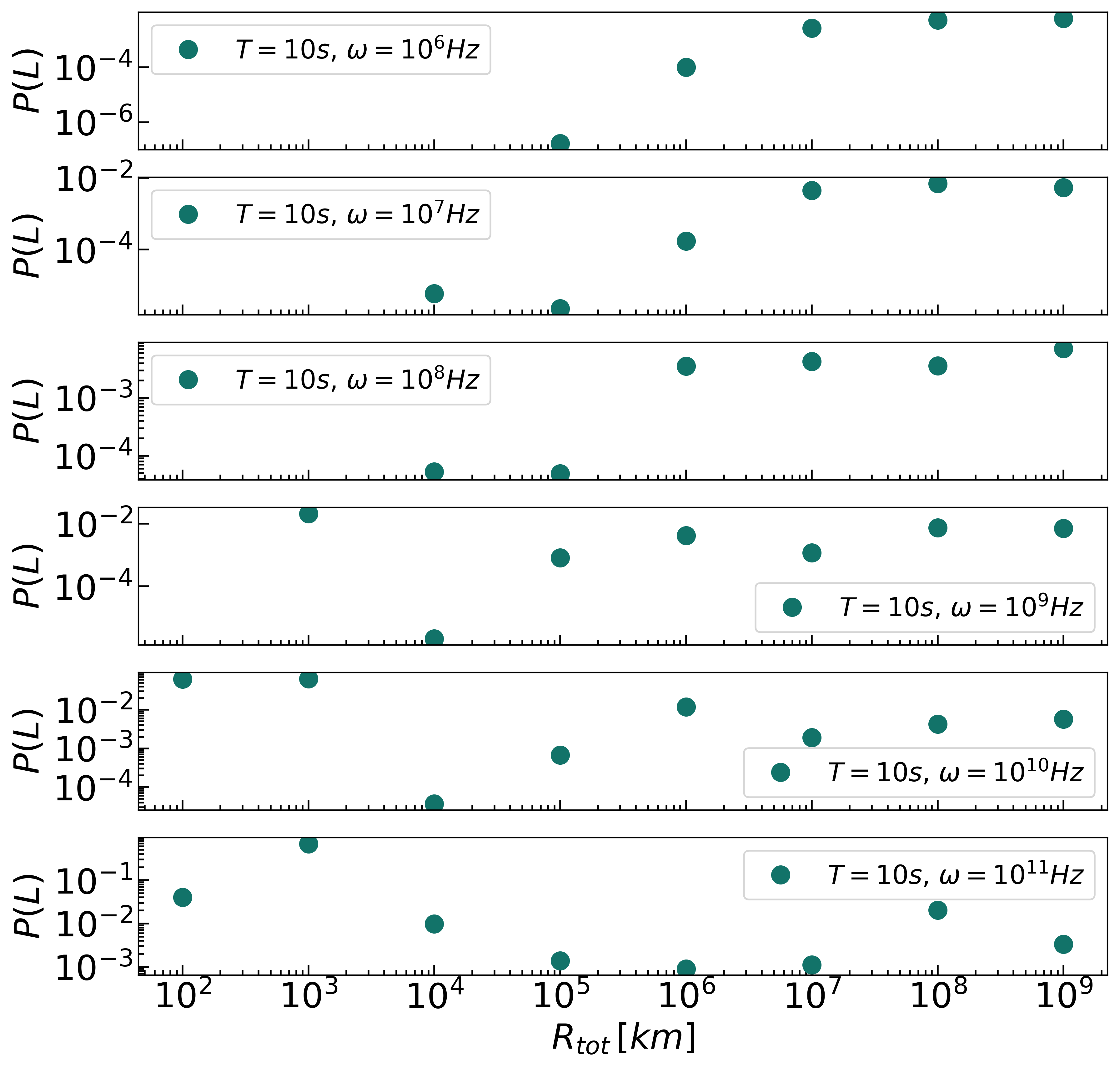}
	\caption{The total conversion probability $P_{g \rightarrow \gamma}^{\mathrm{coherence}}(L)$ of the GWs traveling at different $R_{tot}$ in magnetars and pulsars. We have assumed the following parameters for magnetars and pulsars: (top panel) from left to right: $B=10^{13}\mathrm{~Gauss}$, and $T=1\mathrm{~sec}$; $B=10^{13}\mathrm{~Gauss}$, and $T=5\mathrm{~sec}$; $B=10^{13}\mathrm{~Gauss}$, and $T=10\mathrm{~sec}$. (Middle panel) from left to right: $B=10^{14}\mathrm{~Gauss}$, and $T=1\mathrm{~sec}$; $B=10^{14}\mathrm{~Gauss}$, and $T=5\mathrm{~sec}$; $B=10^{14}\mathrm{~Gauss}$, and $T=10\mathrm{~sec}$. (Bottom panel) from left to right: $B=10^{15}\mathrm{~Gauss}$, and $T=1\mathrm{~sec}$; $B=10^{15}\mathrm{~Gauss}$, and $T=5\mathrm{~sec}$; $B=10^{15}\mathrm{~Gauss}$, and $T=10\mathrm{~sec}$.}
	\label{figapp:conversion-probability-total}
\end{figure*}

\normalem
\bibliography{VHFGW}{}
\bibliographystyle{aasjournalv7}

\end{document}